\documentclass[useAMS,usenatbib]{mn2e}
\usepackage{graphicx}
\usepackage{color}

\newcommand{\Msun}{\ensuremath{\mathrm{M}_{\sun}}}
\newcommand{\Lsun}{\ensuremath{\mathrm{L}_{\sun}}}
\newcommand{\kms}{km~s$^{-1}$}
\newcommand{\hi}{H\,{\sc i}}

\newcommand{\vdivsigma}{\ensuremath{V_{\rm dp} / \sigma}}
\newcommand{\vphidivsigma}{\ensuremath{V_{\varphi} / \langle\sigma\rangle}}

\newcommand{\vmaxsig}{\ensuremath{V_{\rm max} / \sigma_{0}}}
\newcommand{\vmaxsigstar}{\ensuremath{(V_{\rm max} / \sigma_{0})^{\star}}}
\newcommand{\rbd}{\ensuremath{R_{bd}}}
\newcommand{\rbdi}{\ensuremath{R_{bd,i}}}

\defcitealias{rc3}{RC3}

\title[Composite Bulges]{Composite Bulges: The Coexistence of Classical Bulges and Disky
Pseudobulges in S0 and Spiral Galaxies}

\author[P. Erwin et al.]{Peter Erwin$^{1,2,7}$, Roberto P. Saglia$^{1,2}$,
Maximilian Fabricius$^{1,2}$, Jens Thomas$^{1,2}$,
\newauthor
Nina Nowak$^{3}$,
Stephanie Rusli$^{1,2}$,
Ralf Bender$^{1,2}$,
Juan Carlos Vega Beltr{\'a}n$^{4}$, 
\newauthor
and John E. Beckman$^{4,5,6}$ \\
$^{1}$Max-Planck-Insitut f\"{u}r extraterrestrische Physik, Giessenbachstrasse, 85748 Garching, Germany \\
$^{2}$Universit\"{a}ts-Sternwarte M\"{u}nchen, Scheinerstrasse 1, D-81679 M\"{u}nchen, Germany \\
$^{3}$Stockholm University, Department of Astronomy, Oskar Klein Centre, SE-10691 Stockholm, Sweden \\
$^{4}$Instituto de Astrof\'{\i}sica de Canarias, C/ Via L\'{a}ctea s/n, 38200 La Laguna, Tenerife, Spain \\
$^{5}$Departamento de Astrof\'{\i}sica, Universidad de La Laguna, Avda. Astrof\'{\i}sico Fco. S\'{a}nchez s/n, 38200, La Laguna, Tenerife, Spain \\
$^{6}$Consejo Superior de Investigaciones Cient\'ificas, Spain \\
$^{7}$Guest investigator of the UK Astronomy Data Centre}

\begin{document}

\maketitle

\label{firstpage}

\begin{abstract} 

We study nine S0--Sb galaxies with (photometric) bulges
consisting of two distinct components. The outer component is a
flattened, kinematically cool, disklike structure: a ``disky
pseudobulge''. Embedded inside is a rounder, kinematically hot spheroid:
a ``classical bulge''. This indicates that pseudobulges and classical
bulges are not mutually exclusive: some galaxies have both.

The disky pseudobulges almost always have an exponential disk (scale
lengths = 125--870 pc, mean $\sim 440$ pc) with disk-related
subcomponents: nuclear rings, bars, and/or spiral arms. They
constitute 11--59\% of the galaxy stellar mass (mean $PB/T = 0.33$),
with stellar masses $\sim 7 \times 10^{9}$--$9 \times 10^{10}$~\Msun.
Classical-bulge components have S\'ersic indices of 0.9--2.2, effective
radii of 25--430 pc and stellar masses of $5 \times 10^{8}$--$3 \times
10^{10}$~\Msun{} (usually $< 10$\% of the galaxy's stellar mass; mean
$B/T = 0.06$). The classical bulges show rotation, but are kinematically
hotter than the disky pseudobulges. Dynamical modeling of three systems
indicates that velocity dispersions are isotropic in the classical
bulges and equatorially biased in the disky pseudobulges.

In the mass--radius and mass--stellar mass density planes,
classical-bulge components follow sequences defined by ellipticals and
(larger) classical bulges. Disky pseudobulges \textit{also} fall on this
sequence; they are more compact than similar-mass large-scale disks.
Although some classical bulges are quite compact, they are distinct from
nuclear star clusters in both size and mass, and coexist with nuclear
clusters in at least two galaxies.

Since almost all the galaxies in this study are barred, they probably
\textit{also} host boxy/peanut-shaped bulges (vertically thickened inner
parts of bars). NGC~3368 shows evidence for such a zone outside its
disky pseudobulge, making it a galaxy with all three types of ``bulge''.

\end{abstract}


\section{Introduction} 

In the standard picture of galaxy structure, disk galaxies have two main
stellar components. The defining component is the disk: a highly
flattened structure dominated by rotation, often (but not always) with a
radial density profile which is exponential; disks often have
significant substructure, particularly bars and spiral arms. The
secondary component, present in early and intermediate Hubble types, is
the bulge. Traditionally, the bulge has been seen as something very like
a small elliptical galaxy embedded within the disk: more spheroidal than
the disk, with stellar motions dominated by velocity dispersion rather
than rotation, and having a strongly concentrated structure -- e.g.,
having a surface brightness profile similar or identical to the
stereotypical $R^{1/4}$ profile of an elliptical).  In addition, the
stellar populations of bulges were said to resemble those of ellipticals
in being older (and possibly more metal-rich and alpha-enhanced) than
the majority of stars in the disk. (See, e.g., \citealt{wyse97} and
\citealt{renzini99} for reviews.) Taken all together, this seemed to argue for a
bulge formation mechanism similar to that proposed for ellipticals,
either via monolithic collapse or by rapid, violent mergers of initial
subcomponents at high redshift.

The past decade or two has seen the growing realization that this
picture is probably \textit{not} true for many bulges, at least when
bulges are defined as the excess stellar light in the central regions of
the galaxy when compared to the dominant exponential profile of the
disk.\footnote{Very compact central concentrations -- e.g., nuclear star
clusters -- are considered separate objects and are excluded from this
definition.} Instead, bulges are now seen as falling into two rather
different classes: classical bulges (the traditional model) and
\textit{pseudobulges} \citep[e.g.,][]{kormendy93,kormendy04}, which are
conceived of as something much more like disks than spheroids; i.e.,
they are flattened and dominated by rotation, with profiles which are
close to exponential. Added to this complexity is the existence of
so-called ``boxy'' and ``peanut-shaped'' bulges, which are now well
understood as the vertically thickened inner parts of bars
\citep[see][for a discussion of the distinctions]{athanassoula05};
confusingly, these structures are also sometimes called pseudobulges.

Although some authors are careful to point out the possibility that
classical bulges could coexist with pseudobulges \citep[see,
e.g.,][]{athanassoula05,fisher-drory10}, it is common to suggest that
galaxies have one or the other, but not both. For example, in
observational surveys such as those of \citet{fisher-drory08} or
\citet{gadotti09}, photometrically identified bulges are classified as
either classical or pseudobulge. Similarly, in studies of how central
supermassive black holes (SMBHs) relate to their host galaxies, disk
galaxies are divided into those with classical bulges and those with
pseudobulges \citep[e.g.,][]{hu08,greene10,kormendy11}.

In this paper, we present evidence for the \textit{coexistence} in nine
galaxies of both a classical bulge -- that is, a round, kinematically
hot stellar structure which is significantly larger than a nuclear star
cluster -- and a disky pseudobulge -- that is, a flattened stellar
system, distinct from the main disk, whose kinematics are at least partly
dominated by rotation and which (usually) hosts nuclear bars, nuclear
rings, or other disky morphology.  

Two of the galaxies discussed here -- NGC~3368 and NGC~3489 --
were previously discussed, in an abbreviated fashion, in
\citet{nowak10}, using the term ``composite pseudobulges''.  Some
analysis of the morphological substructure in NGC~3945 and NGC~4371 has
been previously presented in \citet{erwin99} and \citet{erwin03-id}.

The outline of this paper is as follows. After some initial discussion
of data sources and reduction (Section~\ref{sec:obs}), we lay out our
terminology in Section~\ref{sec:terms}. We introduce our methodology for
identifying classical bulges by considering in
Section~\ref{sec:simple-classical} two examples of \textit{simple}
classical-bulge-plus-disk systems (galaxies with \textit{only} a
classical bulge in addition to their disk). With this as a reference, we
then consider two galaxies (NGC~3945 and NGC~4371) in some detail in
Section~\ref{sec:n3945n4371}, demonstrating first that much of their
photometrically defined bulges are \textit{not} classical bulges as
previously defined, but something else: (disky) pseudobulges. We then go
on to show that inside each pseudobulge is an additional structure which
\textit{does} resemble a classical bulge. The evidence for composite
bulges in seven other galaxies follows a similar pattern, but is
postponed to the Appendix (Section~\ref{sec:others-full}) so as not to
interrupt the flow of the paper. Section~\ref{sec:dynamics} uses the results of
Schwarzschild modeling for three composite-bulge galaxies to investigate
the 3D stellar dynamics of the classical-bulge and disky-pseudobulge
components. Section~\ref{sec:discussion} considers these composite-bulge
galaxies and their subcomponents, including an analysis of their place
in the mass-radius and surface-density--mass diagrams and a
demonstration that the classical-bulge components, while generally
rather small, are not in the same class as nuclear star clusters.
Finally, Section~\ref{sec:summary} summarizes our findings.

\section{Data Sources} 
\label{sec:obs}

\subsection{Imaging Data and Surface-Brightness Profiles}

Imaging data for this study comes from a variety of sources; for each
galaxy, we specify the individual images used and their origins. Some of
the large-scale, ground-based optical images come from the WIYN Survey
\citep{erwin-sparke03}, the INT-WFC observations of \citet{erwin08}, or
from the Sloan Digital Sky Survey \citep[SDSS;][]{york00}; for the
latter, we use Data Release 7 \citep[DR7;][]{abazajian09}. Other
ground-based image sources include WHT-INGRID $K$-band images from
\citet{knapen03}, available via NED, and images from the European
Southern Observatory (ESO) archive and the Isaac Newton Group (ING)
archive.

For high-resolution imaging of the central regions of galaxies, we rely
on archival images from the \textit{Hubble Space Telescope}, obtained
with the Wide Field Planetary Camera 2 (WFPC2), the Wide Field Channel
of the Advanced Camera for Surveys (ACS-WFC), or the NICMOS2 and NICMOS3
near-IR imagers. In some cases -- i.e., when \textit{HST} imaging data
is lacking or when the centre of a galaxy is particularly dusty -- we
use adaptive-optics $K$-band images derived from our VLT-SINFONI IFU
observations by collapsing the datacubes along the wavelength direction.
The latter observations are described in more detail in \citet{nowak10},
\citet{rusli11} and \citet{erwin14-smbh}.

The surface-brightness profiles we construct and analyze combine
data from multiple images for each galaxy: high-resolution data from
\textit{HST} or AO imaging for the smallest radii, where good spatial
resolution is critical, and non-AO ground-based imaging for larger radii, since
the galaxies extend beyond the high-resolution imaging
fields of view. To combine these data, we must
match the overall intensity scaling, and we must also account for the fact
that the high-resolution images are too small for accurate background
subtraction. (The VLT-SINFONI AO
observations were taken with offsets to blank sky, but these were spaced
too far apart in time for accurate removal of the overall background
levels.) The problem is then to find the correct multiplicative scaling
$k$ to match the background-corrected ``inner'' profile (from the
high-resolution image) with the ``outer'' profile (from the
low-resolution image, which is large enough for its own background to be
properly estimated), while also determining the unknown background level
$B_{i}$ in the high-resolution image, so that we can transform the
observed inner profile $I_{i}$ to a corrected profile $I_{i}^{\prime}$:
\begin{equation} 
I_{i}^{\prime} \; = \; k (I_{i} - B_{i}).
\end{equation}

Fortunately, these two issues can be dealt with in combination, by
taking advantage of the fact that the same galaxy was observed
in both images. The trick is to identify a radial overlap zone between
the two profiles (outside the region where seeing distorts the
low-resolution data) and iteratively fit for the values of $k$
and $B_{i}$ which minimize the difference $I_{i}^{\prime}(r) - I_{o}(r)$
for values of $r$ in the overlap zone (with $I_{o}$ being the outer
profile). The two profiles are then merged at the best-matching radius
in the overlap region. Although the profiles we analyze are usually
major-axis cuts, in practice we determine $k$ and $B_{i}$
using profiles from ellipse fits with fixed position angle and ellipticity,
to increase the signal-to-noise ratio. This approach is ideal when the
high- and low-resolution images were taken with the same filter; in some
cases, we are forced to match and combine profiles from images with dissimilar
filters (e.g., F814W and $R$, or $K$ and $z$).

When decomposing our surface-brightness profiles -- which are
typically cuts along the major axis of the galaxy -- we do not attempt
to correct for PSF convolution, though we do exclude the inner 2--3
pixels of the profile from the fit. For several galaxies with
\textit{HST} images, we estimated the possible effects of neglecting PSF
convolution by also extracting profiles from images which had been
deconvolved using the Lucy-Richardson algorithm (via the \textsc{IRAF} task
\texttt{lucy}) and TinyTim-generated PSFs. (This was not possible for
all galaxies, since for some galaxies we rely on SINFONI data for which
PSFs cannot be determined with nearly the same precision.) Comparison of
fits to both uncorrected and ``deconvolved'' profiles of the same galaxy
showed that parameters for the central (classical) bulges differ by $\la
15$\% in the S\'ersic index $n$ and $\la 3$\% for other parameters.

\subsection{Spectroscopic Data}

Some of the spectroscopic data used for NGC~2859 and NGC~4371 is based
on previously unpublished data obtained with the ISIS double spectrograph
on the 4.2m William Herschel Telescope. Details of the observations are
provided in the Appendix (Section~\ref{sec:data-spec}).

For NGC~3368, NGC~3945 and NGC~4371 we use long-slit data obtained with
the Marcario Low Resolution Spectrograph at the Hobby-Eberly Telescope,
previously presented in \citet{fabricius12}.  For NGC~3368 and NGC~4699,
we also use IFU data from our SINFONI $K$-band SMBH measurement program
\citep{nowak10}; \citet{erwin14-smbh}.

\subsubsection{Other Sources}

We also make use of various published long-slit and IFU kinematic data; the
specific sources are listed in the discussions of each galaxy. We note here
some datasets provided directly to us.
For NGC~1068, this includes both long-slit data from \citet{shapiro03},
provided by Joris Gerssen, and SINFONI data for NGC~1068
\citep{davies07}, provided by Ric Davies. Large-scale SINFONI kinematic data for
NGC~3368 \citep{hicks13} were provided by Erin Hicks. Finally, for NGC~4262 we make
use of OASIS IFU data \citep{mcdermid06}, provided by Richard
McDermid.

\section{Terminology and Definitions: What Do \textit{We} Mean by ``Classical Bulge'' and 
``Pseudobulge''?}\label{sec:terms} 

The terms ``pseudobulge'' and ``classical bulge'' are unfortunately
rather ambiguous at present. Sometimes they are defined in terms of
their presumed formation methods: e.g., pseudobulges are central
concentrations of stars formed from bar- or spiral-driven inflows of gas
in the disk plane, or even by any process that does not explicitly
involve major mergers, while classical bulges are those structures
formed by violent relaxation in major mergers (usually at high
redshifts). The fundamental problem with such approaches is that there
are few if any clear observational predictions for how to distinguish
such formation methods in nearby galaxies.  

For example, the formation of central ``bulges'' by the merger of
massive star-forming clumps in gas-rich, high-$z$ disks can produce
thick, dispersion-dominated central structures with $\sim R^{1/4}$ light
profiles and $\alpha$-enhanced metallicities
\citep[e.g.,][]{immeli04,elmegreen08}, fulfilling most of the
traditional criteria for classical bulges. If one insists on major
mergers as the formation mechanism, then these are not classical bulges
-- but we would have little or no way to distinguish these structures at
$z \sim 0$ from ``proper'' classical bulges. Other theoretical studies
of this formation mechanism argue that the resulting bulges should be
smaller and more exponential-like, with S\'ersic indices of $\sim 2$ or
even $\sim 1$, and significant rotation
\citep[e.g.,][]{hopkins12,inoue12}. What this means is that a
classification of structures in present-day galaxies based on their
supposed formation mechanisms, though desirable, is probably still
premature.

If we turn to the more feasible approach of observationally-based
classification, we still find considerable discord, if not an outright
cacophony. Some surveys classify the central region of a galaxy as
classical or pseudobulge depending on the presence or absence of certain
morphological features: e.g., a smooth light distribution means a
classical bulge, while the presence of dust lanes, spiral arms, rings,
nuclear-scale bars, or so-called boxy/peanut-shaped isophotes means a
pseudobulge \citep[e.g.,]{kormendy04,fisher-drory08}. Other studies
make distinctions based on photometric profiles of the ``bulge''
component that results from a bulge-disk decomposition -- e.g.,
pseudobulges are by definition any central structure with a S\'ersic
index $< 2$, or even any such structure with $n < 4$
\citep{laurikainen09} -- or some combination of mean surface brightness
and size for the bulge component \citep[e.g.,][]{gadotti09}.

Because of this confusion, we feel it is important to be clear about our
terminology and our methods for identifying and classifying different
types of ``bulges''. As part of our analysis, we first identify what we
call \textbf{photometric bulges}. This term refers to the region of a
galaxy where the observed stellar surface brightness is brighter than an
inward extrapolation of the outer disk component, or is brighter than
the inward extrapolation of a previously identified disky pseudobulge. 
We require that the photometric bulge should be more extended  than a
simple nuclear star cluster (i.e., the half-light radius should be $\ga
10$ pc). This is, as the name suggests, a purely photometric
classification, and is used only as a preliminary tool (and for
comparison with the results of purely photometric methodologies).

We then analyse the photometric bulges and classify them into two
categories:
\begin{itemize}
\item \textbf{Classical bulge:} This is a photometric bulge which is some type
of kinematically hot spheroid. That is, it must be clearly
\textit{rounder} in a three-dimensional sense than the main galaxy disk 
(i.e., $(c/a)_{\rm bulge} > (c/a)_{\rm disk}$, where $c$ is the vertical scale
length and $a$ is the radial)
\textit{and} must have stellar kinematics which are dominated by velocity
dispersion rather than rotation.

\item \textbf{Disky pseudobulge:} This is a photometric bulge which is
``disklike'' in \textit{two} main ways: it has a flattening similar or
identical to that of the main galaxy disk, and the stellar kinematics
are dominated by rotation rather than velocity dispersion at least
some point within the photometric bulge region. We also consider the
presence of clear morphological features such as bars, rings and spiral
arms to be additional signatures of a disky pseudobulge, but do not
rely on them alone.
\end{itemize}

It is important to note that our definition of classical bulge does
\textit{not} assume a particular surface-brightness profile shape: we
are not assuming that kinematically hot spheroids must have de
Vaucouleurs $R^{1/4}$ profiles, nor that they must have S\'ersic indices
greater than some minimum value.

We also note that we are not considering several characteristics which
are sometimes, as alluded to previously, discussed as indicative of
``pseudobulges'' \citep{kormendy04}. For example, we are mostly not
concerned with the presence or absence of dust in the centres of these
galaxies, since this can sometimes be due to off-plane or
counter-rotating gas (likely the result of accretion), and in other
cases may merely indicate that the disk and bulge are co-extensive.
Since many of the galaxies we consider are lenticular, we do not require
the presence of current or recent star formation as a pseudobulge
indicator either. Thus, we are explicitly including what
\citet{fisher09} called ``inactive pseudobulges'' (objects in their
sample which had what they considered the morphological and photometric
signatures of pseudobulges but which showed little or no evidence for
recent star formation; see also \citealt{fisher-drory10}).

Finally, we are, for the most part, explicitly excluding 
boxy/peanut-shaped bulges from consideration in this study. As
\citet{athanassoula05} pointed out, these are the vertically thickened
inner parts of bars, the result of a common dynamical instability which
appears to accompany the formation of most bars. As such, they are
\textit{not} the sort of highly flattened, axisymmetric structures we
are most interested in. None the less, we \textit{do} consider the
question of their possible co-existence with disky pseudobulges and
classical bulges later on in the paper (Section~\ref{sec:boxy}).

In Section~\ref{sec:simple-classical}, we present two cases of S0
galaxies with purely classical bulges, as a way of providing both
examples of how we identify classical bulges and some context for the
composite bulges we discuss later. In Section~\ref{sec:n3945n4371} we
then go on to analyze two composite-bulge S0 galaxies in detail; we
start by identifying disky pseudobulges in each galaxy. While we would
ideally like to present an example or two of ``pure disky pseudobulge''
systems before moving on to the composite bulges, we have encountered
difficulty in trying to identify any clear examples in the very nearby
(e.g., $D \la 20$ Mpc), early-type disk galaxy population for which the
necessary data (particularly stellar kinematics with the right
combination of high spatial resolution and radial extent) exist. The
problem is not so much identifying candidate disky pseudobulges in other
nearby galaxies (as numerous others have done), but rather being able to
clearly demonstrate that these are \textit{not also} composite bulge
galaxies: i.e., that there are no classical bulges, however small,
inside these galaxies. The fact that we do not present any examples of
S0--Sb disk galaxies with pure disky pseudobulges should not, however,
be taken as a claim that such systems are absent in the local universe.

\section{Methods and Applications: Simple Classical Bulges}\label{sec:simple-classical}

\subsection{Basic Methodology} 

How does one identify the ``bulge'' of a galaxy, and how does one determine
whether such a structure is more like a classical bulge or a disky pseudobulge?
Our basic approach has three steps. First, we identify the photometric bulge via
a standard S\'ersic + exponential decomposition of the entire galaxy. The outer
boundary of the (photometric) ``bulge-dominated'' region is then identified by
finding the radius \rbd{} where the S\'ersic and exponential components are
equal in brightness: 
\begin{equation} 
\mu_{\rm Ser}(\rbd) \; = \; \mu_{\rm exp}(\rbd) 
\end{equation}
Note that this radius may vary somewhat depending on the wavelengths
of the data being used for the fit.

Second, we analyse the morphology of the photometric bulge region,
focusing especially on the shape of the isophotes. Our working
assumption is that the photometric bulge and the outer disk share a
common equatorial plane (i.e., they have the same line of nodes and,
most importantly, the same inclination to the line of
sight). This means that if the bulge is intrinsically rounder (more
spheroidal) than the disk, its projected isophotes should appear rounder
than those of the outer disk; in the extreme case of a spherical bulge, we 
would expect to see the elliptical isophotes of the (projected) outer disk give way
to circular isophotes at small radii, where the bulge dominates the light.

Finally, we analyse the stellar kinematics in the photometric bulge
region, trying to determine whether they are more dominated by rotation
or velocity dispersion. Traditionally, one way of using stellar
kinematics to discriminate between classical bulges and pseudobulges
\citep[going back to ][]{kormendy82} has been to note the position of
the bulge in question on the $V/\sigma$--$\epsilon$ diagram
\citep{illingworth77}, where $V$ and $\sigma$ are the ``characteristic''
stellar velocity and velocity dispersion and $\epsilon$ is the
ellipticity.\footnote{$V$ is usually taken to be the maximum stellar
rotation velocity $V_{\rm max}$ and $\sigma$ is some ``central'' value;
the ellipticity is sometimes the maximum observed value and sometimes a
mean value.} One can define a curve in this diagram which corresponds to
an ``isotropic oblate rotator'' (IOR), a simple model for a classical
bulge or elliptical with isotropic velocity dispersion and possible
flattening due to modest amounts of stellar rotation
\citep[e.g.,][]{binney78,binney05}.  If the object clearly lies above
the IOR curve, then the argument is that the object is too dominated by
rotation to be considered a classical bulge. (Kinematically hot systems
with little or no rotation but significant anisotropy will tend to lie
below the IOR line.) This is one of the methods by which some of the
original ``pseudobulges'' (avant la lettre) were identified
\citep{kormendy82,kormendy93}. Recent discussion of this diagram,
primarily in the context of elliptical and S0 galaxies and taking
advantage of 2D kinematics, include, e.g.,  \citet{cappellari07},
\citet{spolaor10}, and \citet{emsellem11}.

There are, however, some problems with the $V/\sigma$--$\epsilon$ approach. The
underlying theoretical arguments for the reference IOR models presuppose simple,
coherent stellar systems with unique, unambiguous values for the ellipticity, velocity and
velocity dispersion. The original application envisaged was elliptical galaxies,
where at least the domain (the entire galaxy) is unambiguous. But in the case of
complex systems such as a bulge embedded within a disk containing secondary
structures (nuclear rings, bars, etc.), it is not at all clear how one is
supposed to define ``the'' ellipticity; nor is it clear how to define ``the''
velocity dispersion when the latter can vary significantly with radius. Even the
common technique of choosing the \textit{maximum} stellar velocity as ``the''
velocity runs into trouble if the rotation curve continues to rise throughout
the bulge region and on into the disk-dominated part of the galaxy (or if the rotation
curve has multiple peaks).

Faced with these difficulties, we opt for a different approach: we define a
simple \textit{local} measurement of the relative importance of rotation versus
disperson, by deprojecting the observed rotation to its in-plane value $V_{\rm
dp} = V_{\rm obs} / \sin i$ and then dividing this by the observed velocity
dispersion $\sigma$ \textit{at the same radius}. The resulting quantity --
\vdivsigma{} -- is a continually varying function of the radius, not a
``universal'' value for an entire galaxy (or entire galactic component).
We adopt an admittedly crude and ad-hoc limit of $\vdivsigma = 1$ as the
dividing line between kinematically ``cool'' and kinematically ``hot'' systems,
so that classical bulges should have $\vdivsigma < 1$ within the region
where they dominate the galaxy's light. In the following subsections we
provide some partial justification for this criterion by showing that
galaxies with simple disk + spheroidal bulge morphologies do seem to
have $\vdivsigma < 1$ within their bulge-dominated regions.

Since we identify the photometric bulge region via decomposition of the
major-axis profile, it makes sense to use major-axis value of \vdivsigma.  This
lets us use major-axis long-slit spectroscopy, which is in many cases the only
available stellar kinematic data -- or the only data covering the full radial
range of interest -- for the galaxies we examine.

To show how this approach works in the simple case of disk galaxies \textit{without}
pseudobulges, the following subsections apply our methodology to two S0s
with classical bulges.

\subsection{Simple Classical Bulge Example: NGC~7457}\label{sec:n7457} 

NGC~7457 is a nearby ($D = 12.9$~Mpc)\footnote{Based on the SBF distance of \citet{tonry01}
and the \citet{mei05} correction.} low-luminosity S0 galaxy, seen at moderate
inclination ($i \approx 58\degr$; \citealt{gutierrez11}); it has a
central velocity dispersion of only $\sim 60$--70 \kms{}
\citep[e.g.,][]{trager98,wegner03,ho09}. Although it has been suggested
as a possible pseudobulge host in the past, largely on the basis of
supposed deviations from the Faber-Jackson relation
\citep[e.g.,][]{kormendy93,pinkney03}, more recent analyses clearly identify it
as having a classical bulge
\citep[e.g.,][]{fisher-drory08,fisher-drory10,kormendy11}.

\begin{figure*}
\includegraphics[width=6.8in]{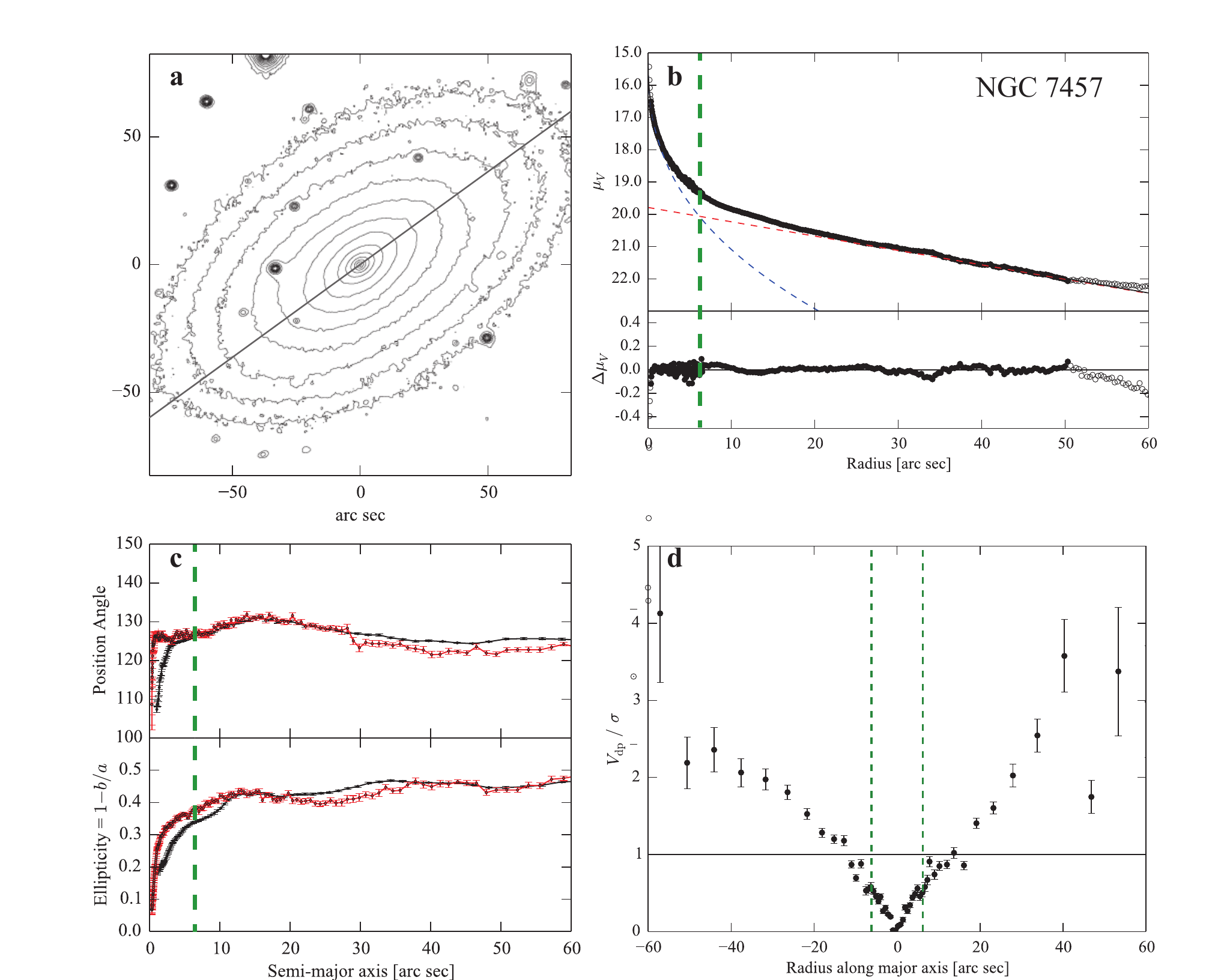}
  
\caption{Evidence for a classical bulge in the low-mass S0 galaxy NGC~7457.
\textbf{a:} log-scaled $V$-band isophotes (JKT image); gray line marks major
axis (PA $= 126\degr$). \textbf{b:} Bulge-disk decomposition of major-axis
profile. Data (black circles) combine major-axis cuts through \textit{HST} WFPC2
F555W image ($r < 6.6$ arcsec) and ground-based $V$-band image. Dashed lines
are S\'ersic + exponential fit to the data for $r = 0.4$--50 arcsec
(filled circles), with residuals in lower sub-panel; data at $r >
50$ arcsec are part of a shallower outer zone in the antitruncated disk profile.
Vertical dashed green line marks ``bulge=disk'' radius \rbd, where S\'ersic
and exponential components are equally bright; this sets the boundary of the
``photometric bulge region''. \textbf{c:} Ellipticity and position angle of
ellipse fits to $V$-band image (black) and \textit{HST} image (red). \textbf{d:}
Plot of deprojected stellar rotation velocity divided by local velocity
dispersion $\vdivsigma$ along the major axis, using long-slit data from
\citet{simien97c}. Vertical dashed lines mark the photometric bulge region $|R|
< \rbd$; $\vdivsigma$ is $< 1$ inside, indicating a kinematically hot region
(i.e., a classical bulge).} \label{fig:n7457-complete}

\end{figure*}

The upper left panel of Figure~\ref{fig:n7457-complete} shows the
$V$-band isophotes for NGC~7457, based on an archival image from the
Jacobus Kapteyn Telescope (JKT) of the ING \citep[see][for
details]{gutierrez11}. The upper right panel shows our B/D decomposition
of a major-axis cut which combines data from the $V$-band image with
data from an HST WFPC2 F555W image (for $r < 6$ arcsec). The fit
excluded both the outer part of the disk \citep[which has an
antitruncated profile;][]{gutierrez11} and the inner nuclear excess at
$r < 0.35$ arcsec, which has been attributed to either AGN emission
\citep{gebhardt03} or a nuclear star cluster \citep{graham09}. The
bulge/disk crossover radius is at $\rbd = 6.2$ arcsec; the disk is
clearly the dominant component at $r \ga 15$--20 arcsec.

Turning to the isophotes shapes, the lower left panel of
Figure~\ref{fig:n7457-complete} shows the results of fitting ellipses to both
images. The ellipticity stays roughly constant in to $a \sim 12$ arcsec, and
then becomes progressively rounder inside. This is consistent with the influence
of a round bulge embedded within a highly elliptical (inclined) disk, so we have
evidence that the photometric bulge identified in the B/D decomposition is a
rounder (and thus more spheroidal) object than the disk.

Finally, the lower right panel of Figure~\ref{fig:n7457-complete} shows
\vdivsigma{} as a function of radius along the major axis, using the long-slit
kinematic data of \citet{simien97c}. Within the
photometric bulge region, $\vdivsigma$ is consistently $< 1$ (in fact, it never
gets above $\sim 0.6$); it increases to larger radii, finally becoming $> 1$ at
$r \ga 15$ arcsec. Note that \vdivsigma{} continues to increase as we move into
the (photometric and morphological) disk region, reaching values $\ga 2$ in the
region which is unambiguously disk-dominated.

As a crude approximation, then, we can identify ``kinematically disklike'' regions
as having $\vdivsigma > 1$.

\subsection{Simple Classical Bulge Example: NGC~1332}\label{sec:n1332} 

With a $K$-band luminosity of $1.56 \times 10^{11}$~\Lsun{} and a central
velocity dispersion of 328 \kms{} \citep{rusli11}, NGC~1332 is one of
the most massive S0 galaxies in the nearby universe ($D = 22.3$~Mpc,
from \citealt{tonry01} and \citealt{mei05}); it also hosts a central
supermassive black hole with a mass of $1.45 \times 10^{9}$~\Msun{}
\citep{rusli11}.

\begin{figure*}
\includegraphics[width=6.0in]{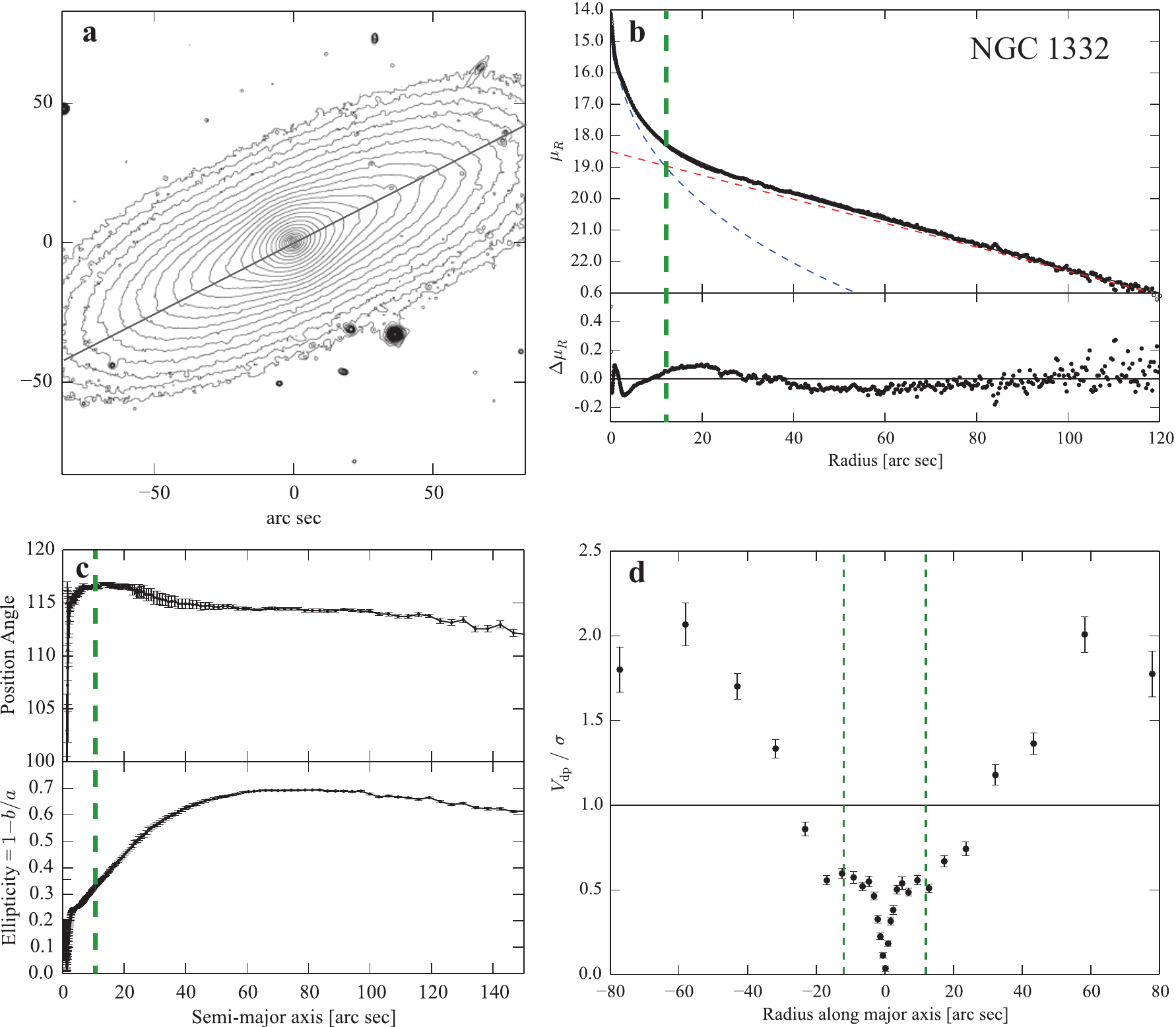}
  
\caption{As for Figure~\ref{fig:n7457-complete}, but now showing evidence for a
classical bulge in the high-mass S0 galaxy NGC~1332. \textbf{a:} log-scaled
$R$-band isophotes (NTT-EMMI image); gray line marks major axis (PA $=
117\degr$). \textbf{b:} Bulge-disk decomposition of major-axis profile. Data
(black circles) combine ellipse fits to SINFONI 100mas and HST-WFPC2 F814W
images ($r < 2.6$ arcsec) and ground-based $R$-band image. Dashed lines are
S\'ersic + exponential fit to the data, with residuals in lower sub-panel.
Vertical dashed green line marks ``bulge=disk'' radius \rbd, where S\'ersic and
exponential components are equally bright; this sets the boundary of photometric
bulge region. \textbf{c:} Ellipticity and position angle of ellipse fits to
$R$-band image. \textbf{d:} Plot of $\vdivsigma$ along major axis, using
long-slit data from \citet{kuijken96} as re-reduced by \citet{rusli11}. Vertical
dashed lines mark the photometric bulge region $|R| < \rbd$; $\vdivsigma$ is $<
1$ inside, indicating a kinematically hot region (i.e., a classical
bulge).}\label{fig:n1332-complete}

\end{figure*}

Figure~\ref{fig:n1332-complete} shows the overall morphology of the galaxy in
panel~a: a highly elliptical outer disk with a distinctly rounder inner zone.
Decomposition of the major-axis profile (panel~b, combining data from a
ground-based $R$-band image with \textit{HST} WFPC2 F814W data at smaller radii
and data from our SINFONI $K$-band datacube at the very smallest radii to reduce
the effects of circumnuclear dust extinction; see \citealt{rusli11} for details)
shows a photometric bulge dominating the light at $r < \rbd = 12$ arcsec. As was the
case for NGC~7457, the isophotes in the bulge-dominated region are clearly
rounder (panel~c) than those of the outer disk.

Previous B/D decompositions for this galaxy (both 1D and 2D) are discussed in
\citet{rusli11}. Of particular note is the fact that their 2D decomposition had
a best fit using a S\'ersic bulge component with ellipticity $= 0.27$, in
contrast to the best-fitting exponential disk ellipticity of 0.73. This is clear
support for the idea that the photometric bulge corresponds to a region which is
significantly rounder than the disk. The slight twisting and rounding of
isophotes between $a \sim 20$ and 40 arcsec may indicate a very weak
bar or lens, but otherwise this galaxy is very close to an ideal exponential
disk + S\'ersic bulge system.

Finally, panel~d of Figure~\ref{fig:n1332-complete} shows the radial trend of
$\vdivsigma$, using data from \citet{kuijken96} as re-reduced by
\citet{rusli11}.  $\vdivsigma$ clearly reaches a plateau value ($\sim 0.5$--0.6)
within the photometric bulge; as was the case for NGC~7457, the ratio only
becomes $> 1$ outside the photometric bulge region. 

As in the case of NGC~7457, we conclude that the photometric bulge in NGC~1332 is a
structure which is clearly rounder than the disk and has stellar kinematics
dominated by velocity dispersion: in other words, a classical bulge.

\section{Composite Bulges: Detailed Examples}\label{sec:n3945n4371} 

In this section, we turn to a set of galaxies whose photometric bulges show
significantly more complex structure than was true for the two S0
galaxies considered in the previous section. Basic parameters for these galaxies
are presented in Table~\ref{tab:galaxies}.

We begin by examining two S0 galaxies -- NGC~3945 and NGC~4371 -- in
detail, as paradigms for the analysis we apply to the whole set.
Individual details and analysis for the other galaxies are discussed in the
Appendix (Section~\ref{sec:others-full}).

\begin{table*}
\begin{minipage}{126mm}
\caption{Galaxies with Composite Bulges}
\label{tab:galaxies}
\begin{tabular}{@{}llrlrrrr}
\hline
Galaxy    & RC3 Type          & D     & Source & $M_{B}$   & PA  & $i$   & \rbd{} \\
          &                   & Mpc   &        &           & deg & deg   & arcsec \\
(1)       & (2)                & (3)  & (4)    & (5)       & (6) & (7)   & (8) \\
\hline
NGC 1068  &  (R)SA(rs)b       & 14.2  & 1      & $-21.23$  &  86   & 31/40 & 24 \\
NGC 1543  &  (R)SB(l)$0^0$    & 20.0  & 2      & $-20.12$  & ---   & 20?   & 26 \\
NGC 1553  &  SA(r)$0^0$       & 18.0  & 2      & $-21.08$  & 152   & 48    & 16 \\
NGC 2859  &  R)SB(r)$0^{+}$   & 24.2  & 1      & $-20.21$  &  85   & 32    & 52 \\
NGC 3368  &  SAB(rs)ab        & 10.5  & 3      & $-20.37$  & 172   & 50    & 52 \\
NGC 3945  &  (R)SB(rs)$0^{+}$ & 19.8  & 1      & $-19.94$  & 158   & 55    & 17 \\
NGC 4262  &  SB(s)$0^-$?      & 15.4  & 4      & $-18.72$  & 155   & 30    & 7.2 \\
NGC 4371  &  SB(r)$0^{+}$     & 16.9  & 4      & $-19.49$  &  90   & 58    & 32 \\
NGC 4699  &  SAB(rs)b         & 18.9  & 1      & $-21.33$  &  35   & 37    & 44 \\
\hline 
\end{tabular}

\medskip
Basic characteristics of galaxies with composite bulges.
Column 1: Galaxy name. Column 2: Hubble type \citepalias{rc3}. Column
3: Distance. Column 4: Source for distance: 1 = Virgocentric-infall-corrected redshift from HyperLeda; 
2 = \citet{tonry01}, with correction from \citet{mei05};
3 = \citet{freedman01}; 4 = \citet{blakeslee09}.
Column 5: absolute $B$ luminosity (HyperLeda $B_{tc}$ +
adopted distance). Column 6: adopted position angle of disk.
Column 7: adopted inclination. Column 8: Radius where luminosity of S\'ersic
component = luminosity of exponential component from initial photometric
decomposition (see text).

\end{minipage}
\end{table*}

\begin{table*}
\begin{minipage}{126mm}
\caption{Imaging and Spectroscopic Data Summary}
\label{tab:data}
\begin{tabular}{@{}llllll}
\hline
Galaxy    & Telescope/Instrument  & Filter     & Source & Telescope/Instrument & Source  \\
          & Imaging               &           &         & Spectroscopy & \\
(1)       & (2)                   & (3)       & (4)     & (5)          & (6)  \\
\hline
NGC 1068  &  VLT-SINFONI (AO)     & $H$   & 1  &   VLT-SINFONI (AO)            & 1 \\
          &  \textit{HST}-NICMOS3 & F200N &    &   Gemini-GMOS                 & 2 \\
          &  2MASS                & $K$   &    &   KNPO-4m-RCFS                & 3 \\
          &  SDSS                 & $i$   &    &   \ldots                      &  \\
NGC 1543  &  \textit{HST}-WFPC2   & F814W &    &   \ldots                      & \\
          &  \textit{Spitzer}-IRAC     & IRAC2 &    &   \ldots                      & \\
NGC 1553  &  \textit{HST}-WFPC2   & F814W &    &   CTIO-4m-RCS                 & 4 \\
          &  \textit{HST}-NICMOS2 & F160W &    &   ESO-1.52m-B\&C              & 5 \\
          &  \textit{Spitzer}-IRAC     & IRAC1 &    &   \ldots                      &  \\
NGC 2859  &  \textit{HST}-ACS/WFC & F814W &    &   WHT-ISIS                    & 6 \\
          &  WIYN-3.5m            & $R$   & 7  &   WHT-SAURON                  & 8 \\
          &  SDSS                 & $i$   &    &   \ldots                      &  \\
NGC 3368  &  VLT-SINFONI (AO)     & $K$   & 9  &   VLT-SINFONI (AO)            & 9 \\
          &  \textit{HST}-NICMOS2 & F160W &    &   HET-MLRS                    & 10 \\
          &  WHT-INGRID           & $K$   & 11 &   \ldots                      &  \\
          &  SDSS                 & $r$   &    &   \ldots                      &  \\
NGC 3945  &  \textit{HST}-WFPC2   & F814W &    &   \textit{HST}-STIS           & 12 \\
          &  WIYN-3.5m            & $R$   & 7  &   HET-MLRS                    & 10 \\
NGC 4262  &  \textit{HST}-ACS/WFC & F850LP &   &   WHT-OASIS                   & 13 \\
          &  SDSS                 & $i$   &    &   WHT-SAURON                  & 14 \\
NGC 4371  &  \textit{HST}-ACS/WFC & F850LP &   &  VLT-SINFONI (AO)             & 15 \\
          &  INT-WFC              & $r$   & 16 &   WHT-ISIS                    & 6 \\
NGC 4699  &  VLT-SINFONI (AO)     & $K$   & 15 &  VLT-SINFONI (AO)             & 15 \\
          &  SDSS                 & $i$,$z$ &  &  Las Campanas-2.5m-MS         & 17 \\
\hline 
\end{tabular}

\medskip

Imaging and spectroscopic data used for the composite-bulge galaxies.
For each galaxy, we list the data in order of decreasing spatial
resolution (e.g., \textit{HST} or AO data, followed by ground-based
data). Column 1: Galaxy name. Column 2: Telescope + instrument or survey
for imaging data (``AO'' = adaptive optics used). Column 3: Filter used
for imaging data. Column 4: Source of imaging data, if not from public
telescope archives. Column 5: Telescope + instrument for
spectroscopic/kinematic data. Column 6: Source of
spectroscopic/kinematic data. References: 
1 = \citet{davies07}; 2 = \citet{gerssen06}; 3 = \citet{shapiro03}; 4 = \citet{kormendy84a}; 
5 = \citet{longo94}; 6 = this paper (Appendix~\ref{sec:data-spec});
7 = \citet{erwin-sparke03}; 8 = \citet{delorenzo-caceres08}; 
9 = \citet{nowak10}; 10 = \citet{fabricius12}; 11 = \citet{knapen03}; 
12 = \citet{gultekin09a}; 13 = \citet{mcdermid06}; 14 = \citet{emsellem04};
15 = Erwin et al., in prep.; 16 = \citet{erwin08}; 
17 = \citet{bower93}.

\end{minipage}
\end{table*}

\subsection{Composite Bulge Example: NGC~3945} 

NGC~3945 is a double-barred S0 galaxy \citep{erwin99,erwin04} which was
originally singled out as an unusual object by \citet{kormendy82}; it
had the largest value of \vmaxsigstar{}\footnote{This is the ratio of
the observed \vmaxsig{} to the value expected for an isotropic oblate
rotator having the same ellipticity; the value for NGC~3945 was 1.82.}
of any of the galaxies in his sample \citep[see also Figure~17
of][]{kormendy04}, indicating an unusually high degree of rotational
support for a ``bulge''.

\citet{erwin99} re-examined this galaxy using \textit{HST} imaging. They
pointed out that much of the inner region (i.e., the photometric bulge)
appeared to be similar to a small exponential disk, complete with a
nuclear bar surrounded by a stellar ring, embedded inside the main body
of the galaxy, and that this region had a flattening roughly consistent
with that of the outer disk. \citet{erwin03-id} revisited this analysis
by performing B/D decompositions and a further analysis of the original
kinematics of \citet{kormendy82}; they noted the existence of a separate
photometric component within the inner 2 arcsec, with isophotes which
were rounder than the main part of the photometric bulge (and the outer
disk). Because NGC~3945 is such a paradigmatic case for the phenomenon
of composite bulges, we repeat most of their analysis here, with the
addition of \textit{HST} STIS stellar kinematics from
\citet{gultekin09a} which enables us to explore the kinematic status of
the classical bulge; we also make use of new, large-scale kinematics
obtained with the HET.

\begin{figure*}
\includegraphics[width=6.0in]{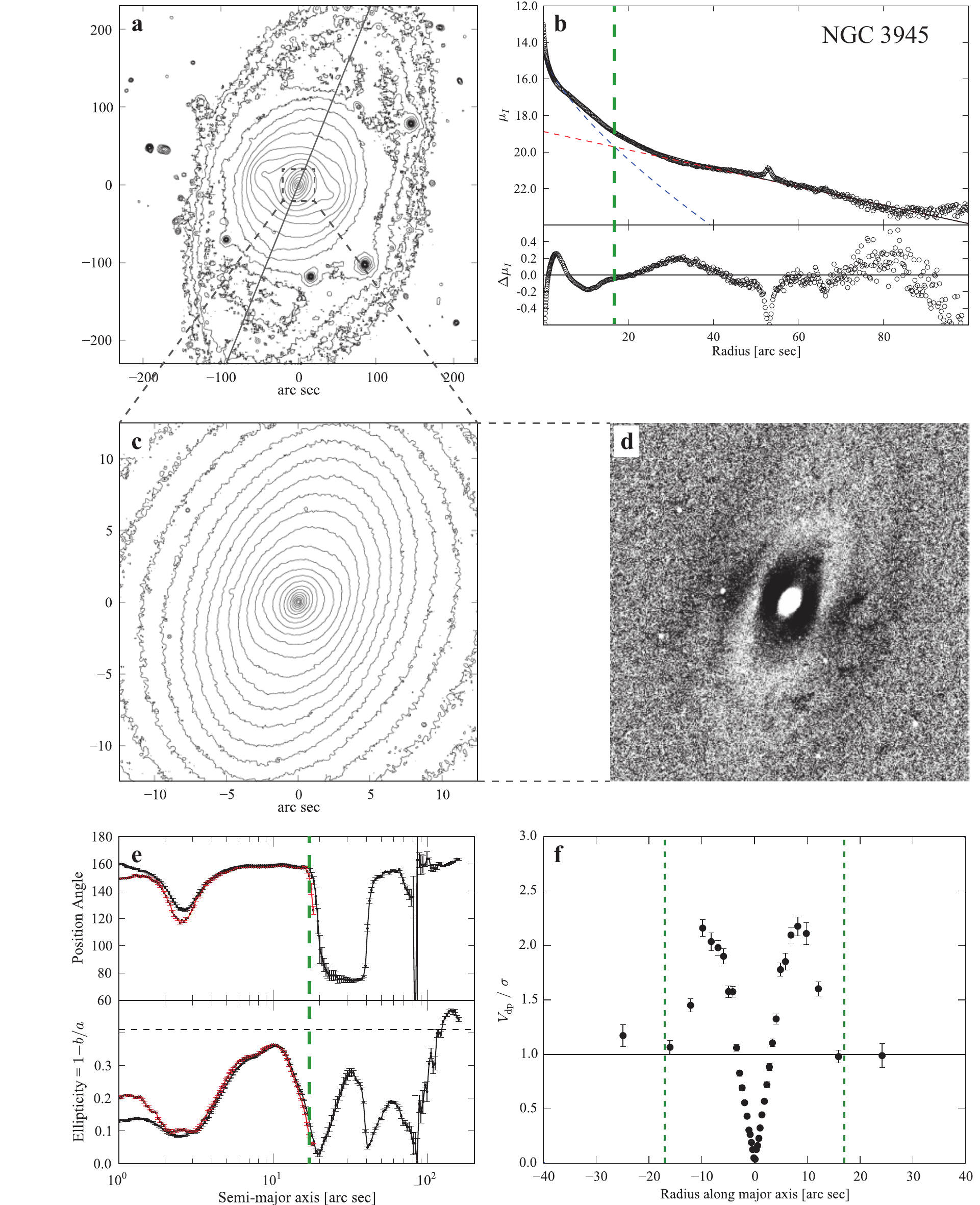}

\caption{Evidence for a disky pseudobulge in the S0 galaxy NGC~3945. \textbf{a:}
log-scaled $R$-band isophotes (WIYN image, smoothed with 15-pixel-wide median
filter); gray diagonal line marks major axis (PA $= 158\degr$). \textbf{b:}
Bulge-disk decomposition of WIYN major-axis profile.  Dashed lines show S\'ersic
+ exponential fit to the data, with residuals in lower sub-panel. Vertical
dashed green line marks ``bulge=disk'' radius \rbd, where S\'ersic and
exponential components are equally bright; this sets the boundary of the
photometric bulge. \textbf{c:} Close-up of photometric bulge region,
(log-scaled isophotes from \textit{HST} F814W WFPC2 image). \textbf{d:} Unsharp
mask (using $\sigma = 15$ pixel Gaussian kernel) of inner region on the same
scale as panel~c, showing stellar nuclear ring. \textbf{e:} Ellipse fits to WIYN
and \textit{HST} images; note that the ellipticity reaches $\sim 0.36$, similar
to outer disk value (horizontal dashed line), in the photometric bulge; this
suggests that at least part of the photometric bulge has flattening similar to
the outer disk. \textbf{f:} Deprojected stellar rotation velocity divided by
local velocity dispersion $\vdivsigma$ along the major axis, using HET long-slit
data. Vertical dashed lines mark the photometric bulge region $|R| < \rbd$;
$\vdivsigma$ rises to $> 2$ in this region, indicating a kinematically cool
region more like a disk.}\label{fig:n3945a}

\end{figure*}

\begin{figure*}
\includegraphics[width=6.0in]{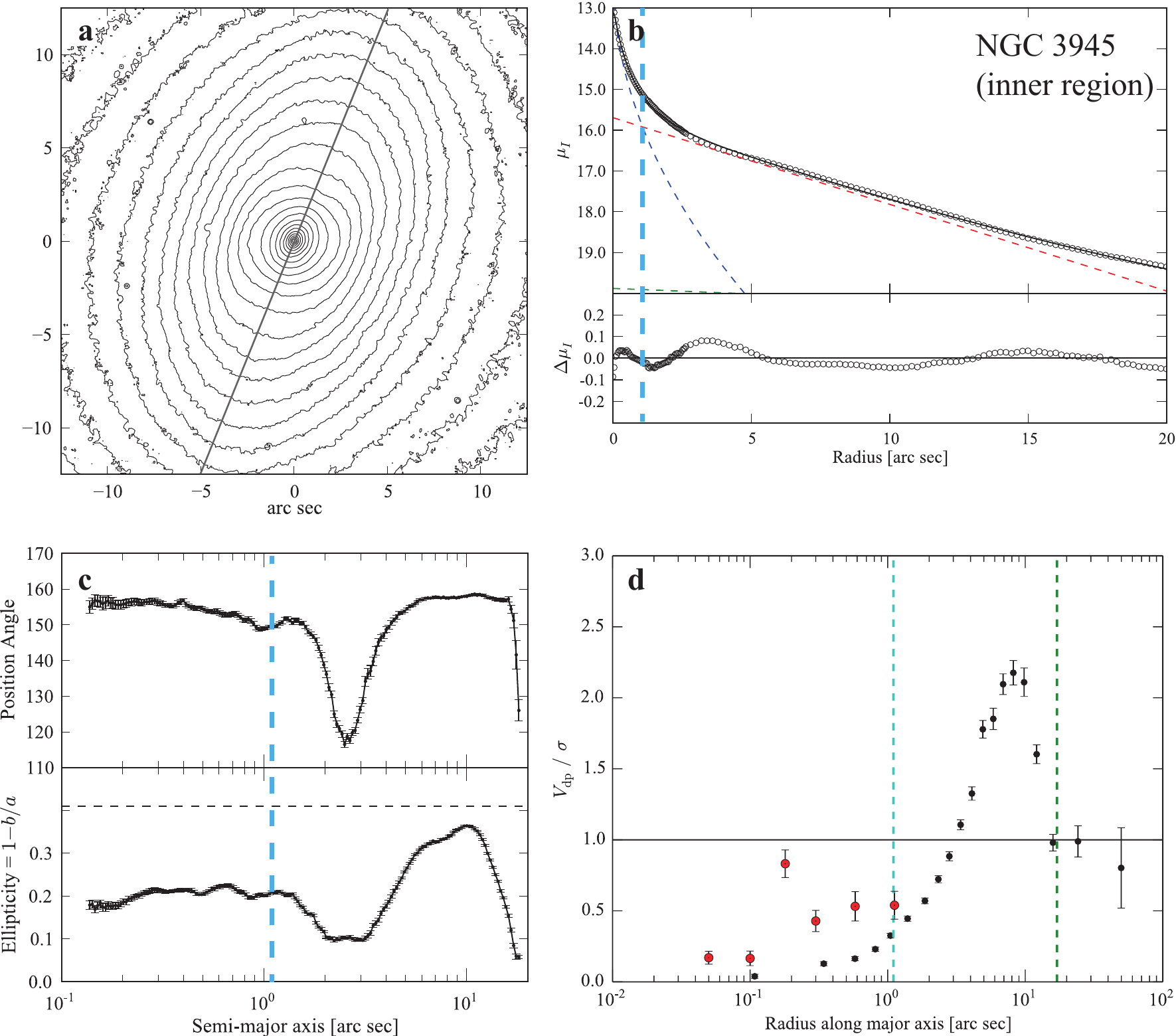}

\caption{Evidence for a small classical bulge inside the disky pseudobulge of
NGC~3945. \textbf{a:} Close-up of main photometric bulge region from
\textit{HST} WFPC2 F814W image, gray line marks major-axis PA. \textbf{b:}
major-axis profile from \textit{HST} F814W WFPC2 image, with S\'ersic +
two-exponential fit (dashed lines) and residuals from fit in lower sub-panel.
Vertical dashed blue line marks inner ``bulge=disk'' radius \rbdi, where the S\'ersic
and exponential components of this fit are equally bright; this sets the
boundary of the \textit{inner} photometric bulge. \textbf{c:} Ellipse fits to
\textit{HST} image; ellipticity in the inner photometric bulge region ($a \la
1.1$ arcsec) is $\sim 0.21$, much less than that of the disky pseudobulge
outside. \textbf{d:} Deprojected, folded stellar rotation velocity divided by
local velocity dispersion $\vdivsigma$ along the major axis, using \textit{HST}
STIS data from \citet{gultekin09a} [red] and HET long-slit data [black].
Vertical dashed blue and green lines mark the inner and main photometric bulge
regions $|R| < \rbdi$, respectively. In the inner region ($r \la 1$ arcsec),
$\vdivsigma$ remains well below 1, suggesting a kinematically hot region.
}\label{fig:n3945b}

\end{figure*}

\subsubsection{Photometric Bulge as Disky Pseudobulge} 

As we did for the pure-classical-bulge S0s in the previous section, we
performed a simple B/D decomposition of the major-axis profile (panel~b
of Figure~\ref{fig:n3945a}). This is not entirely successful, since the
outer disk is \textit{not} a simple exponential; instead, it is
distorted by the presence of a very luminous outer ring, which we
exclude from the fit.  (On the other hand, this profile largely avoids
any contribution from the primary bar, which is oriented close to the
galaxy minor axis.) None the less, there \textit{is} a clear inner
excess which dominates the light at $r \la 20$ arcsec; from our fit, we
find $\rbd = 17$ arcsec.

The fit in Figure~\ref{fig:n3945a} is somewhat different from the global
$B/D$ decomposition presented in \citet{erwin03-id}, which used an
ellipse-fit profile rather than a major-axis cut, and tried to fit the
outer ring beyond 80 arcsec with the same exponential component. That
fit increased the role of the S\'ersic component (in part because it
included light from the bar, which our cut largely avoids because the
bar is oriented almost perpendicular to the major axis), moving \rbd{}
out to $\sim 30$ arcsec; a similar $\rbd$ value can be seen in the
ellipse-fit-based B/D decomposition of this galaxy in
\citet{fisher-drory10}. If we use either of those fits instead, the
photometric bulge region becomes larger, but the conclusions of our
analysis remain unchanged.

Thus far, our analysis of this galaxy has not shown any clear deviation
from the simple classical-bulge systems discussed in
Section~\ref{sec:simple-classical}. Something rather different does
emerge, however, when we look at the isophote shapes in the photometric
bulge region (panels~c and e of Figure~\ref{fig:n3945a}). The
ellipticity of this region is quite large: it reaches a maximum value of
0.36 at $a \approx 10$ arcsec, which is close to that of the outer disk.
Given that these isophotes may still be distorted by light from the
(primary) bar outside, it is possible that the underlying shape is
actually the \textit{same} as the outer disk; in any case, it suggests
that much of the photometric bulge region is nearly as flat as the outer
disk.

Unsharp masking (panel~d of Figure~\ref{fig:n3945a}) shows the signature of a
stellar nuclear ring in this region (with $a \sim 6.5$ arcsec); additional
unsharp masking also reveals the presence of a nuclear bar inside the ring
\citep[see Figure~2 of][]{erwin99}. This inner bar actually produces a
\textit{minimum} in the ellipticity at $a \approx 2.6$ arcsec because it is
oriented close to the minor axis of the galaxy. The presence of these structures
reinforces the idea that the photometric bulge of NGC~3945 is predominantly a
disklike structure.

Finally, panel~f of Figure~\ref{fig:n3945a} shows \vdivsigma, using our
HET kinematics. In contrast to the simple classical-bulge galaxies
studied above, \vdivsigma{} reaches a peak value $> 2$ within the
photometric bulge region. This is clear, dramatic evidence that the
photometric bulge region is \textit{not} a simple, kinematically hot
structure; instead, the local stellar motions are sufficiently dominated
by rotation as to resemble the \textit{disk} regions of NGC~1332 and NGC~7457.
Combined with the previous morphological evidence, this makes NGC~3945
one of the clearest cases of an S0 galaxy with a disky pseudobulge.

\subsubsection{Inner Photometric Bulge as Classical Bulge} 

All of the foregoing is strong evidence that the photometric bulge of NGC~3945
is mostly, if not entirely, a disky pseudobulge. However, there is also good
evidence that the innermost regions of the galaxy are dominated by a separate
component, distinct from the pseudobulge.

The inner part ($r < 20$ arcsec) of the major-axis surface-brightness
profile (panel~b of Figure~\ref{fig:n3945b}) has two interesting
characteristics: first, most of the profile is very nearly a perfect
exponential; second, the inner $r \la 2$ arcsec show a steep central
excess. \citet{erwin03-id} pointed out that this profile appeared eerily
similar to that of an exponential plus an inner S\'ersic component, and
proceeded to treat it in just that fashion. Panel~b of
Figure~\ref{fig:n3945b} produces a revised version of that fit, this
time including the contribution from the lens/outer-disk (dashed green
line in the bottom left corner of the upper panel). Although the details
of the fit are slightly different from \citet{erwin03-id}, the result is
still an excellent match to the profile; the small deviations at $r <
5$ arcsec are due to the presence of the inner bar and its surrounding
nuclear ring. Given this \textit{inner} decomposition, we can identify a
new, much smaller photometric bulge region, with $\rbdi = 1.1$ arcsec (we use
\rbdi{} to indicate an \textit{inner} bulge = disk radius, in contrast
to the \rbd{} value from the global decomposition of the previous subsection).

The inner ellipse fits -- particularly those of the HST image -- show that the
region inside $\rbdi = 1.1$ arcsec is distinctly \textit{rounder} than the main
part of the disky pseudobulge (and rounder than the outer disk), with an
ellipticity $\approx 0.21$ (panel~c of Figure~\ref{fig:n3945b}). Photometrically
and morphologically, then, we have evidence for a distinct central component,
rounder than the outer disk \textit{and} the disky pseudobulge. 

What about the stellar kinematics? Here, we make use of the HST-STIS
kinematics published by \citet{gultekin09a}. Panel~d of
Figure~\ref{fig:n3945b} shows that \vdivsigma{} reaches a plateau value
of $\sim 0.5$ in the region $\la 1.1$ arcsec. Since the stellar
kinematics of this central structure appear dominated by random motions,
much like the larger classical bulges of NGC~1332 and NGC~7457 (above),
we conclude that this inner structure is a (small) classical bulge.

\subsection{Composite Bulge Example: NGC~4371} 

Like NGC~3945, NGC~4371 was noted by \citet{kormendy82} for its rather large
value of \vmaxsigstar{} -- second only to NGC~3945 in his sample. The variation
in ellipticity of the inner isophotes in this barred S0 galaxies led
\citet{wozniak95} to suggest that the galaxy might have a total of
\textit{three} bars: the obvious large outer bar and two more in the photometric
bulge region. Using \textit{HST} images, \citet{erwin99} were able to show that
the apparent signature of the inner two ``bars'' was actually the result of a
bright, stellar nuclear ring distorting the isophotes \citep[see
also][]{erwin01-nr}.

\subsubsection{Photometric Bulge as Disky Pseudobulge} 

\begin{figure*}
\includegraphics[width=6.0in]{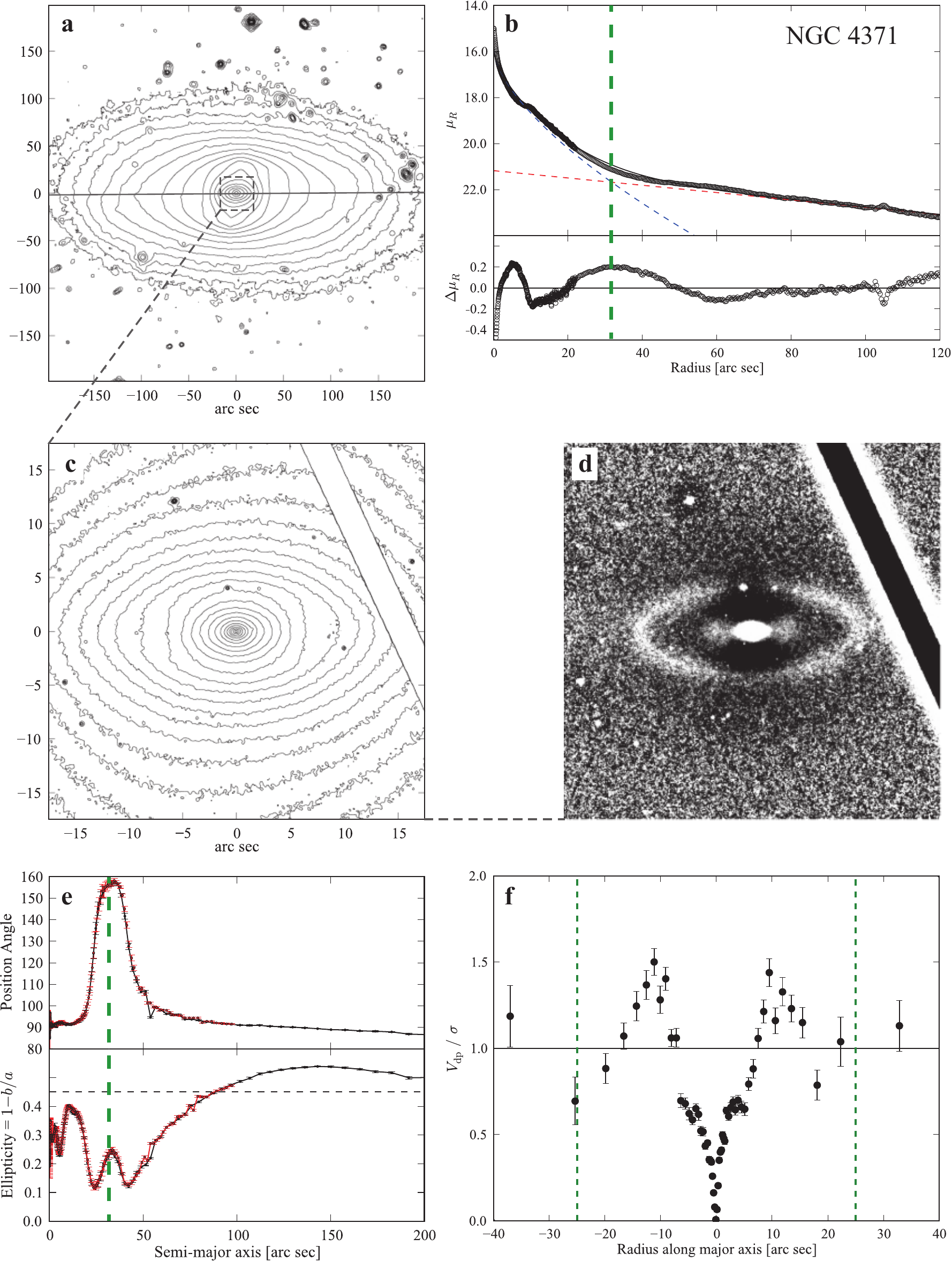}

\caption{Evidence for a disky pseudobulge in the S0 galaxy NGC~4371. \textbf{a:}
log-scaled $R$-band isophotes (INT-WFC image, smoothed with 15-pixel-wide median
filter); gray line marks major axis (PA $= 90.4\degr$). \textbf{b:} Bulge-disk
decomposition of INT-WFC major-axis profile.  Dashed lines represent S\'ersic +
exponential fit to the data, with residuals plotted in lower sub-panel. Vertical
dashed green line marks ``bulge=disk'' radius \rbd, where the S\'ersic and
exponential components are equally bright; this sets the boundary of the
``photometric bulge''. \textbf{c:} Close-up of photometric bulge region
(log-scaled contours from \textit{HST} ACS-WFC F850LP image). \textbf{d:}
Unsharp mask ($\sigma = 20$ pixels) of same region, showing stellar nuclear
ring. \textbf{e:} Ellipse fits to INT-WFC and \textit{HST} (red) images; note that
ellipticity reaches $\sim 0.40$, almost as high as the outer disk value
(horizontal dashed line), in photometric bulge; this suggests that at least part
of the photometric bulge has flattening similar to the outer disk. \textbf{f:}
Deprojected stellar rotation velocity divided by local velocity dispersion
$\vdivsigma$ along major axis, using WHT-ISIS long-slit data. Vertical
dashed lines mark the photometric bulge region $|R| < \rbd$; $\vdivsigma$ rises
to $\sim 1.5$ in this region, indicating a kinematically cool region more like a
disk.}\label{fig:n4371a}

\end{figure*}

Our B/D decomposition of the major-axis profile is shown in panel~b of
Figure~\ref{fig:n4371a}. The fit shows significant residuals, due in
part to the presence of the nuclear ring at $a \sim 10$ arcsec; however,
it does allow us to define the photometric bulge boundary as $\rbd =
25$ arcsec. (This is smaller than the bulge region defined by the
decomposition of \citealt{fisher-drory10}, presumably because their
ellipse-fit-derived profile includes light from the bar, which our
major-axis cut largely excludes.)

As was the case for NGC~3945, a close-up of the photometric bulge region shows
very elliptical isophotes interior to those defining the bar (panels~c and e of
Figure~\ref{fig:n4371a}); the peak ellipticity of $\sim 0.40$ is close to that of
the outer disk. (Note that the plotted ellipticity in panel~e peaks at 0.54 at
$a \sim 140$ arcsec due to the presence of an outer ring; the true outer-disk
ellipticity of 0.45 is determined from isophotes further out; see, e.g., Fig~3b
of \citealt{erwin05}.) Unsharp masking (panel~d) shows the nuclear ring, which
is a purely stellar phenomenon with no signs of gas, dust, or ongoing star
formation, though \citet{comeron10} did find that the ring has a slightly bluer
colour than the surrounding light in their \textit{HST} colour map. Unlike the case of
NGC~3945, there is no evidence for a nuclear bar in this galaxy.

The peculiarly box-shaped isophotes interior to the nuclear ring, which are visible at
$a \sim 5$--7arcsec and which produce the local ellipticity minimum in the ellipse
fits, can be explained as the side effect of adding the isophotes of an elliptical ring
to those of a rounder structure inside \citep[see][]{erwin01-nr}.

Finally, analysis of our major-axis WHT-ISIS spectroscopy shows that the \vdivsigma{}
profile (panel~f) has a peak of $\sim 1.5$ within the photometric bulge region -- in fact, the
peak is more or less at the radius of the nuclear ring. Once again, we have good evidence
that the photometric bulge region is kinematically more like a disk than a classical
spheroid, in addition to the clear morphological evidence for a disky pseudobulge.

\subsubsection{Inner Photometric Bulge as Classical Bulge}\label{sec:n4371b} 

\begin{figure*}
\includegraphics[width=6.0in]{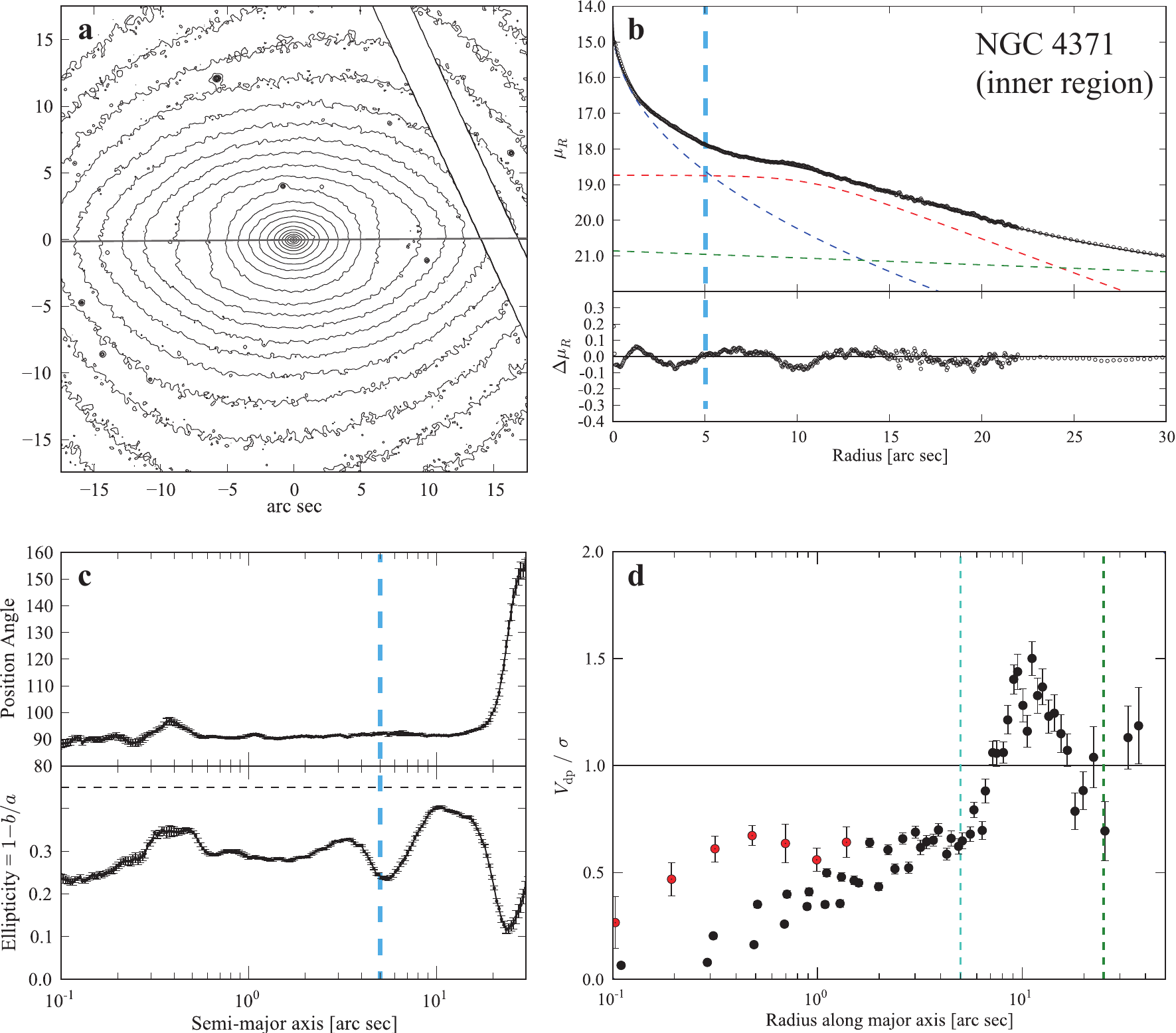}

\caption{Evidence for a classical bulge inside the disky pseudobulge of
NGC~4371. \textbf{a:} Close-up of main photometric bulge region from
\textit{HST} ACS-WFC F850LP image; gray line marks major-axis PA.
\textbf{b:} major-axis profile from \textit{HST} ($r < 22$ arcsec) and
INT-WFC images, with fit (S\'ersic + broken-exponential + outer
exponential; dashed lines) and residuals from fit in lower sub-panel.
Vertical dashed blue line marks inner ``bulge=disk'' radius \rbdi, where the
S\'ersic component has surface brightness equal to the sum of the disk
components; this sets the boundary of the \textit{inner} ``photometric
bulge''. \textbf{c:} Ellipse fits to the \textit{HST} image; note that
ellipticity in the inner photometric bulge region ($a \la 5$ arcsec) is
$\sim 0.28$, significantly less than that of the disky pseudobulge
outside. \textbf{d:} Deprojected, folded stellar rotation velocity
divided by local velocity dispersion $\vdivsigma$ along the major axis,
using WHT-ISIS long-slit data [black] and SINFONI AO data [red].
Vertical dashed blue and green lines mark the inner and main photometric
bulge regions $|R| < \rbdi$, respectively. In the inner region ($r \la
5$ arcsec), $\vdivsigma$ reaches a plateau value of $\approx 0.65$,
suggesting a kinematically hot region (i.e., a classical bulge).
}\label{fig:n4371b}

\end{figure*}

Although the inner profile is not as clean and simple as that of
NGC~3945, we can still identify a clear central excess at $r \la
5$ arcsec. Panel~b of Figure~\ref{fig:n4371b} shows a plausible
decomposition, where we treat the inner disk + nuclear ring as a single
component with a broken-exponential
profile \citep{erwin08}. (See \nocite{erwin14-smbh}Erwin et al.\ 2014 for
a more complex, 2D decomposition which yields similar results in terms
of the classical bulge.) As we did for NGC~3945, we can define an
inner value of $\rbdi = 5$ arcsec, where the S\'ersic component is
brighter than the sum of the outer exponential plus the
nuclear-ring/inner-disk component. 

The ellipticity of the \textit{HST} isophotes interior to this radius
(panel~c of Figure~\ref{fig:n4371b}) is consistently $\approx 0.30$ (the
variation at $a \approx 0.3$--0.6 arcsec is due to a circumnuclear
dust ring, noted by \nocite{comeron10}Comer\'on et al.\ 2010). Moreover,
the stellar kinematics for this region (panel~d of the same figure)
shows that \vdivsigma{} reaches a plateau of $\sim 0.65$ within $\rbdi =
5$ arcsec, so the kinematics of this region are dominated by velocity
dispersion instead of rotation. In other words, the inner $r < 5$ arcsec
of this galaxy appears to be dominated by a classical bulge.

\section{Other Composite Bulges}\label{sec:others} 

Detailed discussion and analysis of the other seven composite-bulge
galaxies can be found in Appendix~\ref{sec:others-full}. We note here
that not all these galaxies are as clear-cut as the two just discussed
(NGC~3945 and NGC~4371). For example, we lack kinematic data with high
enough spatial resolution to properly resolve the interior of the
classical-bulge region for some of those galaxies. In most such cases, however,
\vdivsigma{} is significantly $< 1$ in the (resolved) region just \textit{outside}
the classical bulge, so it is unlikely that $\vdivsigma$ is actually $> 1$ inside
the classical bulge.

For some galaxies, we use profiles from position angles other than the
galaxy major axis, especially when strong nuclear bars are present. This
includes profiles perpendicular to nuclear bars (when present), to minimize its
contribution, as well as alternate cases where we use a profile along
the nuclear-bar major axis, including an extra component to account for
the bar contribution.

The results of these analyses, combined with the previous ones for NGC~3945 and
NGC~4371, are presented in Tables~\ref{tab:disk-features}--\ref{tab:BtoT}.

\begin{table*}
\begin{minipage}{126mm}
\caption{Disky Pseudobulge Characteristics}
\label{tab:disk-features}
\begin{tabular}{@{}lllllllr}
\hline
Galaxy & Exp. Profile? & Max$(\vdivsigma)$ & Disky Features & $\mu_{0}$ & $h$ &  & $\log M_{\star}$ \\
    &     &     &     & (mag arcsec$^{-2}$) & (arcsec) & (pc) &   (\Msun{})   \\
(1) & (2) & (3) & (4) & (5) & (6) & (7) & (8) \\
\hline
NGC 1068  &   Y    &  2.80   &  bar, spirals    & 13.12 ($K$) & 6.80 & 468 & 10.89 \\
NGC 1543  &   Y    &  ---$^{1}$   &  bar, NR    & 17.15 ($I$) & 7.03 & 665 &  9.84 \\
NGC 1553  &   Y?   &  1.38   &  bar, NR/spirals & 15.31 ($I$) & 5.33 & 466 & 10.57 \\
NGC 2859  &   Y    &  1.56   &  bar, NR/spirals & 16.88 ($i$) & 4.45 & 522 & 10.49 \\
NGC 3368  &   Y    &  1.36   &  bar, spirals    & 12.71 ($K$) & 3.05 & 149 &  9.85 \\
NGC 3945  &   Y    &  2.18   &  bar, NR         & 15.69 ($I$) & 5.11 & 491 & 10.43 \\
NGC 4262  &   Y    &  1.58   &  NR              & 14.45 ($i$) & 1.66 & 124 & 10.09 \\
NGC 4371  &   Y[*] &  1.50   &  NR              & ---$^{2}$   & ---$^{2}$ &---$^{2}$ &  9.88 \\
NGC 4699  &   Y    &  1.36$^{3}$ &  bar, spirals & 15.83 ($z$) & 7.61 & 697 & 10.97 \\
\hline
\end{tabular}

\medskip
Column 1: Galaxy name. Column 2: Indicates whether surface-brightness
profile is exponential. Column 3: Maximum value of \vdivsigma{} in the
disky pseudobulge. Column 4: ``Disky'' features found in the pseudobulge
region (NR = nuclear ring); note that for NGC~1068 and NGC~3368, the bars
are only seen clearly in the near-IR. Column 5: Central surface brightness of
exponential component of disky pseudobulge; band is listed in parentheses.
Columns 6--7: Exponential scale length of same, listed in both angular and
linear sizes. Column 8: Logarithm of estimated stellar mass of disky pseudobulge (including
nuclear bar components; see text). Notes: 1 = galaxy too close
to face-on to reliably deproject velocities; 2 = In NGC~4371, the disky
pseudobulge profile is distorted by the strong nuclear ring,
and is not modeled as a simple exponential (see
Section~\ref{sec:n4371b}); 3 = kinematics do not extend to edge of
pseudobulge, so this may be only a \textit{lower} limit on
Max$(\vdivsigma)$.

\end{minipage}
\end{table*}

\begin{table*}
\begin{minipage}{126mm}
\caption{Classical Bulge Parameters}
\label{tab:classical-parameters}
\begin{tabular}{@{}lrrrlrr}
\hline
Name      & $n$     & \multicolumn{2}{c}{$R_{e}$} &  $\mu_{e}$  & $\epsilon$ & $\log M_{\star}$ \\
          &         & (arcsec) & (pc)    & (mag arcsec$^{-2}$) &  &  (\Msun{})       \\
(1)       & (2)     & (3)       & (4)     &  (5)        & (6) & (7)        \\
\hline
NGC 1068  &   0.98  & 0.50      &  34     &  11.05 ($K$)   & 0.14  &  9.47 \\
NGC 1543  &   1.50  & 2.73      & 258     &  17.02 ($I$)   & 0.05  &  9.49  \\
NGC 1553  &   1.66  & 1.48      & 129     &  15.73 ($I$)   & 0.1   &  9.32  \\
NGC 2859  &   1.68  & 1.06      & 124     &  17.00 ($i$)   & 0.15  &  9.47  \\
NGC 3368  &   1.34  & 0.63      &  31     &  13.38 ($K$)   & 0.00  &  8.73  \\
NGC 3945  &   2.02  & 1.24      & 119     &  16.20 ($I$)   & 0.20  &  9.63  \\
NGC 4262  &   0.89  & 0.30      &  23     &  15.02 ($i$)   & 0.05  &  8.70  \\
NGC 4371  &   2.18  & 5.20      & 427     &  18.70 ($z$)   & 0.30  &  9.64  \\
NGC 4699  &   1.43  & 2.15      & 247     &  15.42 ($z$)   & 0.11  & 10.46  \\
\hline
\end{tabular}

\medskip

Characteristics the of the classical bulges in our composite-bulge galaxies.
Column 1: Galaxy name. Columns 2--5: Parameters of S{\'e}rsic fit; effective
radius is given in both angular and linear sizes, and the band of the $\mu_{e}$
value is in parenthesis (these are usually based on the reddest available
images, which vary from galaxy to galaxy).   Column 6: Adopted mean isophotal ellipticity of
classical bulge. Column 7: Logarithm of estimated stellar mass (see text).
 
\end{minipage}
\end{table*}

\begin{table}
\caption{$B/T$ Values}
\label{tab:BtoT}
\begin{tabular}{@{}llrll}
\hline
Name      & $B/T_{\star, \mathrm{cl}}$ & $B/T_{\star, \mathrm{phot}}$   & $B/T_{L}$ (lit.) & Source \\
(1)       & (2)               & (3)                &  (4)         &  (5)   \\
\hline
NGC 1068  &   0.021 &   0.40  & 0.11 &   L10  \\
NGC 1543  &   0.10  &   0.00  & 0.34 &   L10  \\
NGC 1553  &   0.031 &   0.41  & 0.23 &   L10  \\
NGC 2859  &   0.070 &   0.38  & 0.29, 0.47 &   L10, F12  \\
NGC 3368  &   0.024 &   0.54  & 0.26, 0.41 &   F11, F12 \\
NGC 3945  &   0.045 &   0.37  & 0.39, 0.36 &   L10, F12  \\
NGC 4262  &   0.045 &   0.65  & 0.55 &   L10  \\
NGC 4371  &   0.093 &   0.27  & 0.22, 0.38 &   L10, F12  \\
NGC 4699  &   0.089 &   0.71  & 0.34 &   L04  \\
\hline
\end{tabular}

\medskip

Bulge-to-total values for composite-bulge galaxies. Column 1: Galaxy
name. Column 2: Ratio of classical bulge stellar mass to total galaxy stellar mass.
Column 3: Ratio of photometric bulge stellar mass to total galaxy stellar mass, from simple 1-D
decomposition. Column 4: Near-IR $B/T$ luminosity values from
the literature. Column 5: Sources for values in Column 4 -- L04 = 2D
$K$-band decompositions of \citet{laurikainen04b}; L10 = 2D $K$-band decompositions of
\citet{laurikainen10}; F11 = ``near-IR'' 1D decompositions of
\citet{fisher-drory11}; F12 = $H$-band 1D decompositions of \citet{fabricius12}.
 
\end{table}

\section{Stellar Dynamics of Classical Bulges and Disky Pseudobulges}\label{sec:dynamics} 

\begin{figure*}
\includegraphics[width=6.0in]{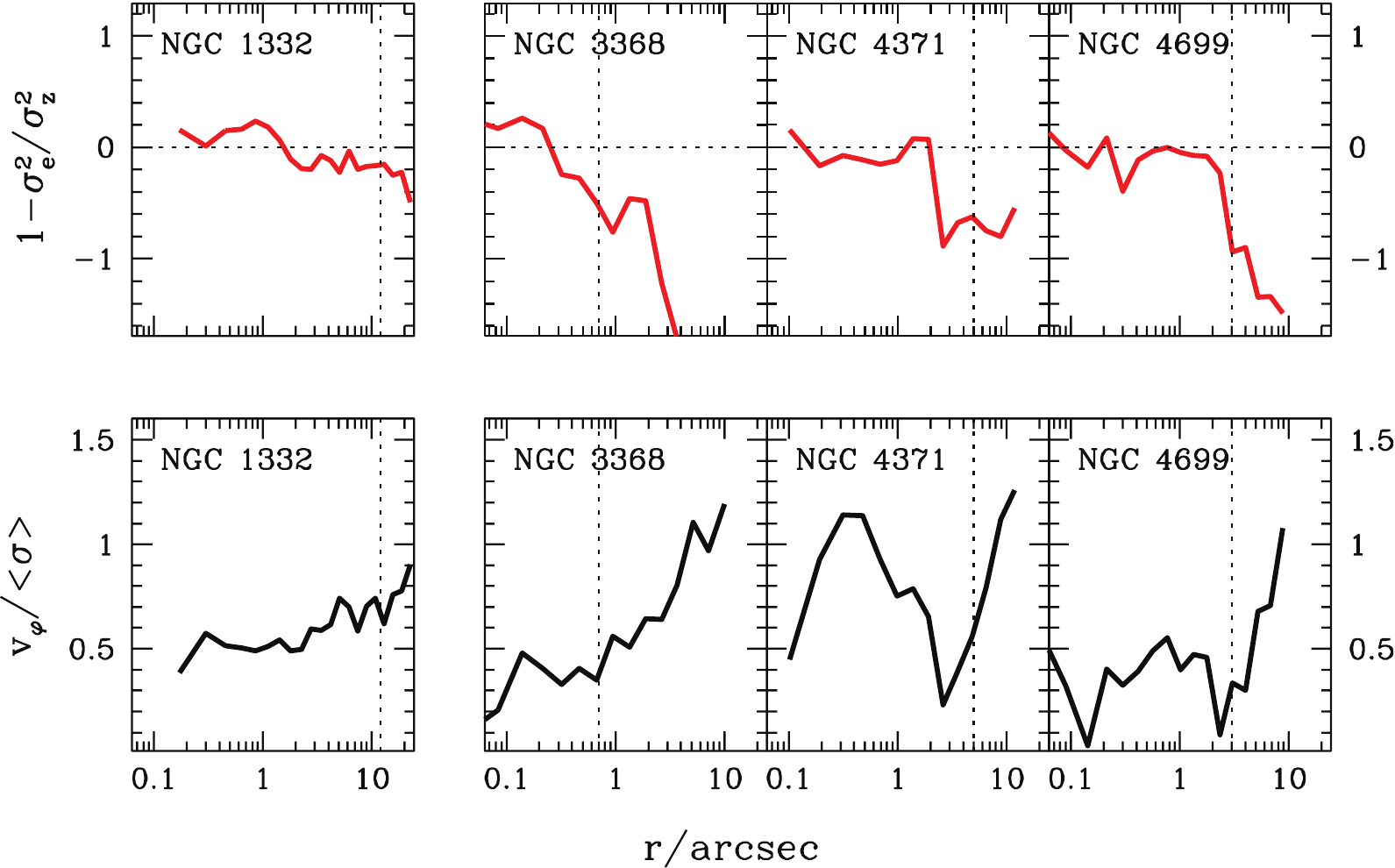}
  
\caption{3D stellar kinematics from dynamical modeling of galaxies
observed with SINFONI, for the classical-bulge galaxy NGC~1332 (left
panels) and three of the composite-bulge galaxies (right set of panels).
Vertical dashed lines mark \rbd{} for NGC~1332 and \rbdi{} for the
classical bulges in the composite-bulge galaxies; regions to the left
of these lines are dominated by the classical bulge components. Top panels:
$\beta_{\rm eq} \: = \: 1 \, - \, \sigma_{e}^{2} / \sigma_{z}^{2}$,
which measures the relative anisotropy of planar ($\sigma_{e}^{2} =
(\sigma_{R}^2 + \sigma_{\varphi}^2)/2$) versus vertical ($\sigma_{z}^{2}$)
velocity dispersion in the best-fitting models; values near zero
indicate isotropic velocity dispersions, while more negative values mean
planar dispersions dominate over vertical dispersions. Lower panels:
local ratio of mean rotation velocity $V_{\varphi}$ to total mean velocity
dispersion in the models.
}\label{fig:aniso_figure}

\end{figure*}

Three of the composite-bulge galaxies, along with one of the example
pure-classical-bulge galaxies (NGC~1332), have been observed with the
SINFONI IFU during a project measuring supermassive black hole masses in
galaxy centres
\citep{nowak07,nowak08,nowak10,rusli11,rusli13a,erwin14-smbh}. As part
of the analysis, we perform Schwarzschild orbit-superposition modeling
in order to reproduce both the stellar light distribution and the
observed stellar kinematics; see \citet{thomas04}, \citet{nowak10} and
\citet{rusli11} for details. Briefly stated, this involves constructing
a galaxy potential from the combination of a central SMBH and one or
more deprojected, 3D luminosity-density distributions (e.g., a
components for the classical bulge and one for the
disky-pseudobulge+outer-disk), converted to 3D stellar mass density
distributions via a stellar mass-to-light ($M/L$) ratio. Trial values of
the SMBH mass and the stellar $M/L$ ratios are assigned, and then
several tens of thousands of sample orbits are integrated in the
potential. The resulting orbit library is weighted to produce the best
match to the observed light distribution and the observed stellar
kinematics, and the process is then iterated with new values of $M/L$
and SMBH mass to map out the $\chi^{2}$ landscape.

The end result, in addition to best-fit values for the stellar $M/L$
ratios and the SMBH mass, is a library of stellar orbits and
corresponding weights which can be used to investigate the phase-space
distribution of the stars and to look for things such as radial trends
in 3D stellar dynamical quantities. We have previously made use of
best-fit orbit libraries to study radial anisotropy trends in core and
non-core elliptical galaxies \citep{thomas14}; here, we use the best-fit
models for the S0 NGC~1332 (which has only a classical bulge) and the
composite-bulge galaxies NGC~3368, NGC~4371 and NGC~4699 to explore how
the models might shed light on the internal 3D kinematics of composite
bulges.

Data and modeling for NGC~1332 and NGC~3368 have already been presented
in \citet{rusli11} and \citet{nowak10}, respectively; full data and
modeling results, including SMBH measurements, for the other galaxies
will be presented elsewhere \citep{erwin14-smbh}. Once the best-fitting
model is determined, stellar-dynamical quantities can be extracted using
the weighted means of orbits in different radial and angular bins.

Working in cylindrical coordinates $(R,\varphi,z)$, we compute the velocity
anisotropy by comparing the vertical dispersion $\sigma_{z}^{2}$ with
the mean velocity dispersion in the equatorial plane $\sigma_{e}^{2}$; the
latter is defined as
\begin{equation}
\sigma_{e}^2 = (\sigma_{R}^2 + \sigma_{\varphi}^2)/2 \, .
\end{equation}
All of these values are averages of the orbits over angular bins running from
$\theta = -23\degr$ to $\theta = +23\degr$ with respect to the equatorial plane
at each radius, using the orbit weights from the best-fitting solution. We 
then define the anisotropy $\beta_{\rm eq}$ as
\begin{equation}
\beta_{\rm eq} \; = \; 1 \: - \: \sigma_{e}^{2} / \sigma_{z}^{2} \, ;
\end{equation}
this is $\sim 0$ for the isotropic case and $< 0$ for planar-biased anisotropy.
For comparison, $\beta_{\rm eq} \approx -1.0$ for the Galactic disk in the
Solar neighborhood \citep[using $z = 0$ values from Table~1 of][]{bond10}, or $-1.2$
for the Galactic thick disk \citep[using dispersions from Table~5 of][]{carollo10}.

Figure~\ref{fig:aniso_figure} shows the radial trend of the anisotropy
for the simple classical-bulge S0 NGC~1332 and for three composite-bulge
galaxies. We also plot the radial trends of $V_{\varphi} / \langle \sigma
\rangle$, where $V_{\varphi}$ is the intrinsic mean rotation velocity
measured in the two bins closest to the equatorial plane and $\langle
\sigma \rangle$ is the total velocity dispersion: 
\begin{equation}
\langle \sigma \rangle = \sqrt{(\sigma_{R}^2 + \sigma_{\varphi}^2 +
\sigma_{z}^2)/3} \, .
\end{equation}

For all galaxies, the anisotropy parameter $\beta_{\rm eq}$ is $\sim 0$
in the classical-bulge region, and decreases outside, indicating a shift
from isotropic velocity dispersion to a dispersion which is dominated by
planar motions. The latter is what we expect for flattened, disklike
structures, and supports the idea that the disky pseudobulges are indeed
dynamically distinct from the classical bulges.

In most cases, the \vphidivsigma{} profiles show a trend similar to what
we have seen in the major-axis \vdivsigma{} profiles:
dispersion-dominated kinematics within the classical-bulge region and
rotation-dominated kinematics in the disky pseudobulge (or main disk in
the case of NGC~1332) outside.  The exception is NGC~4371, where there
is also an \textit{inner} peak in \vphidivsigma{} at $r \sim
0.4$ arcsec, deep within the classical-bulge region. Curiously, the
radius where \vphidivsigma{} reaches its local maximum is also where the
isophotal ellipticity has a local maximum (panel~c of
Figure~\ref{fig:n4371b}), though \textit{HST} colour maps indicate that
this is also a region marked by strong circumnuclear dust
\citep[see][]{comeron10}. This might represent the existence of an
additional disky component with $r \sim 30$~pc deep inside the classical
bulge.

\section{Discussion}\label{sec:discussion} 

In the preceding sections of this paper (and in the Appendix) we have
provided a kind of existence proof demonstrating that at least some
lenticular and early-type spiral galaxies can host \textit{both} disky
pseudobulges \textit{and} compact classical bulges, with the latter
nestled inside the former. This shows that classical bulges and
pseudobulges are not always exclusive phenomena.

The disky pseudobulges can almost always be described with exponential
profiles,\footnote{The main exception is NGC~4371, where the disky
pseudobulge is better fit with a broken-exponential profile; see
Section~\ref{sec:n4371b}.} with the addition of various disk-like
features: nuclear bars, stellar rings, or spiral arms. Our major-axis
decompositions yield exponential scale lengths of 125--870 pc, with a
mean of 440 pc. These are relatively massive structures, with stellar
masses ranging from $7.1 \times 10^{9}$ to $9.4 \times 10^{10}$~\Msun{}
(mean = $3.3 \times 10^{10}$~\Msun), or anywhere from 11 to 59 per cent
of the total galaxy stellar mass (mean fraction = 33 per cent). Thus in
many cases a significant fraction of the galaxy's stars are part of the
disky pseudobulge. (See Section~\ref{sec:pb-plots} for details on the
computation of the disky-pseudobulge stellar masses.)

The classical bulges, in contrast, are relatively compact, low-mass
structures. Our fits yield S\'ersic indices of 0.89--2.18 (mean index =
1.52), half-light radii between 23 and 426~pc (mean $R_{e} = 143$~pc),
and stellar masses ranging from $5.0 \times 10^{8}$ to $2.9 \times
10^{10}$~\Msun. (See Section~\ref{sec:cb-plots} for details of the
stellar-mass estimation.) In only two galaxies is the stellar mass of
the classical bulge more than 10 per cent of the galaxy's total stellar mass
(NGC~1543 and NGC~4699, where it is 13.1 and 11.3 per cent, respectively),
and the mean value is only 5.9 per cent.

\subsection{Hints Concerning the Frequency of Composite Bulges}\label{sec:frequency}

How common are composite-bulge systems? The disadvantage of an
existence-proof study such as this one is that it can do little, by
itself, to answer this question. The galaxies discussed in this paper
are unfortunately not drawn from any well-defined sample which has
been consistently analysed with the same degree of spatial resolution in
both imaging and spectroscopy, primarily because the necessary combined
data (particularly high-resolution spectroscopy for stellar kinematics)
are not available for most nearby galaxies. We \textit{can} note that
most of our galaxies are part of a sample of early-type barred galaxies
originally studied by \citet{erwin-sparke03} and expanded by
\citet{erwin05}, which does allow us to put some very crude lower
limits. Of the 25 barred S0 galaxies in the aforementioned sample with
$i > 30\degr$, we can identify three which are composite bulge hosts
(NGC~2859, NGC~3945 and NGC~4371). Of the 34 barred S0/a--Sb barred
galaxies with the same inclination cutoff, there are three more galaxies
from our composite-bulge set (NGC~1068, NGC~3368 and NGC~4699). This
suggests that \textit{at least} $\sim 10$ per cent of S0--Sb barred
galaxies are probably composite-bulge systems. The true frequency could
be much higher -- though it clearly cannot be 100 per cent, as there are
clearly some galaxies with classical bulges but no disky pseudobulges
(e.g., NGC~1332 and NGC~7457, Sections~\ref{sec:n7457} and
\ref{sec:n1332}).

If we take the presence of nuclear bars as indicators of disky
pseudobulges, as is often done
\citep[e.g.,][]{kormendy04,fisher-drory08}, then the recent review of
\citet{erwin11} suggests that disky pseudobulges (with or without
classical bulges) can be found in at least 20 per cent of early-type (S0--Sab)
disk galaxies. The presence of nuclear bars as indicators of disky
pseudobulges also means that weaker, less luminous disky pseudobulges
may be lurking within more dominant classical bulges. For example, at
least four of the galaxies classified as having classical bulges in the
recent studies of \citet{fisher-drory08} and \citet{fisher-drory10} have
nuclear bars: NGC~3031, NGC~3992, NGC~4548 and NGC~6684
\citep{elmegreen95,erwin04,gutierrez11,erwin14-db}.

\citet{fisher-drory10} argued that galaxies with composite bulges would
not be clearly distinguishable from galaxies with only a disky
pseudobulge when using 1-D surface-brightness profiles, unless the
classical bulge were very luminous; they suggested that some of the
galaxies with what they called ``inactive pseudobulges''\footnote{Low
S\'ersic indices and/or disky morphologies combined with mid-IR colours
indicative of low star-formation rates.} might harbor classical bulges
as well, if the latter had relatively small S\'ersic indices (e.g., $\la
3$). This is exactly what we find: all of our classical-bulge components
have S\'ersic indices $\la 2.2$. Of the four galaxies in common with
their sample (NGC~1543, NGC~3368, NGC~3945 and NGC~4371), all are
classified by Fisher \& Drory as ``inactive pseudobulges'' based on the
mid-IR colours. It is worth noting that their 1-D decompositions produced
S\'ersic indices for the \textit{photometric bulges} of $n < 2$ for all
but one of these galaxies (NGC~4371), so one cannot conclude from the
S\'ersic index of the photometric bulge alone that a galaxy is
\textit{lacking} a classical-bulge component.

\subsection{Embedded Classical Bulges Compared with Other
``Spheroid'' Systems}\label{sec:cb-plots} 

Table~\ref{tab:classical-parameters} presents the structural
characteristics of the classical bulges in our composite-bulge systems.
The listed colours come from aperture photometry on SDSS images \citep[or
\textit{HST} images in the case of NGC~3945;][]{erwin03-id}; we use the
colour-to-$M/L$ calibrations of \citet{bell03} to calculate the resulting
stellar masses. The exception to this is NGC~1068, where the bright AGN
point source makes accurate stellar colour measurements in the inner few
arc seconds very difficult. Instead, we adopted a $K$-band $M/L$ ratio
of 0.7, using the lower end of the age estimate (5--12 Gyr) from the
near-IR spectroscopy of \citet{storchi-bergmann12} and the average $M/L$
ratios of
\citet{longhetti09}.\footnote{http://www.brera.inaf.it/utenti/marcella/
mtol.html}

In this paper, we refer to these compact, inner components
as ``classical bulges'' because they fulfill at least some of the
standard criteria for classical bulges. The fact that they have rounder
isophotes than the outer disk, yet share similar or identical position
angles, suggests they are approximately oblate spheroids (with little
triaxiality) which are rounder (in an edge-on, vertical sense) than the
disk. In addition, we have clear evidence that the classical bulges are
kinematically hot, with \vdivsigma{} always $< 1$, in the cases of
NGC~1068, NGC~3368, NGC~3945, NGC~4371 and NGC~4699, where the
available kinematic data comes from observations with high enough
spatial resolution to resolve the classical bulge region. In other
galaxies, such as NGC~2859.

The one exception to the usual classical-bulge paradigm is in the luminosity
profiles. Traditionally, classical bulges have been claimed to have
$R^{1/4}$ profiles -- i.e., a S\'ersic index of $n = 4$. More recent
formulations have suggested an approximate dividing line of $n = 2$,
with classical bulges having higher values and pseudobulges (of whatever
kind) having lower \citep[e.g.,][]{fisher-drory10}.  Almost all of the
objects we find have $n < 2$; only for NGC~3945 and NGC~4371 is $n \ga
2$. So one \textit{could} argue, purely on the basis of the S\'ersic
indices, that most of our central structures are a peculiar species of
round, kinematically hot ``pseudobulge.''  However, we prefer to
consider the possibility that classical bulges can have a range of
surface-brightness profiles, and that their primary characteristics
remain their overall shape, their lack of disky substructure, and
pressure-dominated stellar kinematics.

There are other kinematically hot, round stellar systems which are not
considered classical bulges, of course. As we have seen, the classical
bulge parts of composite bulges span a range of sizes, with half-light
radii from $\sim 400$ down to $\sim 25$ pc. The lower end of this
range is sufficiently small that one might wonder whether some of these
objects would really be better understood as ``nuclear star clusters''
(NSCs), which are extremely common in late-type spirals and at least
moderately common in earlier-type disks \citep[see, e.g., the review
by][]{boker08}.

\begin{figure*}
\includegraphics[width=6.5in]{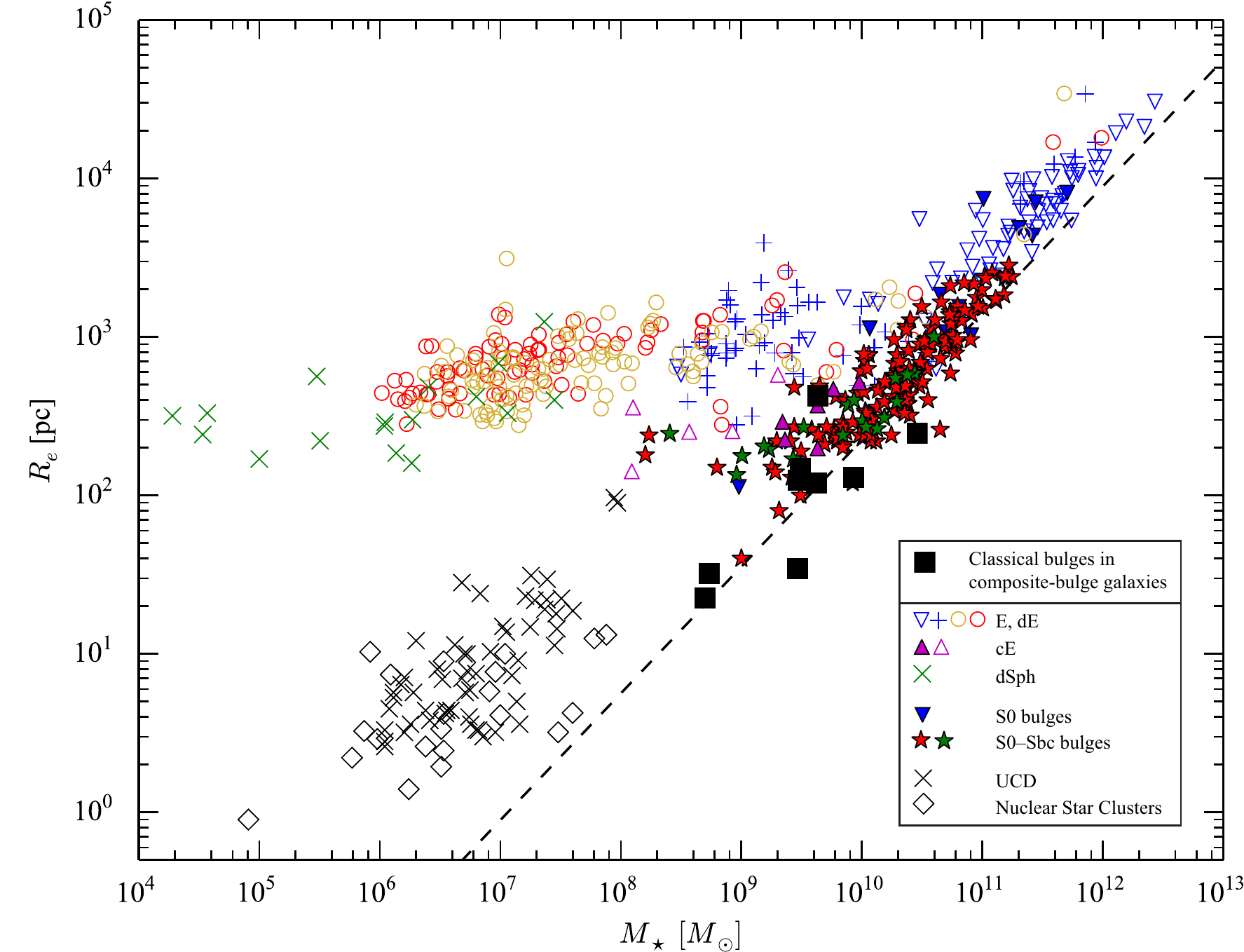}

\caption{Half-light radius versus total stellar mass for a variety of
``spheroids'' or ``hot stellar systems'', based partly on Figure~4 of
\citet{misgeld11}. All except the squares, diamonds and stars are taken
from the their compilation: open and filled blue (inverted) triangles
are ellipticals and bulges of S0 galaxies, respectively, from
\citet{bender93}; blue plus signs and red and orange open circles are E
and dE galaxies from the ACS Virgo Cluster Survey \citep{ferrarese06b},
the Hydra I Cluster \citep{misgeld08} and the Centaurus Cluster
\citep{misgeld09}, respectively; magenta triangles are compact
ellipticals \citep[open triangles are Coma galaxies from][]{price09}; and green and black
$\times$ symbols are dSph galaxies and UCDs, respectively. Red stars are bulges
of S0--Sa galaxies from the 2D decompositions of \citet{laurikainen10};
green stars are bulges of unbarred S0--Sbc galaxies from
\citet{balcells07a}; and open black diamonds are nuclear star clusters
(see text for sources).  Finally, the filled black squares are the
\textit{classical} bulge components of our composite-bulge galaxies;
they appear to fall into the narrow sequence defined by ellipticals and
(early-type disk) bulges. The dashed line is the ``zone of avoidance''
from Eqn.~8 of \citet{misgeld11}.} \label{fig:re-vs-mstar}

\end{figure*}

\begin{figure*}
\includegraphics[width=6.5in]{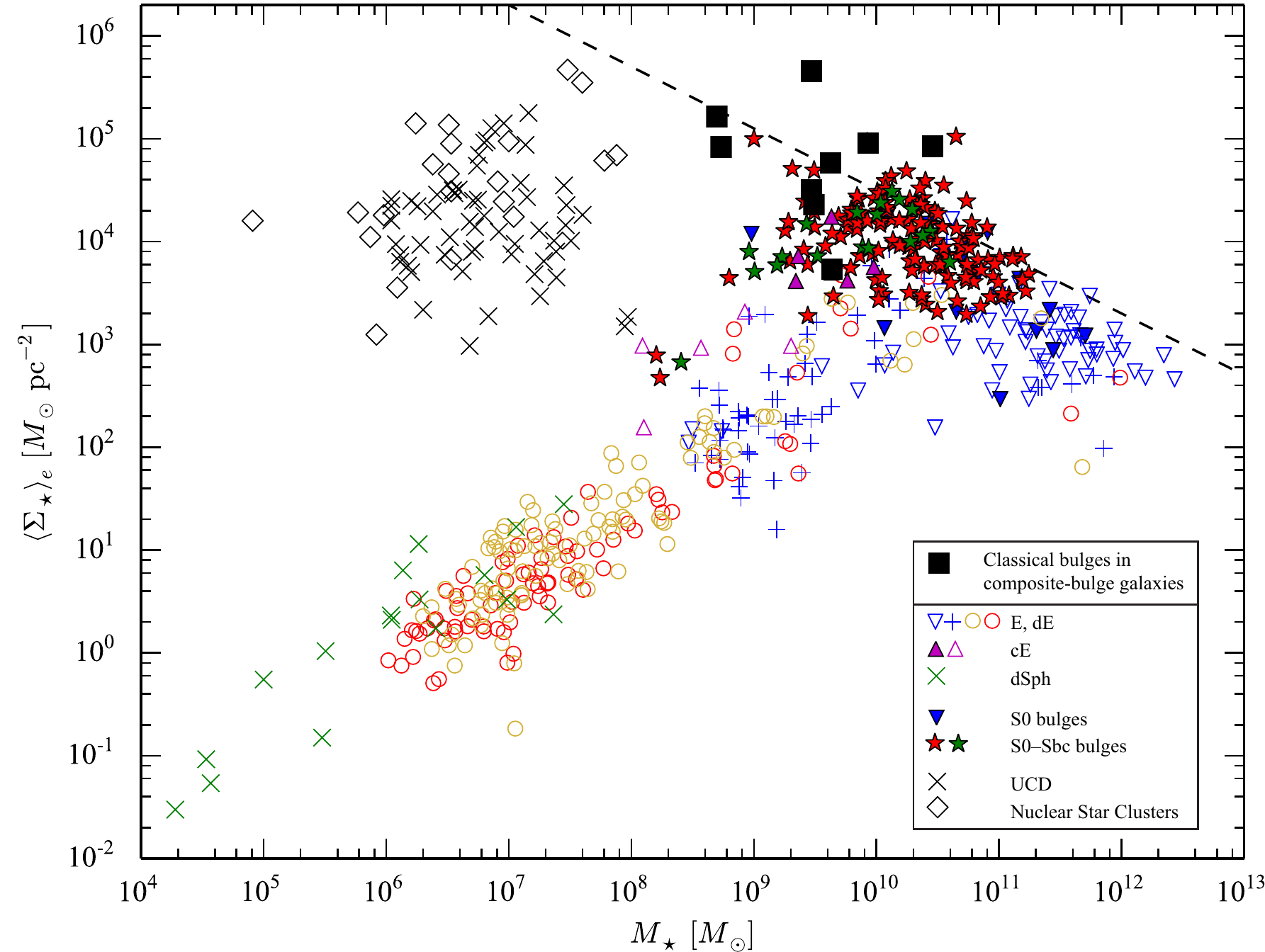}
\caption{As for Figure~\ref{fig:re-vs-mstar}, but now showing mean  stellar mass density 
within the half-light radius versus total stellar mass for the same systems.}
\label{fig:mue-vs-mstar}
\end{figure*}

In Figures~\ref{fig:re-vs-mstar} and \ref{fig:mue-vs-mstar}, we plot the
classical bulge components of our composite bulges in the context of
other ``spheroids'' -- that is, stellar systems dominated by velocity
dispersion -- ranging from NSCs, dwarf spheroidals and ultracompact
dwarfs to giant ellipticals. For the latter three classes of objects, we
use the aggregated data compiled by \citet{misgeld11}. 

To those data we have added the (photometric) bulges of S0--Sa galaxies
from Table~1 of \citet{laurikainen10} (red stars). These are more modern
measurements based on 2D decompositions of near-IR images, which helps
ensure that light belonging to the bar is not counted as part of the
bulge. (We note that these decompositions do not account for the
possibility of composite bulges, and we exclude the seven galaxies in
their sample which are studied in this paper.) We converted their $K$-band
measurements to stellar masses using the $M/L$ ratios of \citet{bell03}
and colours from HyperLEDA. In particular, we used $(B - V)_{e}$ colours,
which are a better approximation to the bulge colours than the total
galaxy colour.\footnote{For two galaxies without $(B - V)_{e}$ colours
(NGC~5636 and NGC~5953), we used $g - r$ colours measured within an
aperture of 5 arcsec radius on SDSS images; we were unable to find or
measure colours for NGC~3892, which is excluded from the plots.} We also
include the bulges of unbarred S0--Sbc galaxies from
\citet{balcells07a}, converting their $K$-band magnitudes to stellar
masses in the same fashion (green stars). Finally, we plot NSCs using a
compilation of masses from \citet{erwin-gadotti12} and half-light radii
primarily from \citet{boker04} and \citet{walcher05}, with some
additional values from \citet{ho96}, \citet{graham09}, \citet{barth09},
\citet{kormendy10b} and \citet{seth10}.

What Figure~\ref{fig:re-vs-mstar} shows is that the structures we
identify as classical bulges in the composite-bulge systems fall into
the ``bulge'' section of the plot, not the ``nuclear cluster'' section. 
Elliptical galaxies\footnote{Here we are excluding dE galaxies
-- roughly, those ``ellipticals'' with $M_{\star} \la 5 \times 10^{9}
\Msun$ -- which tend to form a separate sequence merging into the dwarf
spheroidals; see, e.g., arguments in \citet{kormendy12}, where dE and
dSph are together termed ``spheroidals''.} and classical bulges form a clear,
relatively tight sequence in the $R_{e}$--$M_{\star}$ plane, and our
classical bulges either fall directly into this sequence or form a
natural extension of it. Similarly, in Figure~\ref{fig:mue-vs-mstar},
our classical bulges fall into the same (loose) sequence formed by giant
ellipticals and larger classical bulges.

Although the smallest classical bulges are undeniably small, and begin
to approach the most massive NSCs in \textit{size} (half-light radius),
they are significantly more massive. Moreover, in at least two of the
composite-bulge galaxies (NGC~1543 and NGC~1553) there is evidence for
distinct nuclear star clusters with half-light radii $\sim 4$~pc located
\textit{inside} the classical bulge components. While we cannot rule out
the possibility of an as-yet undetected population of central
spheroids with masses $\sim 10^{8}$~\Msun{} and $R_{e} \sim 10$--20~pc, which
would provide continuity between the classical
bulges and NSCs, for the time being we consider them to be distinct
phenomena.

\subsection{Disky Pseudobulges in the Mass-Size-Density Planes}\label{sec:pb-plots} 

\begin{figure*}
\includegraphics[width=6.5in]{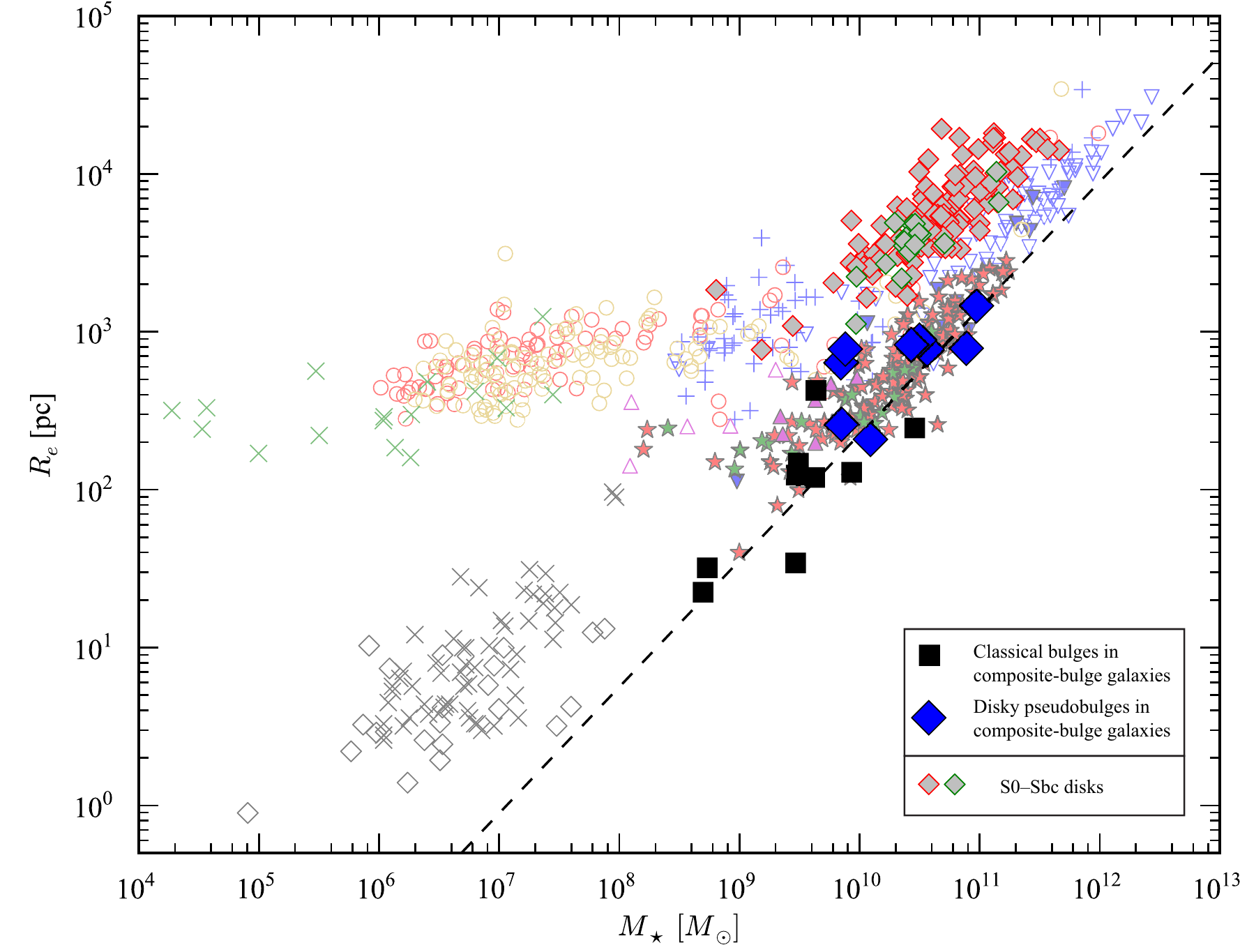}

\caption{As for Figure~\ref{fig:re-vs-mstar}, but now including the disky
pseudobulge components of our composite-bulge galaxies (large blue
diamonds), along with the large-scale disks (smaller grey diamonds) from
\citet[][green borders]{balcells07a} and \citet[][red
borders]{laurikainen10}. The disky pseudobulges are clearly more compact
than large-scale disks of the same mass, and actually fall on the same
bulge/elliptical sequence as the classical-bulge components (filled
black squares).} \label{fig:re-vs-mstar-pseudo}

\end{figure*}

\begin{figure*}
\includegraphics[width=6.5in]{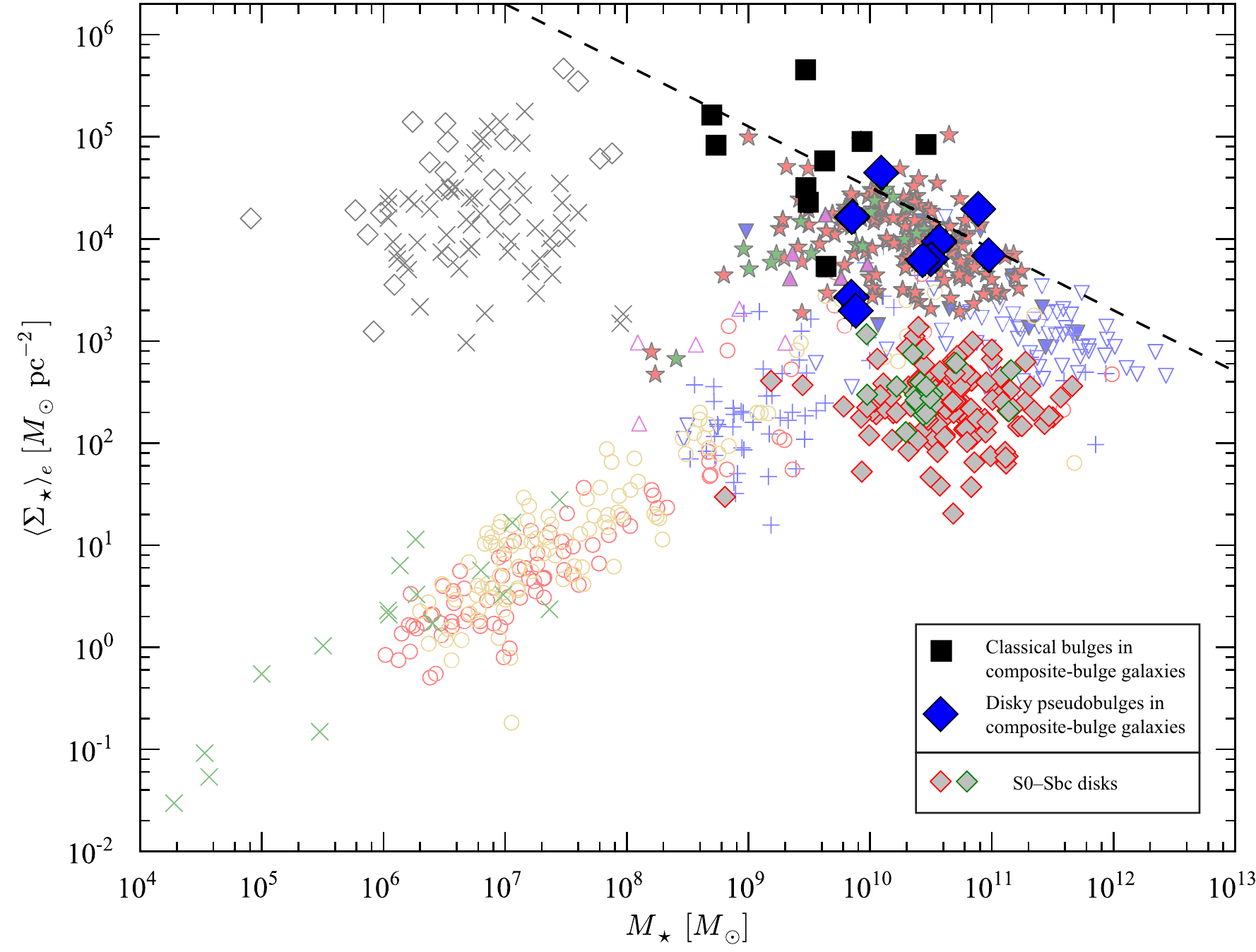}
\caption{As for Figure~\ref{fig:re-vs-mstar-pseudo}, but now showing
mean stellar mass density within the half-light radius versus total
stellar mass for the same systems.} \label{fig:mue-vs-mstar-pseudo}
\end{figure*}

As we have seen, the classical-bulge components of our composite-bulge
systems appear to follow the same trends as (larger) classical bulge and
ellipticals, at least in the $R_{e}$--$M_{\star}$ and $\langle
\Sigma_{\star} \rangle_{e}$--$M_{\star}$ planes. What about the disky
pseudobulges: do they resemble large-scale disks, classical bulges, or
something else? Figure~\ref{fig:re-vs-mstar-pseudo} and
\ref{fig:mue-vs-mstar-pseudo} show where the disky pseudobulges (filled
blue diamonds) fall in the same diagrams used for
Figures~\ref{fig:re-vs-mstar} and \ref{fig:mue-vs-mstar}; we have
desaturated the previously plotted points to allow the pseudobulge data
points to stand out more clearly. We also plot the (large-scale) disk
components from the S0/spiral decompositions of \citet{balcells07a} and
\citet{laurikainen10} as filled grey diamonds.

The stellar masses of our disky pseudobulges -- and the main disks from
the \citet{laurikainen10} and \citet{balcells07a} samples -- plotted in
Figures~\ref{fig:re-vs-mstar-pseudo} and \ref{fig:mue-vs-mstar-pseudo}
come from the same general colour-to-$M/L$ estimates \citep[based
on][]{bell03} as used for the classical-bulge components in the
preceding section. For most of the disky pseudobulges, we use the
exponential component from the inner decomposition (e.g., panel~b of
Figure~\ref{fig:n3945b}) and assume that the disky pseudobulge is an
exponential disk with an ellipticity equal to that of the outer disk.
(For NGC~4371, we use the plotted broken-exponential fit from panel~b of
Figure~\ref{fig:n4371b}.) In the cases of NGC~1068, NGC~1543 and
NGC~2859, where the inner fits specifically exclude the contribution of
bright inner bars, we include an additional flux term to account for the
inner bar itself \citep[e.g.,][]{erwin11}. The half-light radii are
computed using the best-fitting exponential scale lengths from the inner
decompositions (e.g., panel~b of Figure~\ref{fig:n3945b}) -- i.e.,
excluding any contribution from the classical bulge component. (The
half-light radii for the large-scale disks are also estimated this way.)
For disky pseudobulges with significant inner bars, this may
mis-estimate the true half-light radius, but probably by less than a
factor of 2, which means a very small shift on the plots.

What we can see from Figures~\ref{fig:re-vs-mstar-pseudo} and
\ref{fig:mue-vs-mstar-pseudo} is that in terms of size, stellar mass,
and mean density, disky pseudobulges actually fall into the
bulge/elliptical sequence, just like the classical bulges (albeit with
masses and sizes that are on average larger than the classical bulges).
Although some of the disky pseudobulges are as massive as large-scale
stellar disks, they are roughly a factor of $\sim 5$ times more compact.
This suggests that their formation mechanisms are probably not the same
as those involved in large-scale disks. It also is an indication that
otherwise unclassified objects which fall on the
classical-bulge/elliptical sequence in the mass-size and mass-density
planes may not always be kinematically hot spheroidal systems.

(We note in passing that the precise location and distribution of large-scale disks on
plots such as these may depend partly on Hubble type. For example, disks in very late
type spirals may lie further to the left in the $R_{e}$--$M_{\star}$
plane; see, e.g., Fig.~9 of \citealt{laurikainen10} or Figs.~18 and 20 of
\citealt{kormendy12}.)

\subsection{What About Boxy/Peanut-Shaped Pseudobulges?}\label{sec:boxy} 

\citet{athanassoula05} argued that instead of lumping the central regions of
disk galaxies into just two types -- classical bulges versus pseudobulges -- one could, from
a theoretical point of view, distinguish at least three possible structures:
classical bulges, ``disc-like bulges'' (i.e., what we have been calling disky
pseudobulges), and boxy/peanut-shaped bulges, which are actually the vertically
thickened inner parts of bars. She even suggested that some galaxies might
contain all three at the same time.

In this paper, we have identified a number of disk galaxies containing
\textit{both} classical bulges and disky pseudobulges; what about
boxy/peanut-bulges? We have avoided dealing with the latter in order to
concentrate on the contrasting disky and spheroidal nature of the
pseudobulges and classical bulges we are interested in; moreover, the
traditional approach to identifying boxy/peanut-shaped bulges relies on galaxies
which are edge-on, not moderately inclined. None the less, all but one of the
galaxies discussed in this paper are barred. The extensive survey of
edge-on galaxies by \citet{lutticke00a} found that the frequency of boxy
and peanut-shaped bulges was consistent with most if not all barred
galaxies having a vertically thickened inner region (a box/peanut
structure, or B/P structure). \citet{bureau06} identified a number of
edge-on galaxies which appeared to have both B/P structures \textit{and}
disky pseudobulges, so the coexistence of those two structures is
certainly plausible. (Due to low spatial resolution and the presence of strong
dust extinction in the central regions, small embedded classical bulges
would have been difficult to find in their galaxies, if any such were
present.)

\begin{figure}
\includegraphics[width=3.2in]{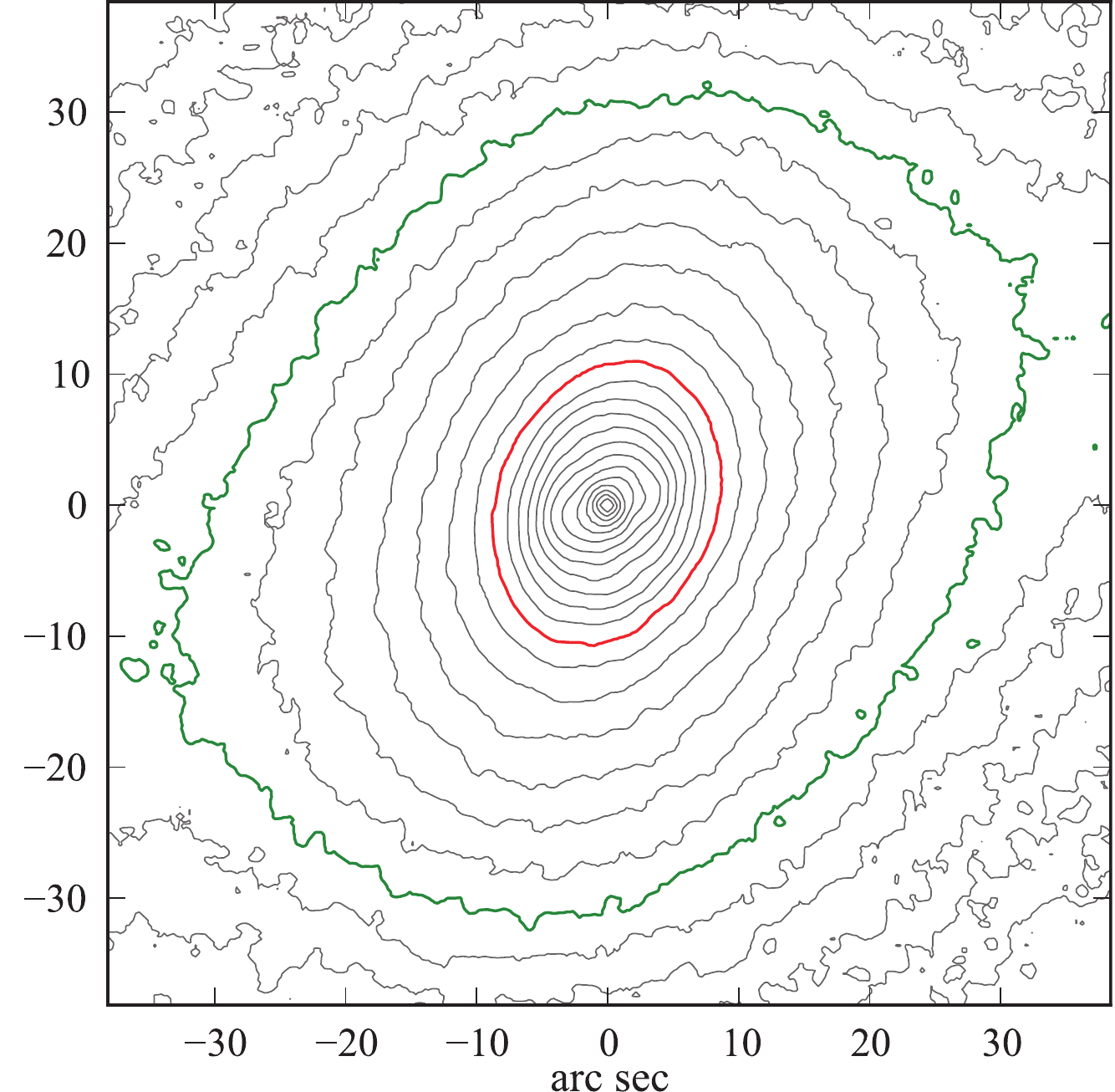}
  
\caption{Disky pseudobulge inside a boxy/peanut-shaped bulge. Median-smoothed
$K$-band isophotes for the interior of the outer bar of NGC~3368 (see
Figure~\ref{fig:n3368a} for larger-scale context). The thicker green contour
indicate the region of boxy isophotes, which are probably the projection of the
inner, vertically thickened part of the bar -- a boxy/peanut-shaped bulge. The
region inside which is dominated by the disky pseudobulge is indicated by the
thicker red contour. See also Fig.~A1 of \citet{erwin-debattista13}.}\label{fig:n3368-boxy}

\end{figure}

More recently, \citet{erwin-debattista13} identified several
morphological signatures of the B/P structure in barred galaxies which
can be seen when the galaxy is only moderately inclined (i.e., $i \sim
40$--70\degr), if the bar is favorably aligned with respect to the disk
major axis. Since NGC~1068, NGC~1543, NGC~2859 and NGC~4262 are close to
face-on (mostly $i \la 30\degr$), we would not expect to see signatures
of the B/P structure. Additionally, the bars in NGC~3945 and NGC~4371
are oriented very close to the minor axes of their respective host
galaxies. The latter orientation is one which minimizes the signature of
the B/P structure; \citeauthor{erwin-debattista13} did not find this
signature in those three galaxies. However, they \textit{did} identify
NGC~3368 as an example of a galaxy with a visible projected B/P
structure; as we noted in Section~\ref{sec:n3368}, this is probably
responsible for the slightly boxy isophotes \textit{outside} the disky
pseudobulge -- see Figure~\ref{fig:n3368-boxy}. Thus, NGC~3368 is a clear
case of a galaxy matching the suggestion of \citet{athanassoula05} that
disk galaxies can simultaneously host classical bulges, disky pseudobulges,
\textit{and} box/peanut bulges.

\subsection{Speculations on Formation} 

\subsubsection{Morphology: Disky Pseudobulges Are (Mostly) Found Inside Bars}\label{sec:pb-formation}

Most discussions of pseudobulges \citep[see, e.g.,][]{kormendy04}
portray them as secondary structures formed late in a galaxy's history,
usually via some form of bar-driven gas inflow and subsequent star
formation in the central regions of the galaxy.  The fact that almost
all the galaxies discussed in this paper are barred is at least broadly
consistent with the hypothesis of bar-driven formation. What is perhaps
not yet clear is whether bar-driven inflow and star formation can
readily create disky pseudobulges as large and as massive as some of
those we identify. A possible additional issue is the fact that
star-formation inside the bars of early-type disks in the local universe
is usually observed taking place in nuclear \textit{rings}. Several of
our disky pseudobulges do feature nuclear rings
(Table~\ref{tab:disk-features}), which argues for a connection; the
question is whether the rest of the disky pseudobulge, both inside and
outside the ring, was formed the same way.

\citet{wozniak09} studied a high-resolution disk galaxy simulation and
noted the formation of an extended ``nuclear disk'' inside the bar, due
to gas-driven bar inflow and star formation.   This disk apparently gave
rise to an independently rotating nuclear bar, with a gaseous nuclear
ring of radius $\sim 400$ pc. The reported size of the disk -- $\sim
500$ pc after 2 Gyr -- is consistent with some of our disky
pseudobulges, and the mass ($7 \times 10^9$~\Msun, 34 per cent of the
galaxy's total stellar mass) is as well.  More recently,
\citet{cole14} have reported on a detailed disk-galaxy simulation in
which a similar large nuclear disk formed inside a bar, and compared it
to the disky pseudobulges in NGC~3368, NGC~3945, and NGC~4371. The
nuclear disk in the simulation was more extended (relative to the size
of the bar) than is the case for the three disky pseudobulges, but did
have a similar mass fraction (29\% of the simulated galaxy's total
stellar mass). So it appears that at least some detailed,
high-resolution simulations \textit{can} produce barred galaxies with
disky pseudobulges similar to those seen in real galaxies.

We note that three of the composite-bulge galaxies do pose some more
general problems for the bar-driven formation scenario, because their
bars are weak, missing, or simply too small. NGC~1068, though formally
unbarred, does at least appear to have a large-scale bar surrounding its
disky pseudobulge \citep[][and references therein]{erwin04}; however,
this bar is quite weak, and is sometimes considered an ``oval disk''
rather than a true bar \citep{kormendy79b,kormendy04}. It is not clear
whether such a large disky pseudobulge (44 per cent of the NGC~1068's stellar
mass) would be formed by such a weak bar. NGC~1553 is classified as
unbarred, though it might be possible that its ``lens'' is either the
remnant of a once-strong bar \citep[as argued by, e.g.,][]{kormendy04},
or is the projection of the B/P structure of a still-existing,
albeit rather weak, bar (see Section~\ref{sec:n1553}).  The most
difficult case is NGC~4699, where the disky pseudobulge is
\textit{larger} than the sole bar in the system. This implies that not
\textit{all} disky pseudobulges can be formed by bar-driven gas inflow.
The alternative would be to postulate a previous, large bar in NGC~4699
which has since vanished. Unfortunately, theories of bar destruction
either require unnaturally high central mass concentrations (which
should prevent the existence of the \textit{current} small bar in this
galaxy), or don't yield testable predictions that would allow us to
unambiguously identify whether a galaxy once had a bar, as opposed to
never having had one. 

A different formation scenario, which could potentially account for
cases like NGC~4699, might be that some disky pseudobulges form
\textit{early}, possibly before any large-scale bar, as part of a
general inside-out process.  For example, \citet{guedes13} have
presented a cosmologically motivated disk-galaxy simulation which formed
with a ``pseudobulge'' inside a larger disk; their analysis showed that
most of the stars in the pseudobulge formed \textit{in situ} and
\textit{prior to} most of the outer disk.

One way of distinguishing between these scenarios would be to look for
evidence of significant differences in ages between the stars making up
the disky pseudobulge and those making up the bar (if present) or disk
outside. Younger ages in the disky pseudobulge would argue for the
bar-driven formation mechanism, while a predominantly \textit{older}
population in the disky pseudobulge would suggest something more like
the inside-out scenario of \citet{guedes13}.

\subsubsection{Possible Clues from Stellar Populations} 

Unfortunately, we are rather lacking in detailed stellar population analyses
for most of our composite-bulge galaxies. Published analyses tend to be
patchy and limited, either spatially or in terms of models with multiple
populations.

\nocite{delorenzo-caceres12,delorenzo-caceres13}De Lorenzo-C{\'a}ceres et
al.\ (2012, 2013) presented an analysis of stellar populations near the
centres of five double-barred galaxies, one of which is the
composite-bulge system NGC~2859. They found minimal age differences
between the stars of the inner and outer bars, with the former being
either coeval with or slightly younger than the latter. This seems to
suggest that the inner bars -- and by implication the rest of the disky
pseudobulges -- in these galaxies could indeed be (somewhat) younger
than the outer bars, and thus formed after the outer bar formed, in
agreement with the bar-driven formation model. 

For the classical bulges, the existing data are (perhaps) contradictory.
In the case of NGC~2859, \citet{delorenzo-caceres13} found that they
youngest part of the central region of NGC~2859 was what we identify as
the classical bulge. While this could be interpreted as evidence that
classical bulges form \textit{after} the disky pseudobulges, one should
note that de Lorenzo-Caceres et al.\ measured luminosity-weighted ages
from comparisons with single stellar population (SSP) models, so there is
the possibility that the measured ages are contaminated by more recent
stellar populations.

The relevance of this concern was shown by \citet{sarzi05}, who used
\textit{HST} STIS spectroscopy to estimate the stellar populations in
the central $0.2 \times 0.25$ arcsec of a number of galaxies, including
the composite-bulge system NGC~3368. Their best-fitting SSP models
indicated a mean age of $\sim 1$~Gyr for the nuclear region (dominated
by the classical bulge) of NGC~3368, but their best-fitting
\textit{multiple}-population models included both younger and older
components, with a 10~Gyr population contributing 27 per cent of the nuclear
light (and the $\la 1$~Gyr components accounting for only 1.4 per cent of the
nuclear stellar mass).

\citet{storchi-bergmann12} modeled the near-IR spectra of the central
$4.5 \times 5.0$ arcsec of NGC~1068 with multiple-age populations and
found that the \textit{oldest} component (5--13 Gyr in age) was
concentrated in the inner $r < 1$ arcsec, corresponding to what we
identify as the classical bulge (Section~\ref{sec:n1068}).

So the limited stellar-population evidence is, overall, broadly consistent with
the disky pseudobulges being younger than both the bars outside and the
classical bulges inside, but there is clearly room for further work.

\subsubsection{Classical Bulge Formation?}

In the  standard picture of galaxy formation, classical bulges are
formed at high redshift from the violent merger of smaller galaxies or
proto-galactic sub-clumps, resulting in a compact, centrally
concentrated object, with stellar kinematics dominated by dispersion and
an old, metal-rich stellar population. Although we know of no reason why
this couldn't be the case for our classical bulges, the latter
\textit{do} tend to be smaller and lower in mass (and certainly in
$B/T$) than what is usually assumed for classical bulges. It is unclear
whether this process would produce such \textit{small} structures, or
ones with S\'ersic indices of $\sim 1$--2.5.

Another possibility is the more recent scenario of clump-cluster
mergers at high redshift
\citep[e.g.,][]{noguchi00,immeli04,elmegreen08}. In this model, massive
star clusters form in turbulent, gas-rich disks and then, due to
dynamical friction, spiral in towards the center, where they merge in a
fashion not entirely unlike the standard classical-bulge
merger-formation. Although some simulations have suggested that the
resulting bulge would be relatively massive, with high S\'ersic indices
\citep[e.g.,][]{elmegreen08}, it seems possible that smaller,
shorter-lived versions of this scenario might form lower-mass,
lower-S\'ersic-index classical bulges like those we find in the
composite-bulge galaxies.

There is also the possibility that the structures we identify
as classical bulges in these galaxies are actually byproducts of secular
evolution -- i.e., that they are produced in some fashion \textit{after}
the main disk forms, or even after the disky pseudobulge forms. We know
that nuclear star clusters can host multiple episodes of star formation
\citep[e.g.,][]{walcher06}, and the center of the galaxy is where gas that loses angular
momentum will tend to accumulate, which might suggest that compact classical
bulges form through a similar process. However, the classical bulges in our
galaxies tend to be significanly larger than -- and even coexist with --
nuclear star clusters (Section~\ref{sec:cb-plots}), which argues against that idea.

Our classical bulges might conceivably form directly out of disky
pseudobulges via some kind of instability, though realistic models for
such a process are lacking. We \textit{can} rule out standard bar
instabilities, since these produce elongated, boxy/peanut-shaped
structures elongated along the parent bar \citep[e.g.,][and references
therein]{erwin-debattista13}, in contrast to the round, approximately axisymmetric
structures we see, and also because some of the strongest cases, such as
NGC~4371, do not have nuclear bars. (The B/P structures of the
\textit{large-scale} bars would be far larger than the classical bulges
we find, as is clearly the case for at least NGC~3368; see
Section~\ref{sec:boxy}). Similarly, the presence of nuclear bars in many
of these galaxies (e.g., NGC~1068, NGC~1543, NGC~1553, NGC~2859,
NGC~3945, and possibly NGC~4699) rules out scenarios in which a nuclear
bar is ``destroyed'' in order to form a classical bulge. We also note
that the apparent high frequency of pseudobulges in late-type spirals 
\citep[e.g.,][]{fisher-drory11} suggests that pseudobulges by themselves probably
do not \textit{always} give rise to classical bulges.

\section{Summary}\label{sec:summary} 

We have presented a morphological and kinematic analysis of nine disk
galaxies (S0 or spiral) in which the photometric bulge region -- that is,
the excess stellar light above an inward extrapolation of the outer disk
profile -- is composed of at least two distinct components:
\begin{enumerate}
\item A \textit{disky pseudobulge}: a structure with flattening similar to
that of the outer disk, one or more morphological features
characteristics of disks (nuclear rings, bars and/or spirals) and a
stellar kinematics which is dominated by rotation.  
\item A \textit{classical
bulge}: a component which is rounder (more spheroidal) than the disky
pseudobulge and which has kinematics dominated by velocity dispersion.
These components do, however, tend to have profiles best fitted
with S\'ersic indices of $n = 1$--2.2, so they are not classical in the
sense of having $R^{1/4}$ profiles.
\end{enumerate}
In at least one galaxy there is also
evidence for a boxy/peanut-shaped bulge (the vertically thick inner part
of a bar), demonstrating that all three types of ``bulge'' can coexist,
as suggested by, e.g., \citet{athanassoula05}.

Using Schwarzschild orbit-superposition modeling for three of these
galaxies (NGC~3368, NGC~4371 and NGC~4699), taken from
\citet{erwin14-smbh}, we investigated the 3D stellar orbital structure
of the best-fitting models for each galaxy. We found that the stellar
velocity dispersion is approximately isotropic within the
classical-bulge regions but is equatorially biased in the disky
pseudobulges (as expected for a highly flattened system), and also that
the ratio of azimuthal velocity to total velocity dispersion
(\vphidivsigma) is typically $\la 0.5$ in the classical bulge and
increases towards values $\ga 1$ in the disky pseudobulge. 

Although we are currently unable to put strong limits on the frequency
of composite-bulge galaxies, they are probably present in \textit{at least}
$\sim 10$ per cent of barred S0 and early-type spiral galaxies.

Plotting the classical-bulge components of the composite-bulge galaxies in
the $R_{e}$--$M_{\star}$ and $\langle \Sigma_{\star}
\rangle_{e}$--$M_{\star}$ planes shows that they fall into the same
general sequence as elliptical galaxies and the (larger) bulges of disk
galaxies. Even though some of the classical-bulge components are quite
small, with half-light radii $\sim 30$~pc, they remain distinct from
nuclear star clusters (and some in fact harbor nuclear star clusters in
their centres). Curiously, the disky pseudobulge components
\textit{also} lie along the elliptical/classical-bulge sequences in
these planes. While some disky pseudobulges are as massive as the main
(outer) disks of S0 and spiral galaxies, they are considerably more
compact.

Since almost all of our composite-bulge galaxies are barred, the disky
pseudobulge components could plausibly be the result of bar-driven gas
inflow and star formation. We note that the ``nuclear disks'' formed in
this fashion in the simulations of \citet{wozniak09} and \citet{cole14} are similar in
size and relative stellar mass to the disky pseudobulges in our
galaxies. Cases where disky pseudobulges are found in \textit{unbarred}
galaxies (or galaxies where the only bar is \textit{smaller} than the
disky pseudobulge) are more difficult to explain, unless disky
pseudobulges can form early in a galaxy's history as part of a general
inside-out process \citep[e.g.,][]{guedes13}.

\section*{Acknowledgements} 

We enjoyed helpful and interesting conversations with a number of
people, including Niv Drory, Witold Maciejewski and Victor Debattista.
We are particularly grateful to Rick Davies, Joris Gerssen, Karl
Gebhardt, Erin Hicks and Richard McDermid for supplying us with
kinematic data, both long-slit and IFU, and to Jakob Walcher for help
with nuclear star cluster data. We would also like to thank Bego{\~n}a
Garc{\'i}a-Lorenzo for faithfully executing our WHT-ISIS service
observations of NGC~4371, and the anonymous referee for comments
which helped improve the manuscript.

P.E. was supported in part by DFG Priority Programme 1177 (``Witnesses of Cosmic
History:  Formation and evolution of black holes, galaxies and their
environment'').

This research is (partially) based on data obtained with the William
Herschel Telescope; the WHT and its service programme are operated on the
island of La Palma by the Isaac Newton Group in the Spanish Observatorio
del Roque de los Muchachos of the Instituto de Astrof{\'i}sica de Canarias.

Based on observations made with ESO Telescopes at the La Silla Paranal
Observatory under programme IDs 080.B-0336 and 082.B-0037.

Funding for the creation and distribution of the SDSS
Archive has been provided by the Alfred P. Sloan Foundation, the
Participating Institutions, the National Aeronautics and Space
Administration, the National Science Foundation, the U.S. Department of
Energy, the Japanese Monbukagakusho and the Max Planck Society. The
SDSS Web site is \texttt{http://www.sdss.org/}.

The SDSS is managed by the Astrophysical Research Consortium (ARC) for
the Participating Institutions.  The Participating Institutions are
The University of Chicago, Fermilab, the Institute for Advanced Study,
the Japan Participation Group, The Johns Hopkins University, the
Korean Scientist Group, Los Alamos National Laboratory, the
Max-Planck-Institute for Astronomy (MPIA), the Max-Planck-Institute
for Astrophysics (MPA), New Mexico State University, University of
Pittsburgh, University of Portsmouth, Princeton University, the United
States Naval Observatory and the University of Washington.

This research made use of the NASA/IPAC Extragalactic Database (NED)
which is operated by the Jet Propulsion Laboratory, California
Institute of Technology, under contract with the National Aeronautics
and Space Administration.  It also made use of the Lyon-Meudon
Extragalactic Database (LEDA; part of HyperLeda at
http://leda.univ-lyon1.fr/).


\appendix

\section{Other Composite Bulges}\label{sec:others-full}

\subsection{NGC~1553}\label{sec:n1553} 

NGC~1553 is a nominally unbarred S0 galaxy, notable for its prominent
lens \citep[e.g.,][]{freeman75,kormendy84a}. \citet{kormendy04} called
attention to it as an example of a pseudobulge in an unbarred galaxy,
based on its rather high \vmaxsig{} value.

\begin{figure*}
\includegraphics[width=6.0in]{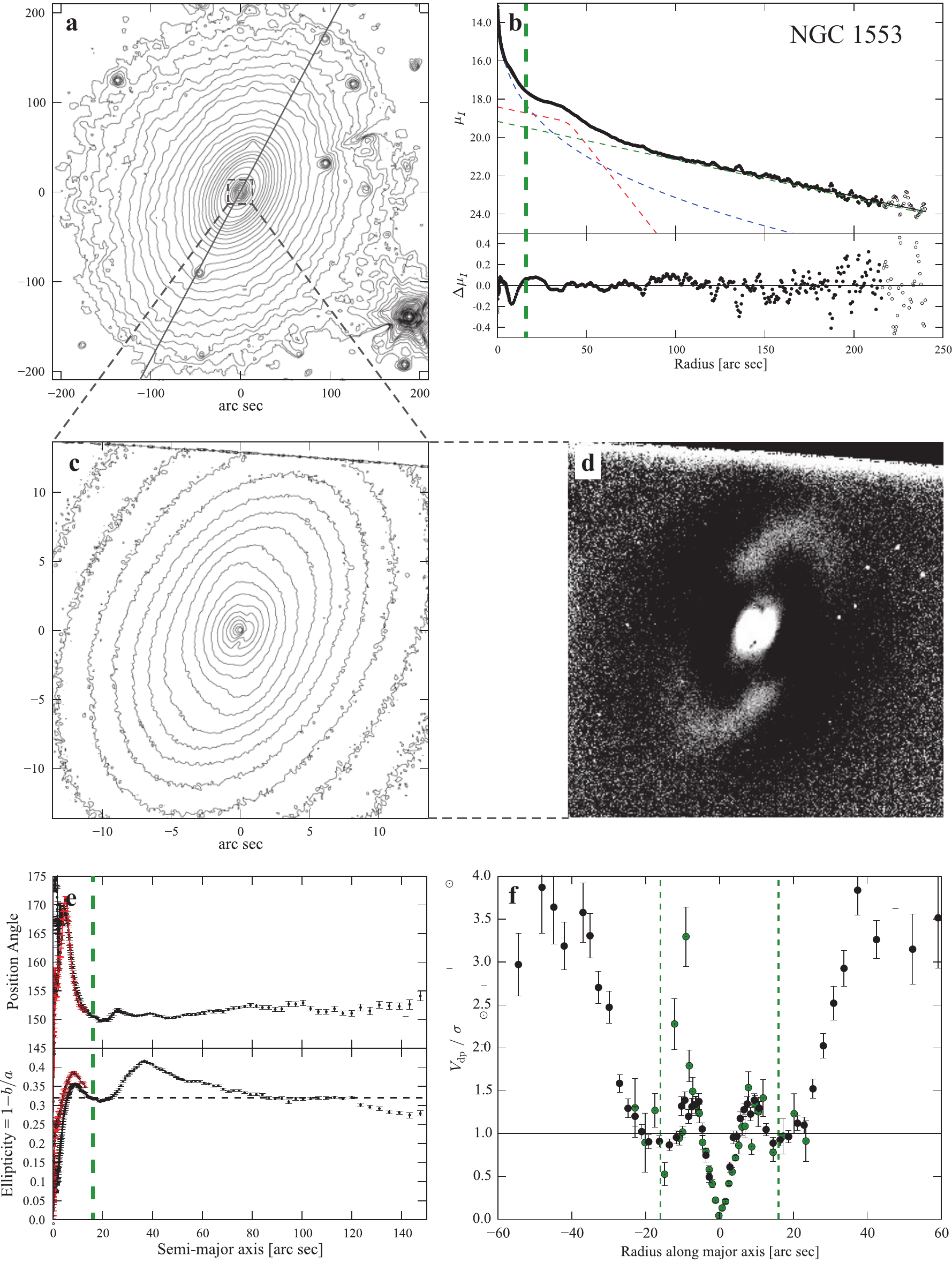}

\caption{The disky pseudobulge in the S0 galaxy NGC~1553. \textbf{a:}
log-scaled \textit{Spitzer} IRAC1 isophotes (smoothed with 15-pixel-wide
median filter); gray line marks major axis (PA $= 152\degr$).
\textbf{b:} Bulge-disk decomposition of $I$-band major-axis profile
(\textit{HST} WFPC2 F814W for $r < 4.1$ arcsec, IRAC1 for larger radii).
Dashed lines represent S\'ersic + broken-exponential (lens) +
exponential fit to the data, with residuals plotted in lower sub-panel.
Vertical dashed green line marks ``bulge=disk'' radius \rbd. \textbf{c:}
Close-up of photometric bulge region (log-scaled contours from WFPC2
F814W PC image). \textbf{d:} Unsharp mask of the same image ($\sigma =
30$ pixels), showing the inner bar and partial spiral arms surrounding
it. \textbf{e:} Ellipse fits to the IRAC1 [black] and WFPC2 F814W [red]
images. \textbf{f:} Deprojected stellar rotation velocity divided by
local velocity dispersion $\vdivsigma$ along the major axis, using data
from \citet{kormendy84a} [black] and \citet{longo94} [green]. Vertical
dashed lines mark the photometric bulge region $|R| < \rbd$. Both
datasets show that $\vdivsigma$ rises to well over 1 within this region,
indicating a kinematically cool region more like a
disk.}\label{fig:n1553a}

\end{figure*}

As \nocite{kormendy04}Kormendy \& Kennicutt noted, this galaxy's
surface-brightness profile looks very similar to that of many barred
galaxies: an extended, shallow-surface-brightness ``shelf''  with a much
steeper falloff at its outer edge (located in between the central bulge
and the outer exponential disk). The lens \textit{is} more elliptical
than the outer disk, and the presence of weak boxyness in the isophotes
(strongest at $a \sim 25$ arcsec) might suggest we are seeing the
projected box/peanut structure of a (weak) bar, rather than a flattened
structure. This would imply an unusually large ratio of B/P size to bar
size, since we see little or no sign of the ``spurs'' which are the
projection of the vertically thin outer part of the bar
\citep[see][]{erwin-debattista13}. Regardless of its true nature, the
presence of the lens leads us to model it as a separate,
broken-exponential component in our global decomposition; the result is
a relatively good fit to the major-axis profile
(panel~b of Figure~\ref{fig:n1553a}).

The photometric bulge region ($r < 16$ arcsec) shows clear evidence for
a disky pseudobulge, as previously suggested by \citet{kormendy04} and
\citet{laurikainen06}: the ellipticity of the isophotes is similar to or
greater than that of the main disk (panels~c and e of
Figure~\ref{fig:n1553a}), and the stellar kinematics are dominated by
rotation, with \vdivsigma{} reaching a maximum of $\sim 1.4$ at $r \sim
8$ arcsec (panel~f). Analysis of the \textit{HST} images reveals a
previously unrecognized nuclear bar, with surrounding stellar spiral
arms (panel~d), further evidence that the photometric bulge is disklike.

\begin{figure*}
\includegraphics[width=6.0in]{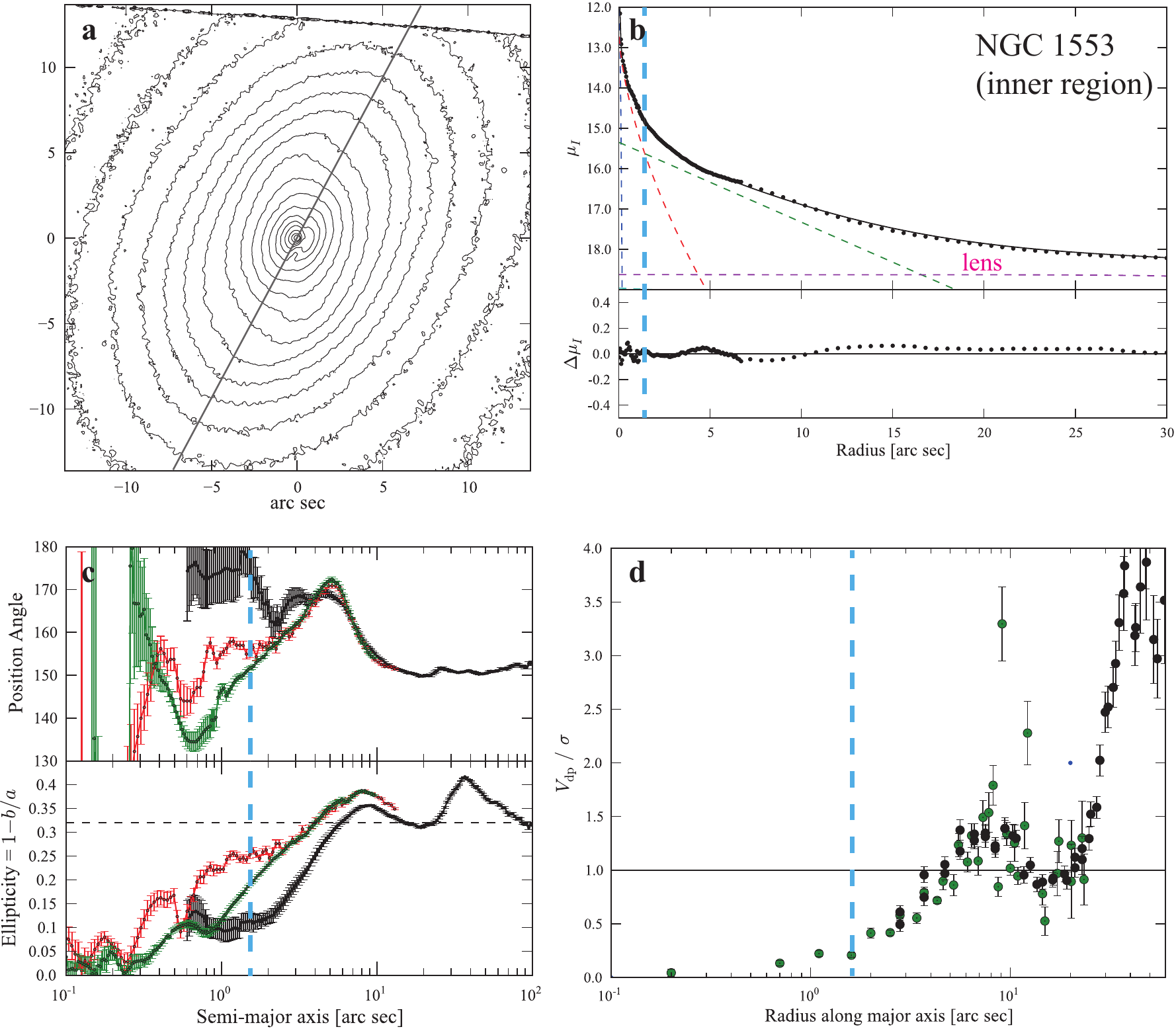}

\caption{The classical bulge inside the disky pseudobulge of NGC~1553.
\textbf{a:} Close-up of main photometric bulge region (see
Figure~\ref{fig:n1553a}); gray line marks the major axis. \textbf{b:}
major-axis profile, with fit (S\'ersic + exponential + broken-exponential; dashed blue,
green and red lines, respectively) and residuals from fit
in lower sub-panel. Vertical dashed blue line marks inner ``bulge=disk''
radius \rbdi. \textbf{c:} Ellipse fits (see Figure~\ref{fig:n1553a};
additional green fit is from NICMOS2 F160W image); note that the
ellipticity in the inner photometric bulge region ($a \la 1.5$ arcsec)
is $< 0.25$ (and $\sim 0.05$--0.15 for $a \la 0.6$ arcsec), clearly less
than that of the disky pseudobulge outside, or the outer disk.
\textbf{d:} Deprojected, folded stellar rotation velocity divided by
local velocity dispersion $\vdivsigma$ along galaxy major axis; see
Figure~\ref{fig:n1553a}. Vertical dashed blue and green lines mark the
inner and main photometric bulge regions $|R| < \rbdi$, respectively. In
the inner region ($r \la 1.4$ arcsec) and immediately outside,
$\vdivsigma$ remains well below 1, suggesting a kinematically hot region
(i.e., a classical bulge). }\label{fig:n1553b}

\end{figure*}

Inside the nuclear bar, the isophotes are clearly rounder than the outer
disk. We also find a clear central excess in the surface-brightness
profile, and are able to fit the inner $r < 30$ arcsec major-axis
profile quite well using the sum of an inner exponential + S\'ersic,
added to the contribution from the lens component we used in the global
decomposition (panel~b of Figure~\ref{fig:n1553b}). As was the case for
NGC~3945 and NGC~4371, the inner photometric bulge (where the inner
S\'ersic component dominates, at $r < 1.5$ arcsec) corresponds to the
round central isophotes. Although the ground-based stellar kinematics of
\citet{kormendy84a} and \citet{longo94} which we use to determine
\vdivsigma{} may not fully resolve our proposed classical-bulge region,\footnote{Note,
however, that Longo et al.\ reported seeing of $< 1$ arcsec for their
observations of this galaxy} the fact that $\vdivsigma < 0.5$ for $r
\la 3$ arcsec suggests this region is indeed dominated by velocity
dispersion, allowing us to classify the inner photometric bulge as a
classical bulge.

\subsection{NGC~2859}\label{sec:n2859} 

NGC~2859 is a strongly double-barred galaxy, as originally noted by
\citet{kormendy79a}; \citet{kormendy82} observed a relatively high degree of
stellar rotation in the photometric bulge region, with $\vmaxsigstar =
1.2$. This makes the galaxy potentially rather similar to NGC~3945.

\begin{figure*}
\includegraphics[width=6.0in]{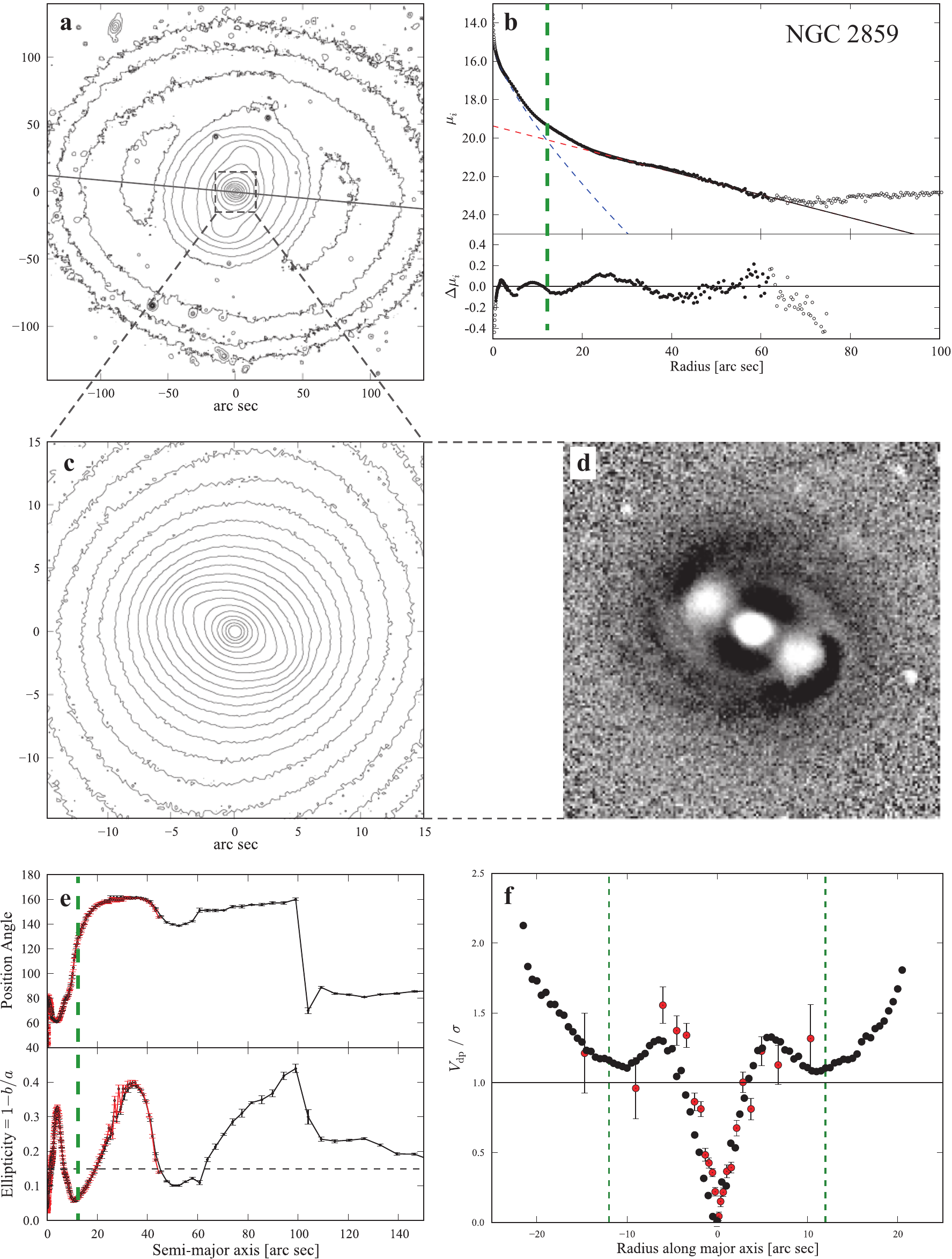}

\caption{The disky pseudobulge in the S0 galaxy NGC~2859. \textbf{a:} log-scaled
$R$-band isophotes (WIYN image, smoothed with 15-pixel-wide median filter); gray
line marks major axis (PA $= 86\degr$). \textbf{b:} Bulge-disk decomposition of
$i$-band major-axis profile (\textit{HST} ACS-WFC F814W for $r < 3.4$ arcsec,
SDSS $i$-band for larger radii).  Dashed lines represent S\'ersic + exponential
fit to the data, with residuals plotted in lower sub-panel. Vertical dashed
green line marks ``bulge=disk'' radius \rbd, where the S\'ersic and exponential
components are equally bright; this sets the boundary of the ``photometric
bulge''. \textbf{c:} Close-up of photometric bulge region (log-scaled contours
from $R$-band image). \textbf{d:} Unsharp mask of the $R$-band image ($\sigma =
5$ pixels), showing the inner bar and partial spiral arms surrounding it.
\textbf{e:} Ellipse fits to the SDSS $i$-band [black] and ACS-WFC F814W [red]
images. \textbf{f:} Deprojected stellar rotation
velocity divided by local velocity dispersion $\vdivsigma$ along the major axis,
using WHT-ISIS long-slit data [red] and SAURON data [black]. Vertical dashed
lines mark the photometric bulge region $|R| < \rbd$. Both datasets show that $\vdivsigma$
rises to well over 1 within this region, indicating a kinematically
cool region more like a disk.}\label{fig:n2859a}

\end{figure*}

The similarity with NGC~3945 begins, in fact, with a luminous outer ring
(panel~a of Figure~\ref{fig:n2859a}) creating a strongly non-exponential
outer-disk profile. We choose to fit the region \textit{interior} to the
outer ring ($r \la 70$ arcsec, panel~b). This produces a reasonable
photometric bulge-disk decomposition, with $\rbd \approx 12$ arcsec. If
we had tried including the outer ring, the photometric bulge region
would only have become larger; e.g., the decomposition in
\citet{fabricius12} gives $\rbd = 30$ arcsec.

Interior to this we find a disky pseudobulge, as evidenced
morphologically by the strong inner bar (Figure~\ref{fig:n2859a}, panels
c and d) and kinematically by the \vdivsigma{} curve (panel~f), which
reaches a maximum of $\sim 1.3$--1.5 at $r \sim 7$ arcsec. The kinematic
data combines our WHT-ISIS long-slit observations
(Appendix~\ref{sec:data-spec}) and a pseudo-longslit constructed from
the SAURON data of \citet{delorenzo-caceres08}; the latter profile was
previously presented in \citet{fabricius12}.  We note that
\citet{delorenzo-caceres13} have also argued for the existence of a
distinct inner disk in this galaxy from their 2D analysis of the SAURON
kinematics.

Unlike the case of NGC~3945, where the inner bar was a small
perturbation within the disky pseudobulge and could thus be ignored in
the fitting process (particularly as it was oriented almost
perpendicular to the major axis), the inner bar in NGC~2859 is much
larger and stronger relative to the disky pseudobulge; indeed, it could
perhaps be argued that the disky pseudobulge is largely just the inner
bar plus its surrounding lens. However, we can take advantage of this
fact by using the inner bar itself to guide our decomposition. A more
detailed discussion of this approach, along with an analysis of the
inner bar's structure, will be presented elsewhere \citep{erwin14-dissect}.

\begin{figure*}
\includegraphics[width=6.0in]{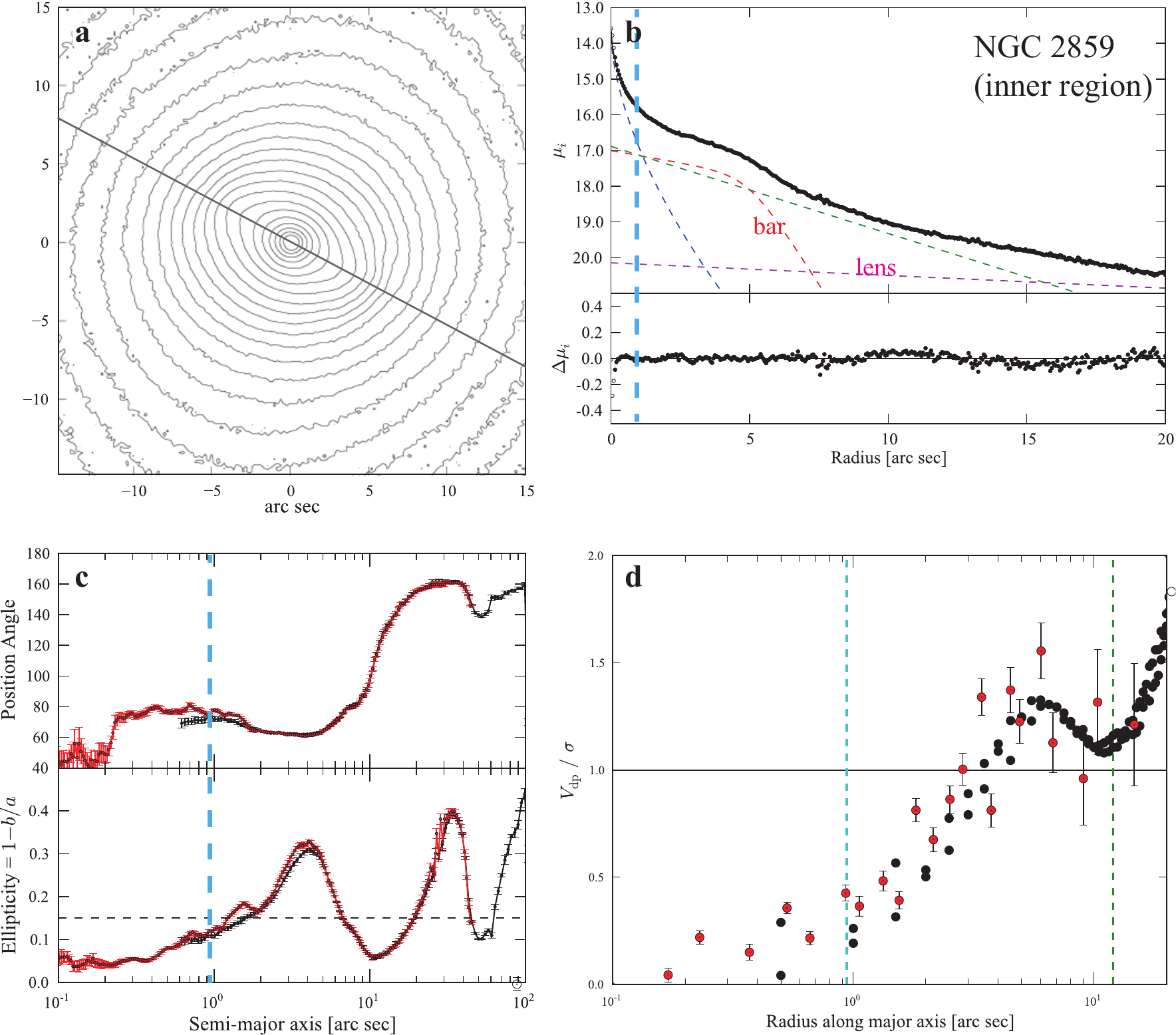}

\caption{The classical bulge inside the disky pseudobulge of NGC~2859.
\textbf{a:} Close-up of main photometric bulge region (see
Figure~\ref{fig:n2859a}); gray line marks the major axis of the inner bar.
\textbf{b:} inner-bar major-axis profile (from ACS F814W image), with fit
(S\'ersic + broken-exponential + exponential; dashed lines) and residuals from
fit in lower sub-panel. Vertical dashed blue line marks inner ``bulge=disk'' radius
\rbdi, where S\'ersic component is as bright as the sum of the other two
components. \textbf{c:} Ellipse fits (see Figure~\ref{fig:n2859a}); note that
the ellipticity in the inner photometric bulge region ($a \la 1$ arcsec is $\sim
0.05$--0.1, less than that of the disky pseudobulge outside. \textbf{d:}
Deprojected, folded stellar rotation velocity divided by local velocity
dispersion $\vdivsigma$ along galaxy major axis; see Figure~\ref{fig:n2859a}.
Vertical dashed blue and green lines mark the inner and main photometric bulge
regions $|R| < \rbdi$, respectively. In the inner region ($r \la 1$ arcsec) and
immediately outside, $\vdivsigma$ remains well below 1, suggesting a
kinematically hot region (i.e., a classical bulge). }\label{fig:n2859b}

\end{figure*}

As Figure~\ref{fig:n2859b} shows, we can use a cut along the inner bar's
major axis and decompose this into an underlying exponential, a
broken-exponential profile for the bar itself and a central S\'ersic
component. The latter defines the inner photometric bulge, with $\rbdi =
0.94$ arcsec. This region is associated with a position angle close to
that of the outer disk and an ellipticity of $\sim 0.05$ (panel~c of
Figure~\ref{fig:n2859b}), so we have morphological evidence for a
classical bulge. Finally, the stellar kinematics (panel~d of the figure)
show that $\vdivsigma$ remains $< 1$ for $r \la 3$ arcsec, in both the
SAURON and the (higher-resolution) WHT-ISIS data.

\subsection{NGC~3368}\label{sec:n3368} 

\begin{figure*}
\includegraphics[width=6.0in]{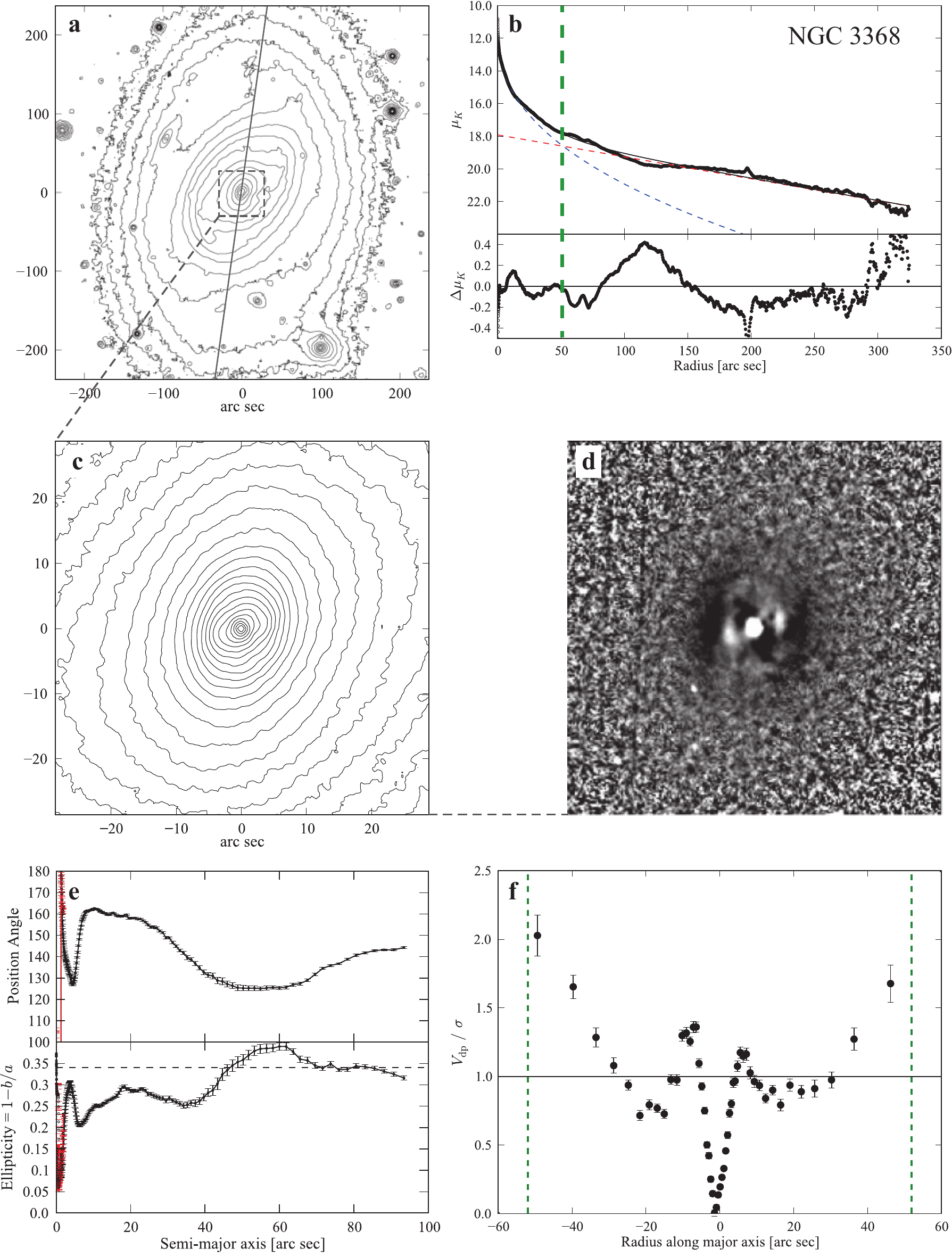}

\caption{The disky pseudobulge in the Sab galaxy NGC~3368. \textbf{a:}
log-scaled SDSS $r$-band isophotes (smoothed with 15-pixel-wide median filter);
gray line marks major axis (PA $= 172\degr$). \textbf{b:} Bulge-disk
decomposition of major-axis profile (combines NICMOS2 F160W, \citet{knapen03}
$K$-band and SDSS $r$-band data).  Dashed lines represent S\'ersic +
exponential fit to the data, with residuals plotted in lower sub-panel. Vertical
dashed green line marks ``bulge=disk'' radius \rbd, where the S\'ersic and
exponential components are equally bright; this sets the boundary of the
``photometric bulge''. \textbf{c:} Close-up of photometric bulge region, using
\citet{knapen03} $K$-band image (smoothed with 5-pixel-wide median filter,
log-scaled contours). \textbf{d:} Unsharp mask of the same image ($\sigma = 5$
pixels), showing the inner bar and dust lanes in the vicinity. \textbf{e:}
Ellipse fits to the $K$-band and SINFONI (red) images. \textbf{f:} Deprojected
stellar rotation velocity divided by local velocity dispersion $\vdivsigma$
along the major axis, using HET long-slit data. Vertical dashed lines mark the
photometric bulge region $|R| < \rbd$. $\vdivsigma$ rises to $\sim 1.36$ within
this region, indicating a kinematically cool region more like a
disk.}\label{fig:n3368a}

\end{figure*}

The case of NGC~3368 was originally considered in \citet{nowak10}; our
analysis here is very similar. The large-scale major-axis decomposition
(panel~b of Figure \ref{fig:n3368a}) shows a fairly plausible S\'ersic +
exponential fit, with $\rbd = 52$ arcsec. We note that this is rather
larger than the value of  $\rbd = 23.3$ arcsec which \citet{fabricius12}
found from their 1-D decomposition, which may reflect the effects of
masking intermediate parts of the profile in the latter study. A similar
fit to a partially masked 1D profile of this galaxy by
\citet{fisher-drory08} resulted in $\rbd \sim 35$ arcsec. Adopting
either of these smaller radii as the outer boundary of the photometric
bulge region has no significant effect on our analysis, however.


The isophotes and ellipse fits (panels~c and e) show two regions of high
ellipticity, corresponding to the ``inner disk'' and inner bar of
\citet{erwin04}; unsharp masking (panel~d) shows the inner bar more
clearly (panel~d). The stellar kinematics from \citet{fabricius12}
clearly shows a peak in the \vdivsigma{} profile (panel~f), which
reaches a maximum of $\sim 1.3$ at $r \approx 7$ arcsec; the improved
S/N of the HET spectrum shows this more clearly that the older
kinematics from \citet{vega-beltran01} used by \citet{nowak10}.

Thus we appear to have reasonable evidence for a disky pseudobulge, as
argued in \citet{nowak10}. As in the cases of NGC~1553 and NGC~2859, the
ellipticity profile (panel~e of Figure~\ref{fig:n3368a}) is more
complicated than those of NGC~3945 and NGC~4371, both because of the
strong contribution from the inner bar (which is indirect evidence for
disk-like structures, but is not a disk itself) and because the
intermediate ellipticity peak (at $a \sim 21$ arcsec) is probably
associated with the structure of the outer bar -- specifically, with the
projection of the vertically thick inner part of the bar (see
Section~\ref{sec:boxy}).

\begin{figure*}
\includegraphics[width=6.0in]{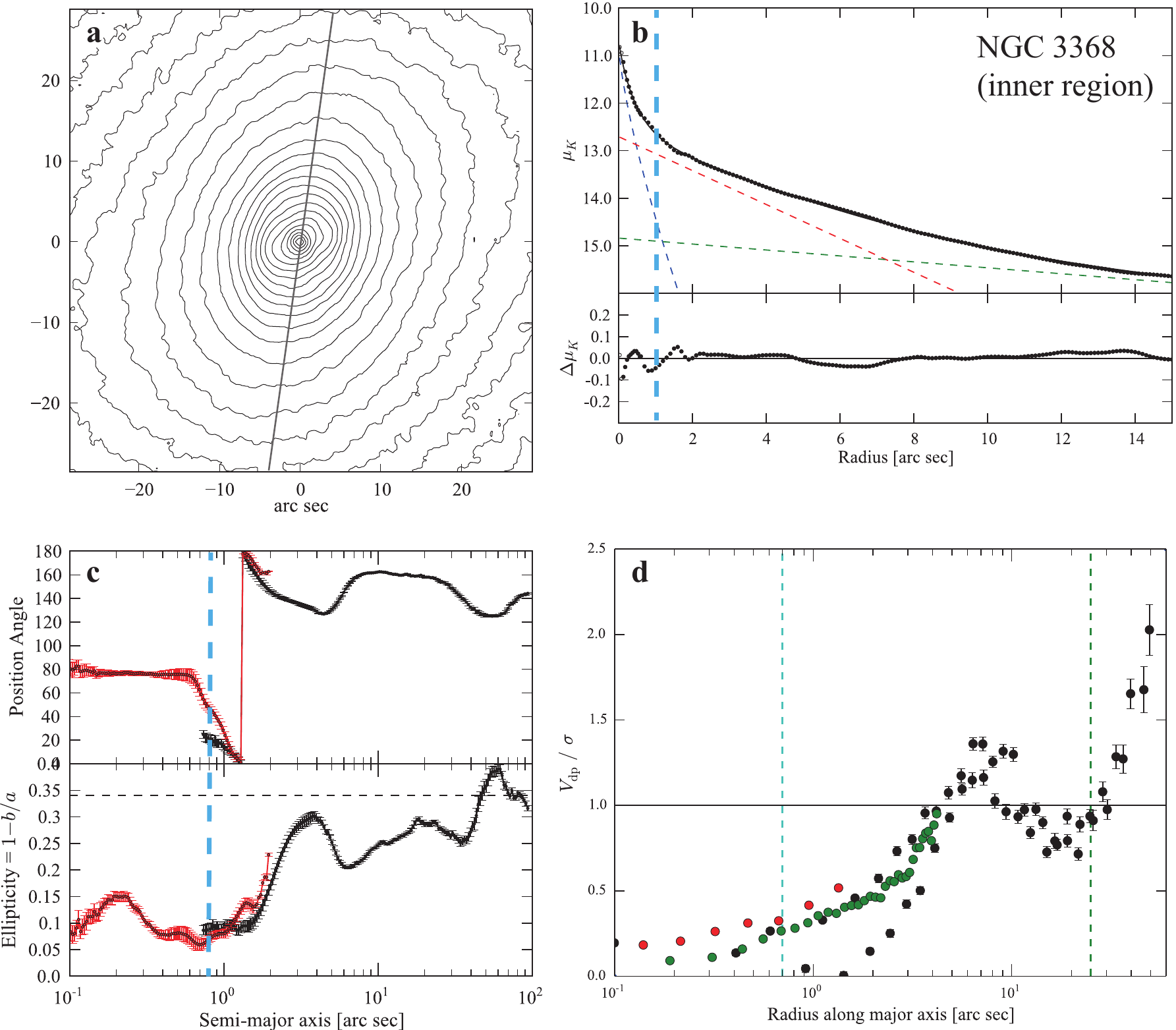}

\caption{The classical bulge inside the disky pseudobulge of NGC~3368.
\textbf{a:} Close-up of main photometric bulge region from the $K$-band
image; gray line marks major-axis PA. \textbf{b:} major-axis profile
combining SINFONI, NICMOS2 and ground-based $K$-band data, with fit
(S\'ersic + exponential + exponential; dashed lines) and residuals from
fit in lower sub-panel. Vertical dashed blue line marks inner
``bulge=disk'' radius \rbdi. \textbf{c:} Ellipse fits to the $K$-band
and SINFONI (red) images.; note that the ellipticity in the inner
photometric bulge region ($a \la 1$ arcsec is $\sim 0.1$, less than that
of the disky pseudobulge outside. \textbf{d:} Deprojected, folded
stellar rotation velocity divided by local velocity dispersion
$\vdivsigma$ along the major axis, combining AO SINFONI (red), non-AO
SINFONI (green) and HET (black) data. Vertical dashed blue and green
lines mark the inner and main photometric bulge regions $|R| < \rbdi$,
respectively. In the inner region ($r \la 1$ arcsec), $\vdivsigma$
remains well below 1, suggesting a kinematically hot region (i.e., a
classical bulge).}\label{fig:n3368b}

\end{figure*}

Figure~\ref{fig:n3368b} shows the evidence for a classical bulge inside
the disky pseudobulge of this galaxy. The inner decomposition presented
here is somewhat different from that of \citet{nowak10}, because we are
now including an additional (exponential) component to represent the
contribution of the (outer) bar and the lens; consequently, the
parameters of the best-fitting inner exponential and S\'ersic components
are somewhat different. None the less, our main conclusions are
unchanged: we find evidence that the light at $r \la 1$ arcsec is due to
a separate photometric component with much rounder isophotes than either
the disky pseudobulge or the outer disk.  The \vdivsigma{} plot
incorporates major-axis values from the SINFONI datacubes of
\citet{nowak10} and \citet{hicks13} and the lower-resolution major-axis
long-slit spectroscopy of \citet{fabricius12}. The innermost SINFONI
kinematics from \citet{nowak10}, obtained with an AO-corrected seeing of
$\sim 0.17$ arcsec FWHM, clearly show that this inner component has
dispersion-dominated stellar kinematics, with \vdivsigma{} significantly
$< 1$.

\subsection{NGC~4262} 

NGC~4262 is a low-inclination barred S0 galaxy in the Virgo Cluster, classified
by \citet{erwin04} as possessing an inner disk inside the bar, based in part on
observations by \citet{shaw95}. Although the galaxy is quite close to face-on,
we do find evidence for both a (possibly) disky pseudobulge and a very small
classical bulge.

\begin{figure*}
\includegraphics[width=6.0in]{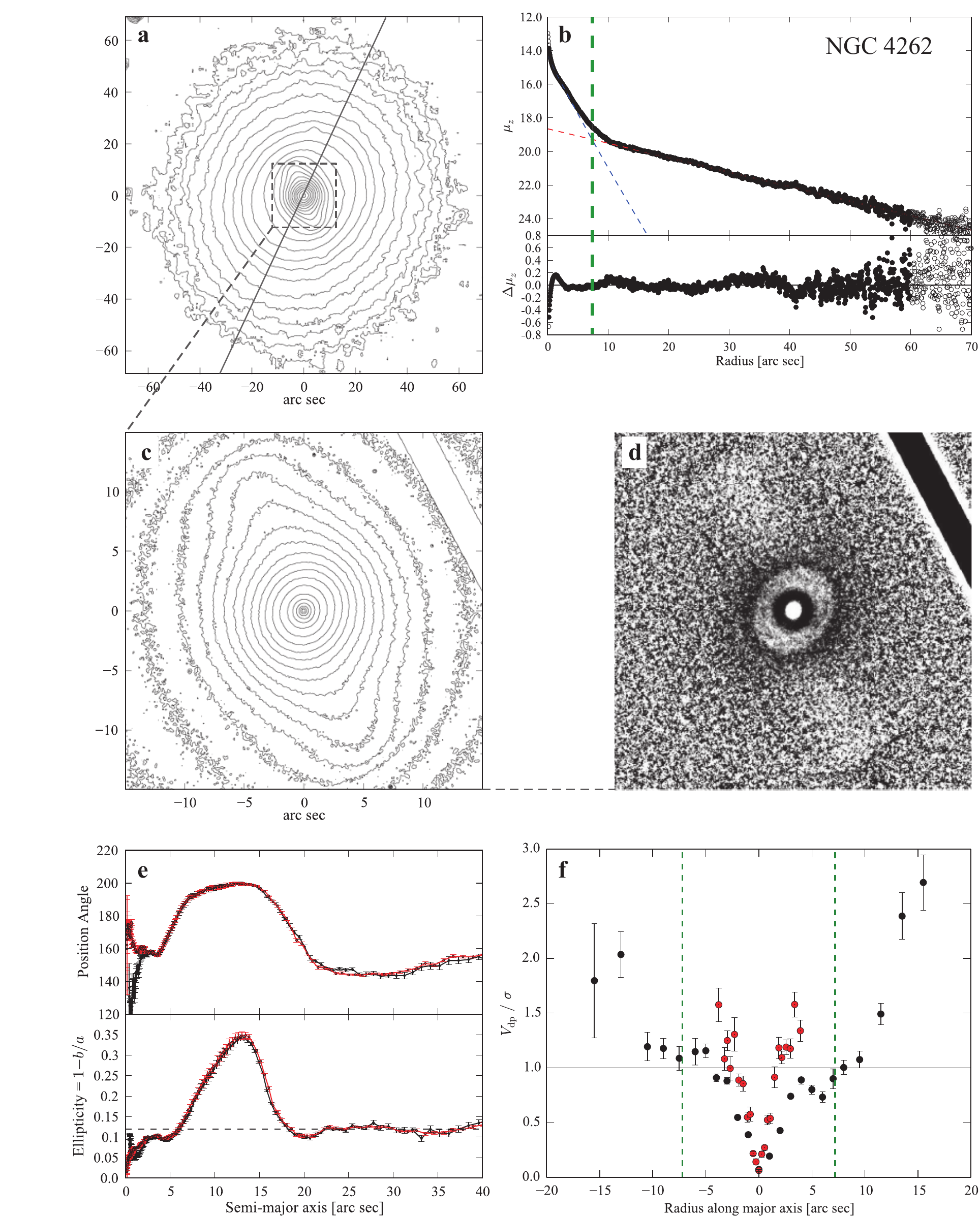}

\caption{The disky pseudobulge in the S0 galaxy NGC~4262. \textbf{a:} log-scaled
SDSS $i$-band isophotes (smoothed with 15-pixel-wide median filter); gray line
marks major axis (PA $= 155\degr$). \textbf{b:} Bulge-disk decomposition of
major-axis profile, from SDSS $i$-band and \textit{HST} ACS-WFC F850LP images.
Dashed lines represent S\'ersic + exponential fit to the data, with residuals
plotted in lower sub-panel. Vertical dashed green line marks ``bulge=disk''
radius \rbd, where the S\'ersic and exponential components are equally bright;
this sets the boundary of the ``photometric bulge''. \textbf{c:} Close-up of
photometric bulge region (log-scaled contours from ACS-WFC F850LP image,
smoothed with 3-pixel-wide median filter). \textbf{d:} Unsharp mask of the
\textit{HST} image ($\sigma = 15$ pixels), showing the nuclear ring. \textbf{e:}
Ellipse fits to the SDSS [black] and \textit{HST} [red] images; note that the
ellipticity interior to the bar is only slightly below that of the outer disk.
\textbf{f:} Deprojected stellar rotation velocity divided by local velocity
dispersion $\vdivsigma$ along the major axis, using OASIS data (red) from
\citet{mcdermid06} and SAURON data (black) from \citet{emsellem04}; the former has much higher
spatial resolution (FWHM = 0.6 versus 2.6 arcsec). Vertical
dashed lines mark the photometric bulge region $|R| < \rbd$. $\vdivsigma$ in the
high-resolution OASIS data rises to $\sim 1.57$ within this region, indicating a
kinematically cool region more like a disk.}\label{fig:n4262a}

\end{figure*}

\begin{figure*}
\includegraphics[width=6.0in]{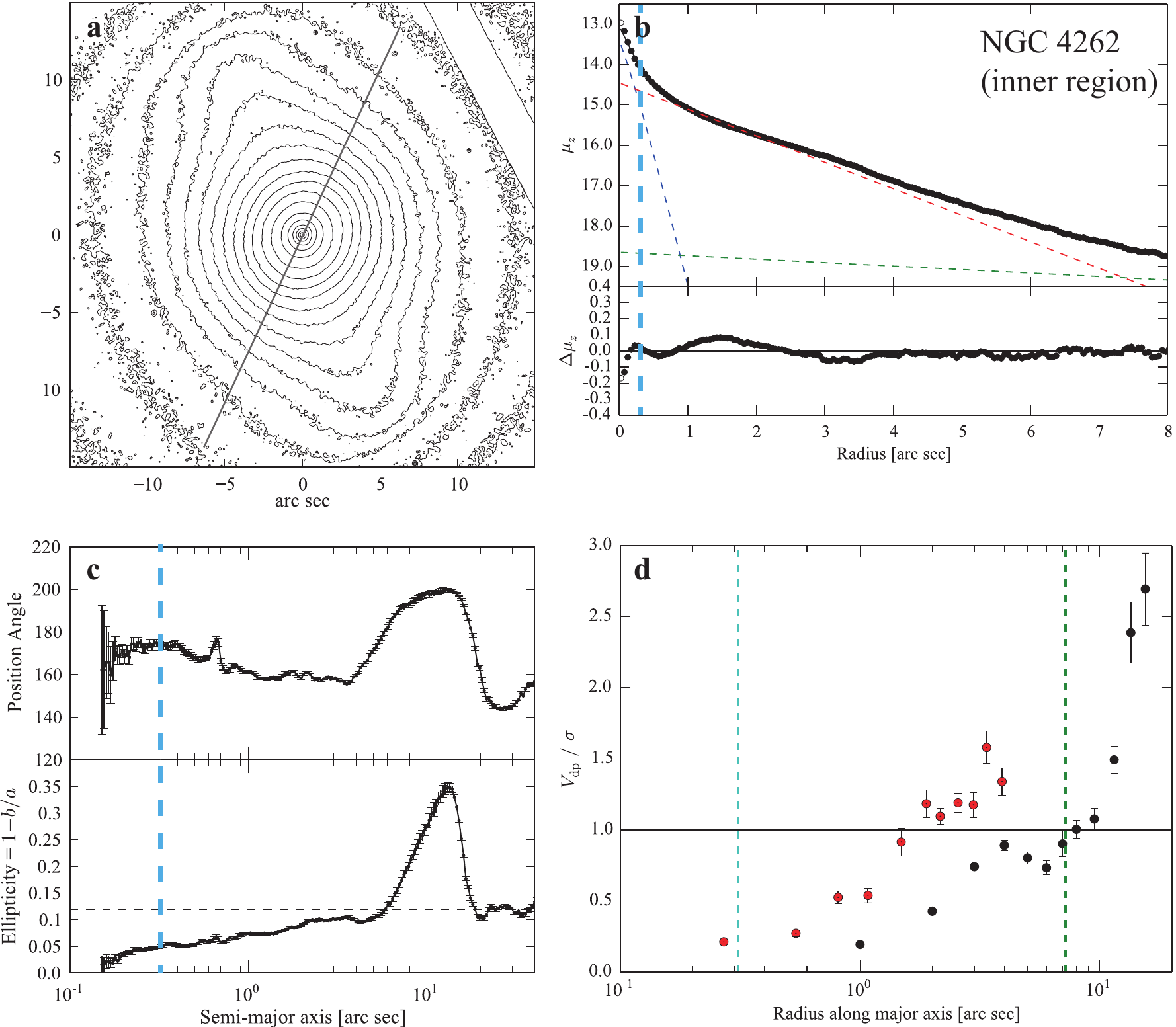}

\caption{The classical bulge inside the disky pseudobulge of NGC~4262.
\textbf{a:} Close-up of main photometric bulge region from \textit{HST}
ACS-WFC F850LP image; gray line marks major-axis PA. \textbf{b:} major-axis profile
from the \textit{HST} image, with fit (S\'ersic + exponential; dashed
lines) and residuals from fit in lower sub-panel. Vertical dashed blue
line marks inner ``bulge=disk'' radius \rbdi. \textbf{c:} Ellipse fits to the
\textit{HST} image; note that the ellipticity in the inner photometric
bulge region ($a \la 0.32$ arcsec is $\sim 0.05$, less than that of the
disky pseudobulge outside. \textbf{d:} Deprojected, folded stellar
rotation velocity divided by local velocity dispersion $\vdivsigma$
along the major axis. Vertical dashed blue and green lines mark the
inner and main photometric-bulge regions $|R| < \rbdi$, respectively.
Although there is essentially no kinematic data for the inner
photometric-bulge region, the fact that  $\vdivsigma$ is below 1 for $r
< 1.8$ arcsec suggests that the latter is almost certainly
kinematically hot (i.e., a classical bulge). }\label{fig:n4262b}

\end{figure*}

The decomposition of the major-axis profile (Figure~\ref{fig:n4262a},
panel~b) is relatively simple and clean, with the resulting photometric bulge
region defined by $r < \rbd = 7.2$ arcsec. The main morphological evidence for a disky
pseudobulge is a distinct stellar nuclear ring with $a \sim 2.9$ arcsec (panel~d),
also noted by \citet{comeron10}; the isophotes in this region have an ellipticity
only marginally less than that of the outer disk, and greater than that of the
inner $a \la 1$ arcsec (see also panel~c of Figure~\ref{fig:n4262b}). Kinematically,
we find that \vdivsigma{} reaches a peak value of almost 1.6 at approximately
the same radius as the nuclear ring. The stellar kinematic data are from major-axis cuts
through SAURON IFU data from the SAURON public
archive,\footnote{http://www.strw.leidenuniv.nl/sauron/} originally published in
\citet{emsellem04}, and higher-resolution OASIS IFU data from \citet{mcdermid06}.

Closer inspection of the ACS-WFC isophotes shows that the isophotes
become gradually rounder interior to the stellar nuclear ring, reaching
values  of $\sim 0.05$ before PSF effects dominate
(Figure~\ref{fig:n4262b}). Decomposition of the inner major-axis profile
(panel~b of Figure~\ref{fig:n4262b}) shows a small central excess,
reasonably well fit by a S\'ersic component with $n \sim 0.9$. Stellar
kinematics with the necessary resolution for this inner
photometric-bulge region ($r < 0.32$ arcsec) are unavailable; however,
the fact that \vdivsigma{} is $< 1$ for radii $\la 2$ arcsec suggests
that the inner stellar kinematics are probably dispersion-dominated, and
that the central photometric excess corresponds to a (very compact)
classical bulge.

\subsection{NGC~4699} 

\begin{figure*}
\includegraphics[width=6.0in]{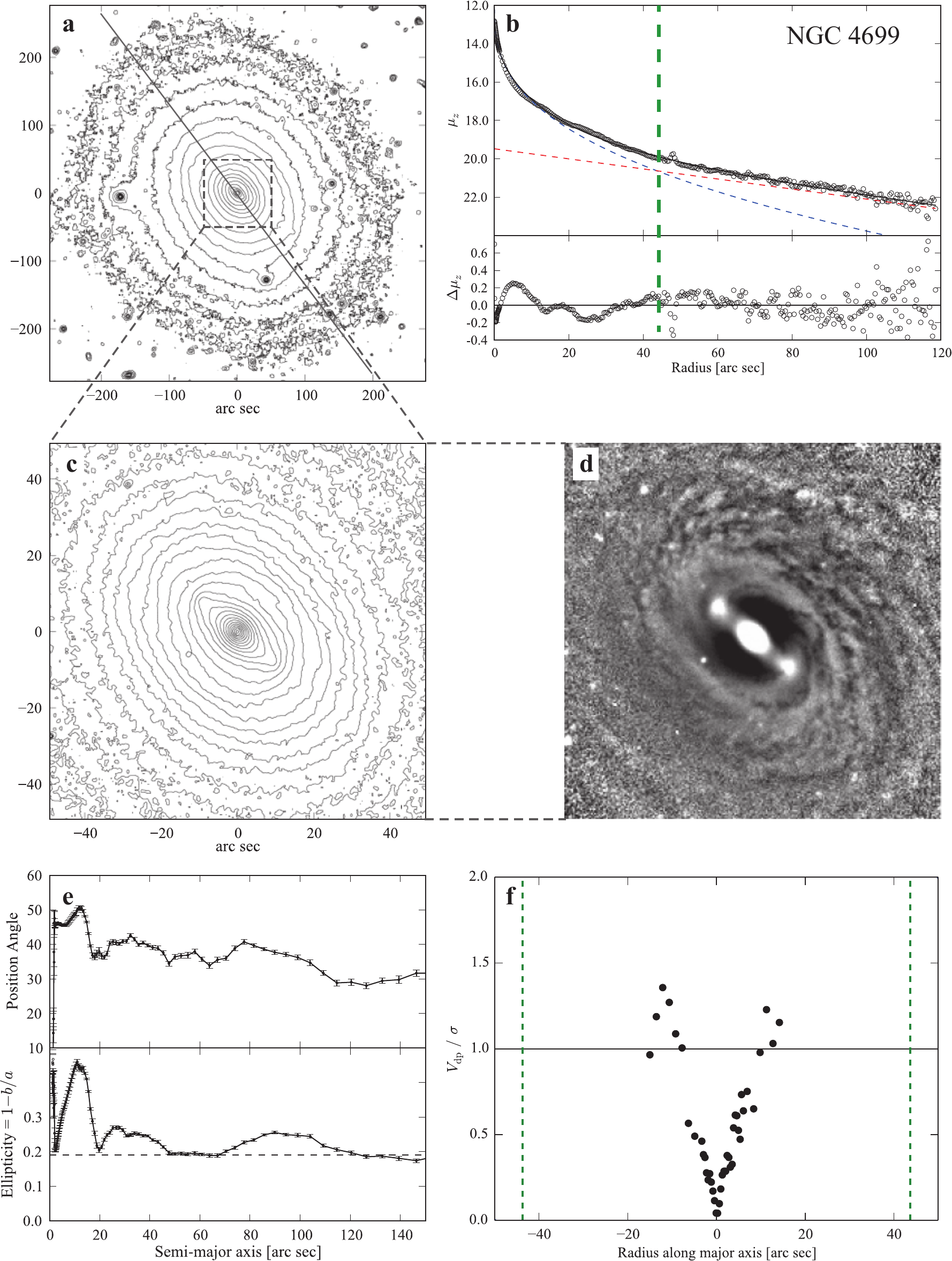}

\caption{The disky pseudobulge in the Sb galaxy NGC~4699. \textbf{a:}
log-scaled $i$-band isophotes (SDSS image, smoothed with 15-pixel-wide median
filter); gray line marks major axis (PA $= 37\degr$). \textbf{b:} Bulge-disk
decomposition of SDSS $z$-band major-axis profile.  Dashed lines represent
S\'ersic + exponential fit to the data, with residuals plotted in lower
sub-panel. Vertical dashed green line marks ``bulge=disk'' radius \rbd, where
the S\'ersic and exponential components are equally bright; this sets the
boundary of the ``photometric bulge''. \textbf{c:} Close-up of photometric bulge
region (log-scaled contours from SDSS $z$-band image, smoothed with 3-pixel-wide
median filter). \textbf{d:} Unsharp mask of the $z$-band image ($\sigma = 20$
pixels), showing the bar and spiral immediately structure outside it.
\textbf{e:} Ellipse fits to the SDSS $i$-band image; note that the ellipticity
in the photometric bulge region remains essentially unchanged from the outer
disk value (horizontal dashed line), except for the peak associated with the
bar, indicating the photometric bulge has flattening similar to the outer disk.
\textbf{f:} Deprojected stellar rotation velocity divided by local velocity
dispersion $\vdivsigma$ along the major axis, using long-slit data from
\citet{bower93}. Vertical dashed lines mark the photometric bulge region $|R| <
\rbd$. Although the kinematics do not extend very far out from the centre,
$\vdivsigma$ \textit{does} rise to $\sim 1.35$ within this region, indicating a
kinematically cool region more like a disk.}\label{fig:n4699a}

\end{figure*}

This is a luminous, intermediate-type (Sb) spiral galaxy with a very
extended, Type~III outer disk \citep{erwin08}. In our analysis, we
ignore the very outer part of this disk (visible in the azimuthally
averaged profile of Erwin et al.\ at $r \ga 120$ arcsec), and so our
``outer'' decomposition (panel~b of Figure~\ref{fig:n4699a}) is
restricted to $r < 120$ arcsec. This fit yields a significant
photometric bulge which dominates the light at $r \la 45$ arcsec; this
clearly corresponds to what \citet{carnegie} call ``the bulge'' of this
galaxy.

Inspection of the photometric bulge region shows that it is as
elliptical as the outer disk, or even more so (panels c and e of
Figure~\ref{fig:n4699a}), and is dominated by tightly wrapped spiral
arms and a relatively strong bar (panels c and d of the same figure).
Although the long-slit stellar kinematics of \citet{bower93} only extend
to $r \sim 20$ arcsec, they clearly show that \vdivsigma{} reaches
values $> 1$ in the photometric-bulge region, so we are confident in
identifying a disky pseudobulge in NGC~4699.

\begin{figure*}
\includegraphics[width=6.0in]{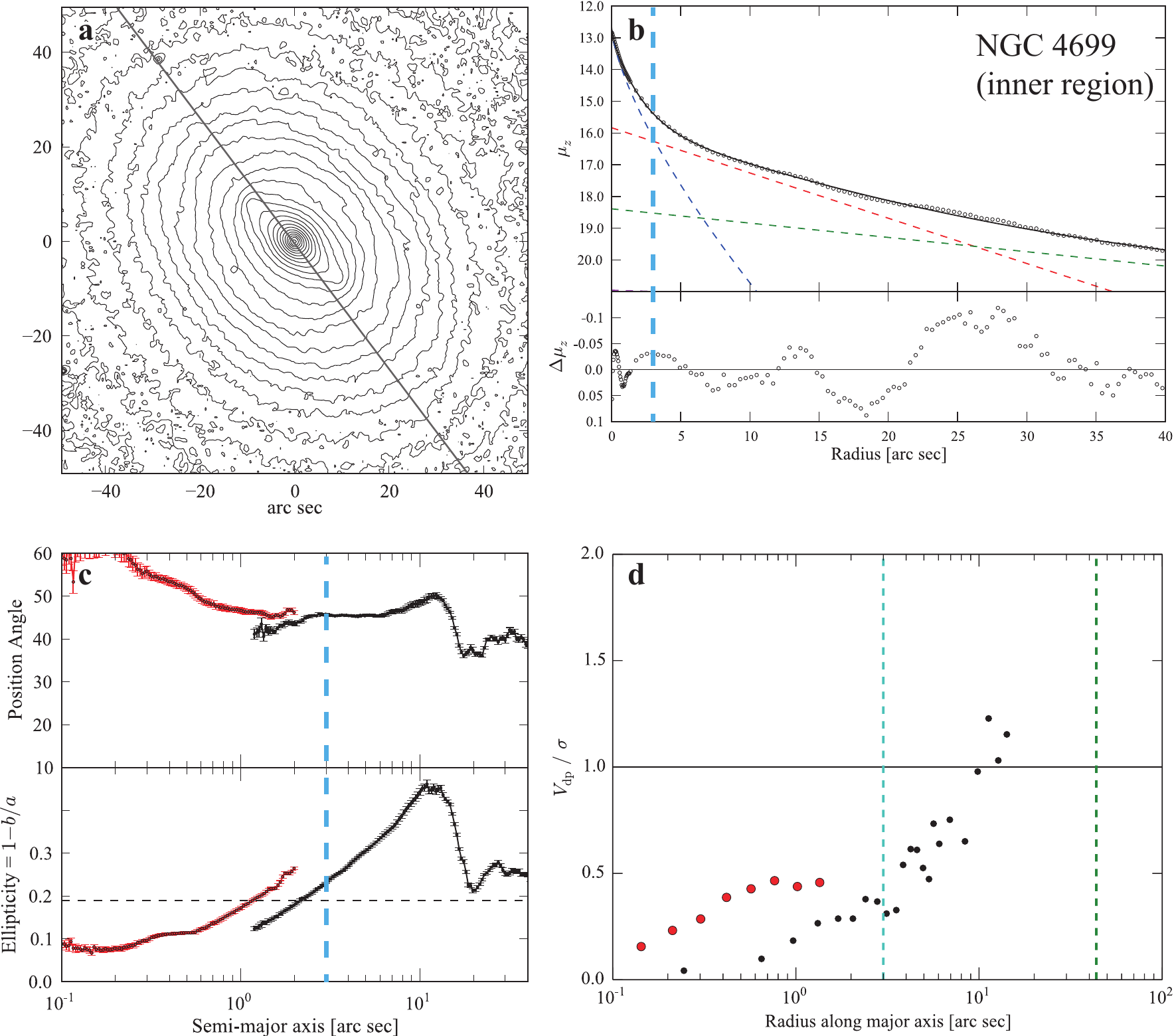}

\caption{The classical bulge inside the disky pseudobulge of NGC~4699.
\textbf{a:} Close-up of main photometric bulge region from SDSS $z$-band image;
gray line marks major-axis PA. \textbf{b:} major-axis profile combining data
from SINFONI $K$-band image ($r < 1.3$ arcsec) and SDSS $z$-band and $i$-band images, with fit
(S\'ersic + multiple exponentials; dashed lines) and residuals from
fit in lower sub-panel. Vertical dashed blue line marks inner ``bulge=disk'' radius
\rbdi. \textbf{c:} Ellipse fits to the SINFONI $K$-band [red] and SDSS
$z$-band images; note that the ellipticity in the inner photometric bulge region
($a \la 3$ arcsec) is $\la 0.15$, less than that of the disky pseudobulge
outside. \textbf{d:} Deprojected, folded stellar rotation velocity divided by
local velocity dispersion $\vdivsigma$ along the major axis, using long-slit data from
\citet{bower93} [black] and SINFONI AO data [red]. Vertical dashed
blue and green lines mark the inner and main photometric bulge regions $|R| <
\rbdi$, respectively. In the inner region ($r \la 3$ arcsec), $\vdivsigma$
remains well below 1, suggesting a kinematically hot region (i.e., a classical
bulge).}\label{fig:n4699b}

\end{figure*}

Figure~\ref{fig:n4699b} shows the previously identified photometric
bulge region. In the case of this galaxy, the major axis is close to the
bar's major axis, but this produces only a very weak bump in the
major-axis profile, similar to that due to spiral structure
further out, so we do not include a separate component for the bar. The
resulting fit (including two exponential components for the outer disk)
is rather good, with a clear inner excess (modeled as a S\'ersic
component with $n = 1.4$) creating an inner photometric-bulge zone
with $\rbdi = 2.8$ arcsec. The stellar kinematics from the ground-based
data of \citet{bower93} suggest a relatively low value of \vdivsigma{}
within \rbdi, and the SINFONI AO data appear to confirm this, since the
SINFONI \vdivsigma{} values reach a plateau of only $\sim 0.5$.
Strictly speaking, we cannot rule out the possibility that \vdivsigma{}
might reach values $\sim 1$ between $r \approx 1.5$ and 3 arcsec,
since this is beyond the range of our SINFONI data but within a region
where seeing effects might reduce \vdivsigma{} from the ground-based
data. None the less, we suggest that the inner photometric bulge region of
NGC~4699, associated with very round isophotes (panel~c of
Figure~\ref{fig:n4699b}) and \vdivsigma{} values $\la 0.5$, is most
likely another example of a compact classical bulge.

Our decomposition and identification of the classical bulge is similar
to that of \citet{weinzirl09}, who performed a 2D bulge/bar/disk
decomposition of an $H$-band image of NGC~4699 from the OSU Bright
Spiral Galaxy Survey \citet{eskridge02}. (Inspection of the OSU-BSGS
$H$-band image shows that the main disk of the galaxy is not visible in
that image, so Weinzirl et al.\ treated the disky pseudobulge as the
galaxy's only disk.)  They found a bulge with $n = 2.08$ and $r_e =
2.62$ arcsec -- values within $\sim 35$ per cent of the corresponding values
from our fit.

\subsection{Ambiguous Case: NGC~1068}\label{sec:n1068} 

In most respects, the Sb galaxy NGC~1068 appears to be similar to the previous
composite-bulge galaxies: it has a large, disky pseudobulge combined with a much
smaller central component. The sole ambiguity concerns the morphology of the central
component -- is it rounder than the galaxy disk, or is it more of a nuclear disk? --
and stems from uncertainty about the galaxy's inclination.

Both \citet{schinnerer00}, using \hi{} data, and \citet{davies07}, using stellar
kinemetry analysis of their SINFONI data, suggest an inclination of $\approx
40\degr$, which implies an observed ellipticity of $\approx 0.22$ for an
early-type disk. On the other hand, \citet{gutierrez11} used SDSS images and
found an outer-disk ellipticity of $\sim 0.14$, implying an inclination of $\sim
31\degr$. Since, as we will see below, the inner component has an ellipticity
of $\sim 0.15$, the question of whether the inner component is disklike or spheroidal
depends on the adopted galaxy inclination -- thus, we consider this galaxy a
somewhat ambiguous case, though we include it with the other composite-bulge
galaxies in our analysis.

\subsubsection{Disky Pseudobulge} 

Our global analysis of NGC~1068's surface-brightness profile uses the
light at $r \la 100$ arcsec and is very similar to that carried out by
\citet{shapiro03} for this galaxy (their Fig.~2),  As both \citet{pt06}
and \citet{gutierrez11} showed, there is considerable stellar light
further out in a shallower profile, much of which is due to the outer
pseudoring. The global fit for this galaxy is thus similar to those 
for NGC~3945 and NGC~2859, where we deliberately exclude light due to
extended outer rings.

\begin{figure*}
\includegraphics[width=6.0in]{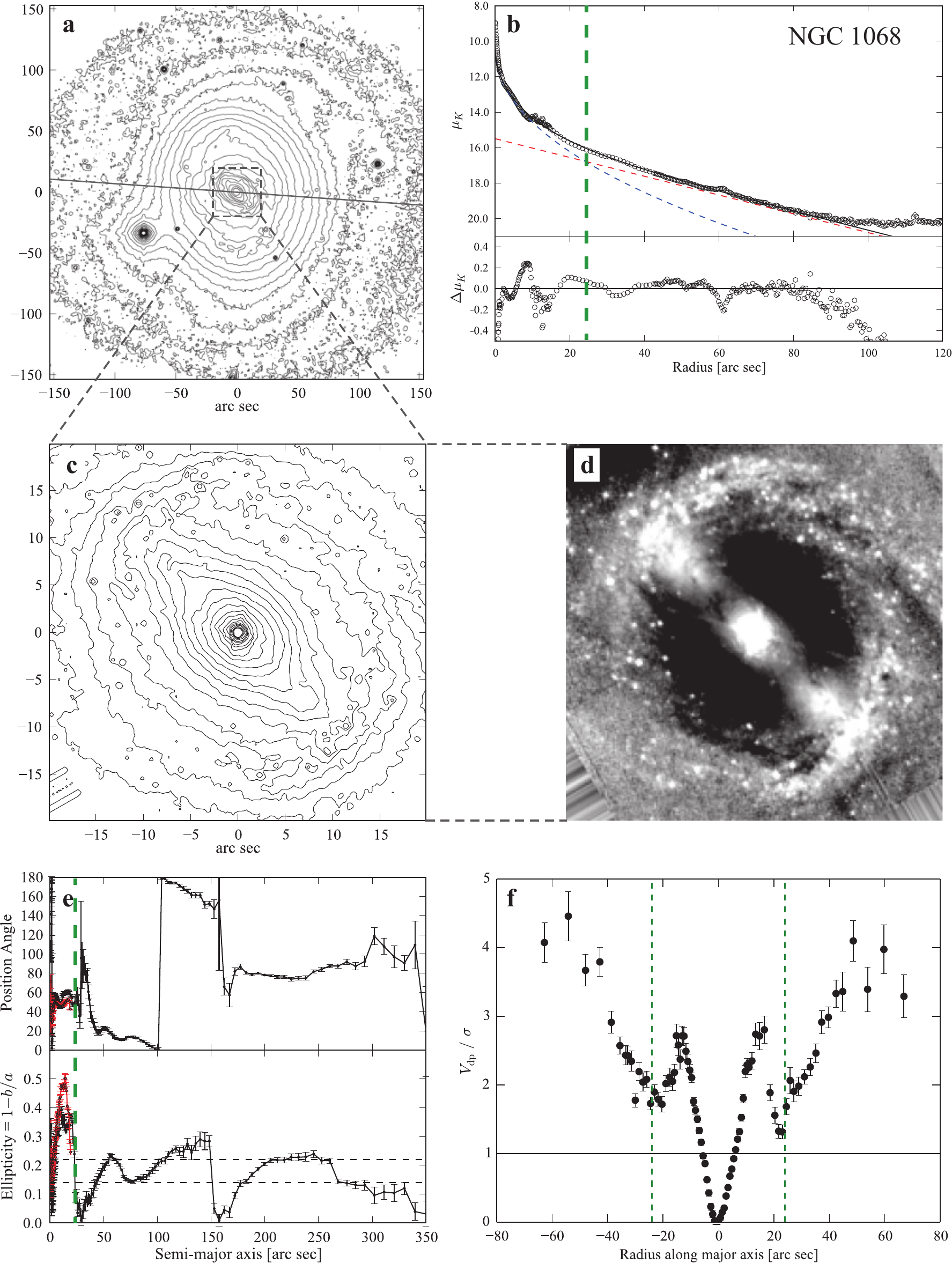}

\caption{The disky pseudobulge in the Sb galaxy NGC~1068. \textbf{a:}
log-scaled $i$-band isophotes (SDSS image, smoothed with 7-pixel-wide
median filter); gray line marks major axis (PA $= 86\degr$). \textbf{b:}
Bulge-disk decomposition of major-axis profile (profile combines
AGN-subtracted SINFONI data from \citet{davies07} for $r < 1.3$ arcsec,
\textit{HST} NICMOS3 F200N for $r = 1.3$--15 arcsec, 2MASS $K$-band for
$r = 15$--42 arcsec, SDSS $i$-band for $r > 42$ arcsec).  Dashed lines
represent S\'ersic + exponential fit to the data, with residuals plotted
in lower sub-panel. Vertical dashed green line marks ``bulge=disk''
radius \rbd, where the S\'ersic and exponential components are equally
bright; this sets the boundary of the ``photometric bulge''. \textbf{c:}
Close-up of photometric bulge region (log-scaled contours from NICMOS3
image). \textbf{d:} Unsharp mask ($\sigma = 20$ pixels) of same region,
showing bar and spiral arms. \textbf{e:} Ellipse fits to SDSS $i$
(black) and NICMOS (red) images. \textbf{f:} Deprojected stellar
rotation velocity divided by local velocity dispersion $\vdivsigma$
along major axis, using long-slit data from \citet{shapiro03}. Vertical
dashed lines mark the photometric bulge region $|R| < \rbd$;
$\vdivsigma$ rises to almost 3 in this region.}\label{fig:n1068a}

\end{figure*}

The decomposition (Figure~\ref{fig:n1068a}, panel~b) is relatively
straightforward, and the resulting photometric bulge has $\rbd =
24$ arcsec. The morphology interior to this radius is strikingly
disklike: the near-IR stellar light is dominated by the very strong
(inner) bar first noted by \citet{scoville88}, with star-forming spiral
arms forming a pseudoring just outside the bar (panels~c and d); due to
this bar, the isophotes have an ellipticity much higher than that of the
outer disk (panel~e).

The long-slit stellar-kinematic data of \citet{shapiro03} show
$\vdivsigma > 1$ at all radii $\ga 1$ arcsec, and in fact the ratio
reaches a value of almost $\sim 3$ at $r \sim 15$ arcsec (panel~f), so this
an even more extreme case of a photometric bulge with disklike
kinematics than NGC~3945. We are certainly not the first to
suggest that the kinematics in this part of the galaxy are more disklike
than bulgelike: in particular, \citet{emsellem06} pointed this out using
2D stellar kinematics from the SAURON instrument, in conjunction with
$N$-body/hydrodynamical modeling.

\subsubsection{Nuclear Disk or Classical Bulge?} 

\begin{figure*}
\includegraphics[width=6.0in]{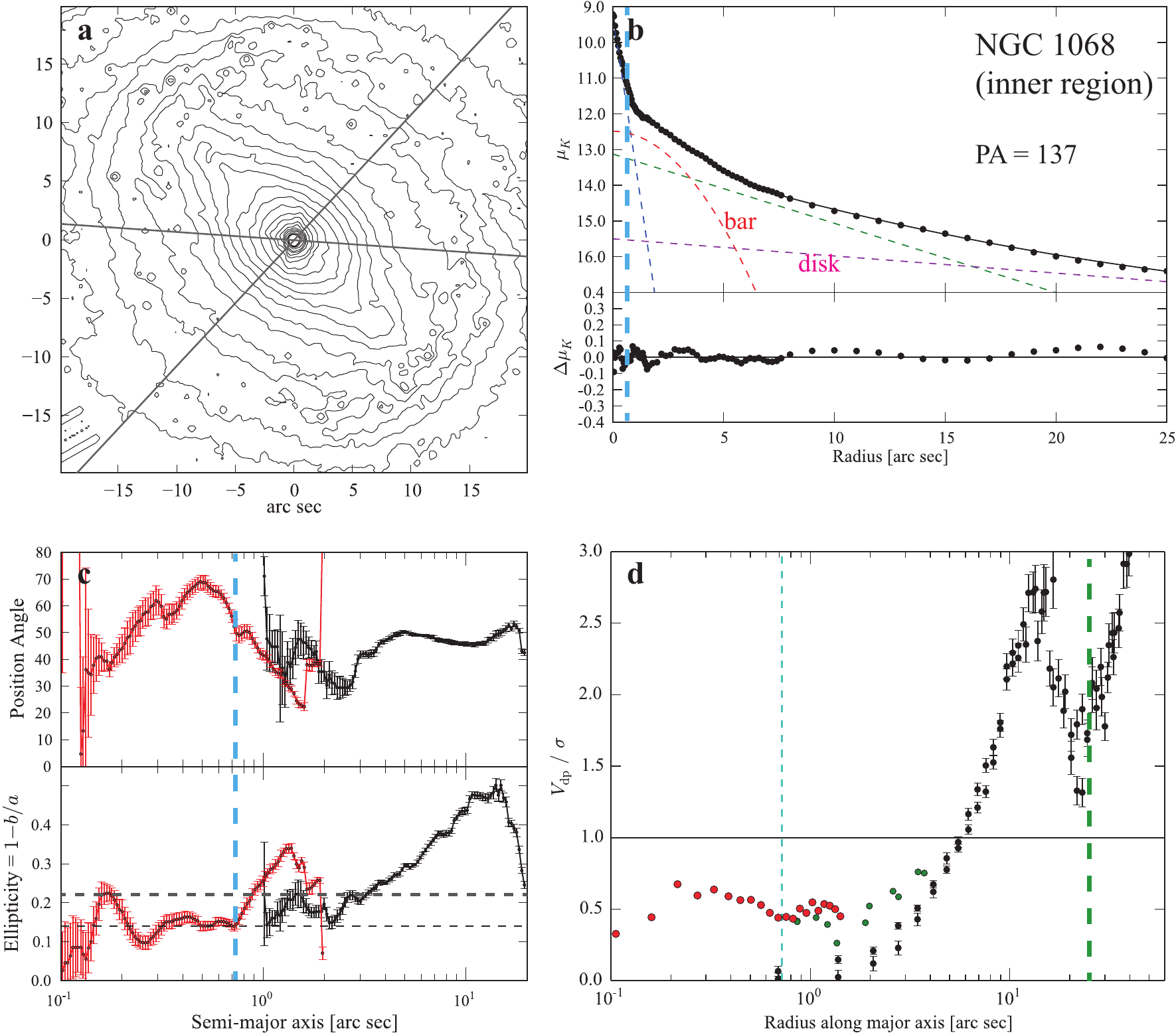}

\caption{Evidence for a possible classical bulge inside the disky
pseudobulge of NGC~1068. \textbf{a:} Close-up of main photometric bulge
region from \textit{HST} NICMOS3 F200N image; gray lines marks
major-axis PA (86\degr) and PA perpendicular to inner bar (137\degr).
\textbf{b:} Profile at 137\degr{} from AGN-subtracted SINFONI $H$-band
image of \citet{davies07} ($r < 1.7$ arcsec) and NICMOS3 F200N image,
with fit (S\'ersic + Gaussian + exponential + exponential; dashed lines)
and residuals from fit in lower sub-panel. Vertical dashed blue line
marks inner ``bulge=disk'' radius \rbdi. \textbf{c:} Ellipse fits to the
SINFONI [red] and NICMOS3 [black] images; the two estimates for the
outer-disk ellipticity (0.13 and 0.22) are marked by the horizontal
black and gray lines.  \textbf{d:} Deprojected, folded stellar rotation
curve divided by local velocity dispersion \vdivsigma{} along the galaxy
major axis. Data from \citet[][black]{shapiro03}, GMOS IFU kinematics of
\citet[][green]{gerssen06} and SINFONI IFU kinematics of
\citet[][red]{davies07}. }\label{fig:n1068b}

\end{figure*}

Optical and near-IR images of NGC~1068 are strongly contaminated by
light from the bright AGN. To get around this problem, we make use of an
$H$-band image from the 100mas SINFONI observations (1.5 arcsec field
of view) of \citet{davies07}, where the AGN emission (assumed to come
mostly from hot dust in the circumnuclear torus) has been
spectroscopically modeled and removed, leaving behind a ``pure stellar''
continuum which is then collapsed along the wavelength axis to form an
image. We matched the surface-brightness profile extracted from this
image to a profile from a larger-scale \textit{HST} NICMOS3 F200N
image.

For the stellar kinematics, we supplement the large-scale (but
low-spatial-resolution) data of \citet{shapiro03} with data from the
Gemini GMOS IFU observations of \citet{gerssen06}, which had a seeing of
FWHM $= 0.5$ arcsec, and with the VLT-SINFONI kinematics of
\citet{davies07}, observed in AO mode with a corrected FWHM of
0.10 arcsec.

We perform our inner decomposition by choosing a position angle which
avoids most of the inner bar (PA $= 137\degr$, perpendicular to the
bar). Since the galaxy is seen at a relatively low inclination, the
projection effects are relatively small; however, this does mean that
our decomposition does not exactly mirror the major-axis stellar
kinematics.

Comparison of the NICMOS3 image with the PA $= 137\degr$ profile shows
that the increase in surface brightness starting at $r \sim 6$ arcsec is
actually due to the profile crossing into the bright part of the bar,
along its minor axis. Since we consider bars to be disk phenomena, this
excess light should be treated as part of the disky pseudobulge. There
is a second, much steeper increase in the profile which sets in for $r
\la 1$ arcsec, associated with rounder isophotes; this is the best
candidate for an embedded bulge.

Our best fit to the surface-brightness profile thus uses \textit{three}
components: an exponential for the main part of the pseudobulge outside
the bar itself (in analogy with large-scale bars, one could perhaps call
this a ``lens''); a Gaussian for the minor axis profile of the
bar;\footnote{Evidence that the minor-axis profiles of at least some
bars are Gaussian can be found in \citet{ohta90} and \citet{prieto97}.}
and a S\'ersic for the innermost component. The innermost (S\'ersic)
component corresponds to the ``extra emission'' at $r < 1$ arcsec noted
by \citet{davies07}, although we identify their 1--5 arcsec
``$R^{1/4}$ bulge'' profile as being due to the inner bar, not to a
larger-scale bulge.  The final $R_{e}$ value recorded in
Table~\ref{tab:classical-parameters} for this S\'ersic component has
been corrected to its major-axis value assuming an ellipticity of 0.15,
based on the ellipse fits (Figure~\ref{fig:n1068b}).

As in the other galaxies, the \vdivsigma{} profile shows
dispersion-dominated stellar kinematics in the inner $r < \rbdi$ region:
\vdivsigma{} reaches a maximum of only $\sim 0.7$.

Is the innermost component a compact classical bulge, or is it something
more disklike? \citet{davies07} argued for a disk, and suggested (based
on indirect arguments about the estimated $M/L$ ratio of the inner
component) that its stellar population was relatively young: $\sim 300$
Myr. If we adopt an inclination of 31\degr{} for the galaxy, then the
inner ellipticity of $\sim 0.15$ is consistent with that of a disk
having a flattening similar to the outer disk's.

However, \citet{crenshaw00} and \cite{storchi-bergmann12} have both
presented high-resolution spectroscopic evidence -- including fits to
the near-nuclear spectra -- indicating that the $r < 1$ arcsec region is
actually dominated by \textit{old} stars, with an age of at least 2 Gyr
according to Crenshaw \& Kraemer's optical STIS spectroscopy, and an age
of 5--15 Gyr according to the near-IR spectroscopy of Storchi-Bergmann
et al. If we combine this with the dispersion-dominated kinematics
\textit{and} assume an inclination of 40\degr{} for the galaxy, then the
central stellar component is rounder than a disk, kinematically hot,
\textit{and} made up of old stars -- a good case for a classical bulge,
albeit one with a nearly-exponential profile.

\subsection{Possible Case: NGC~1543} 

NGC~1543 is something of a borderline case, primarily because we do not
have a good handle on the stellar kinematics. The galaxy is so close to
face-on that it is difficult to determine the major axis with any
accuracy, and thus long-slit data cannot really be used. (We need to
know the major axis accurately in order to properly deproject the
observed velocities to their major-axis, edge-on values.) The only
available stellar kinematics are two long-slit observations by
\citet{jarvis88}. Since both their ``major-axis'' and ``minor-axis''
profiles (at position angles of 90\degr{} and 0\degr, respectively) show
stellar rotation, all we really can say is that the major axis is not
close to 0\degr{} or 90\degr. Thus we are unable to perform a proper
kinematic analysis for NGC~1543.

\begin{figure*}
\includegraphics[width=6.0in]{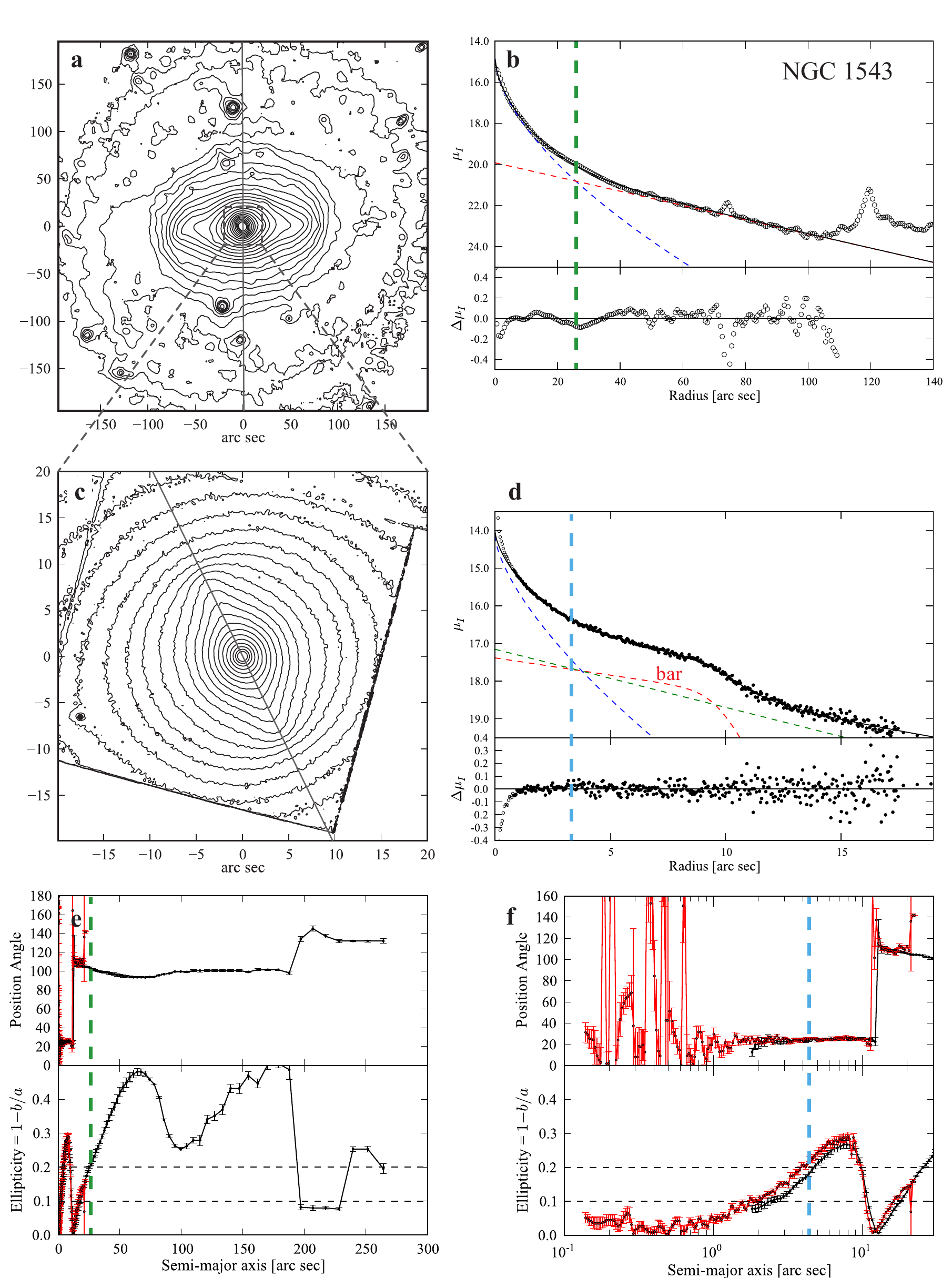}

\caption{Photometric and morphological evidence for a possible composite
bulge in the SB0 galaxy NGC~1543. \textbf{a:} Logarithmically scaled
isophotes from Spitzer IRAC2 image. \textbf{b:} Bulge-disk decomposition
of major-axis IRAC2 profile. Dashed lines represent S\'ersic +
exponential fit to the data, with residuals plotted in lower sub-panel.
Vertical dashed green line marks ``bulge=disk'' radius \rbd. \textbf{c:}
Close-up of photometric bulge region (WFPC2 F814W image). \textbf{d:}
Profile from WFPC2 image ($r < 18$ arcsec) and IRAC2 image ($r >
18$ arcsec) along the inner bar's major axis (PA $= 26\degr$), fit using
S\'ersic + broken-exponential (red, inner bar) + exponential fit (fit
also includes outer-disk exponential component, not visible here);
vertical dashed blue line marks inner ``bulge=disk'' radius \rbdi.
\textbf{e:} Ellipse fits to Spitzer IRAC2 data [black] and HST-WFPC2
F814W data [red]. \textbf{f:} Same as panel~e, but on a logarithmic
semi-major-axis scale. }\label{fig:n1543}

\end{figure*}

None the less, the \textit{morphology} of the galaxy is very suggestive of
a composite-bulge system (Figure~\ref{fig:n1543}). Full details of the
disky pseudobulge decomposition, including analysis of the structure of
the inner bar, will be presented elsewhere \citep{erwin14-dissect}; see
\citet{erwin11} for a preliminary discussion. Because the the major axis
of the inner bar is relatively close to the minor axis of the outer bar,
we use a profile along the position angle of the inner bar for our
decomposition (as in the case of NGC~2859; see Section~\ref{sec:n2859}).

Figure~\ref{fig:n1543} summarizes both the global, ``naive'' bulge/disk
decomposition (panel~b) and the inner, composite-bulge decomposition
(panel~d). We construct our surface-brightness profile using an
\textit{HST} WFPC2 F814W image at small radii and a Spitzer IRAC2 image
at large radii (we use the IRAC2 image instead of the IRAC1 image
because the galaxy position on the former is slightly better for
purposes of measuring the sky background). Excluding the broad outer
ring (similar to those of NGC~2859 and NGC~3945), a decomposition of the
inner $\sim 100$ arcsec (panel~a) yields a photometric bulge with $n =
1.6$, $r_e = 9.4$ arcsec and $\rbd \approx 26$ arcsec; this is similar
to, but slightly larger than, the ``pseudobulge'' reported by
\citet{fisher-drory10} from their 1-D IRAC1 decomposition ($n = 1.51$,
$r_e = 6.5$ arcsec). The photometric bulge region thus defined ($r \la 26$ arcsec)
is dominated by a very strong inner bar, with a very weak stellar
nuclear ring surrounding it \citep{erwin14-dissect}. We take this as strongly
suggestive evidence for a disky pseudobulge, though as noted above we
lack the necessary kinematic data for full confirmation.

In the central regions, the isophotes become very round (panels~c and f
of Figure~\ref{fig:n1543}), and there is a central excess of light above
the inner bar. Simultaneous decompositions along the major and minor
axes of the inner bar support the presence of an inner S\'ersic
component with $n = 1.5$, $r_e = 2.7$ arcsec and ellipticity $\la 0.05$
(panel~d of the figure shows the decomposition along the inner-bar major
axis). So there is morphological and photometric evidence for a
distinct, classical-bulge-like component in the centre of this galaxy as
well. There is in addition evidence for a very compact, distinct feature
inside the classical bulge, which is probably a nuclear star cluster
(see \citealt{erwin11} and \citealt{erwin14-dissect}).

\subsection{The Uncertain Case of NGC~3489}\label{sec:n3489} 

\begin{figure*}
\includegraphics[width=6.0in]{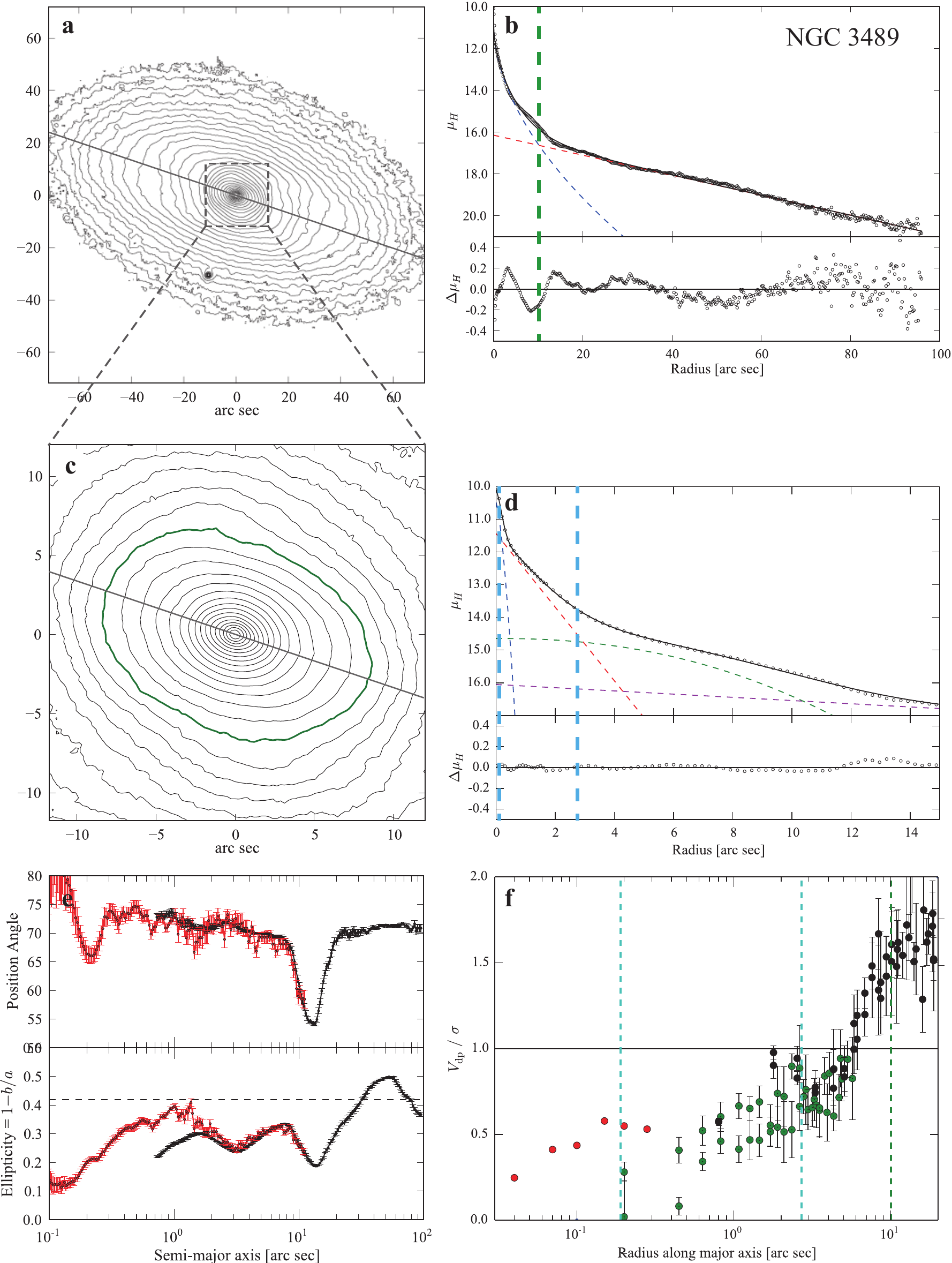}

\caption{Possible kinematically hot, disky pseudobulge in the S0 galaxy
NGC~3489. \textbf{a:} log-scaled INGRID $H$-band isophotes (smoothed
with 9-pixel-wide median filter); gray line marks major axis (PA $=
71\degr$). \textbf{b:} Bulge-disk decomposition of major-axis profile,
from $H$-band and WFPC2 F814W images. Dashed lines show S\'ersic +
exponential fit, with residuals in lower sub-panel. Vertical dashed
green line marks \rbd. \textbf{c:} Close-up of photometric bulge region;
thick green contour shows approximate location of bar's projected
box/peanut structure. \textbf{d:} Decomposition of inner major-axis
profile, using outer exponential + S\'ersic (for B/P structure) + inner
exponential + inner S\'ersic; residuals from fit are in lower sub-panel.
Vertical dashed blue lines indicate \rbd{} for the inner exponential and
the inner S\'ersic components. \textbf{e:} Ellipse fits to $H$-band
image (black) and WFPC2 F814W image (red); horizontal dashed line
indicates outer-disk ellipticity. \textbf{f:} Deprojected stellar
rotation velocity divided by local velocity dispersion $\vdivsigma$
along the major axis, using SAURON data (black) from \citet{emsellem04},
OASIS data (green) from \citet{mcdermid06} and SINFONI AO data from
\citet{nowak10}. Vertical dashed lines mark the previously identified
photometric bulge regions $|R| < \rbd$.}\label{fig:n3489}

\end{figure*}

\citet{nowak10} considered two cases of composite-bulge galaxies: the
double-barred spiral NGC~3368 and the barred S0 galaxy NGC~3489.
Although we confirm the previous classification of NGC~3368 (see
Section~\ref{sec:n3368}, above), NGC~3489 has proved to be more
ambiguous (Figure~\ref{fig:n3489}).

Careful analysis of the isophotes of NGC~3489 suggests that the broad,
slightly boxy region inside the bar may well be an example of a
projected box/peanut (B/P) structure (outlined isophote in panel~a of
Figure~\ref{fig:n3489}), even though \citet{erwin-debattista13} did not
classify this galaxy as such. The presence of a B/P structure certainly
does not prevent the simultaneous existence of a composite bulge in this
galaxy -- indeed, Section~\ref{sec:boxy} argues that NGC~3368 has a
clear B/P structure in addition to its composite bulge --  but it does
mean that the major-axis decomposition in \citet{nowak10} needs to be
revised, since what they considered to be the inner-disk component,
dominating the major-axis light from $\approx 4$--12 arcsec, is more
likely to be a combination of the B/P structure and the lens region
immediately outside the bar.

For our revised inner decomposition (panel~d of Figure~\ref{fig:n3489}),
we include the contribution from the outer disk component (panel~b) and treat the
B/P structure as a separate component with a S\'ersic profile; we also
include an additional S\'ersic component for the nuclear luminosity
excess which \citet{nowak10} modeled with a Gaussian.  The
resulting fit has an inner exponential with scale length $\sim 55$~pc,
corresponding to the original ``classical bulge'' component in
\citet{nowak10}, as well as an extremely compact S\'ersic component with
$n = 0.7$ and $R_{e} = 0.19$ arcsec. The \rbd{} radii for these two
components are 2.7 and 0.2 arcsec, respectively. 

As shown by the ellipticity plot (panel~e), the inner exponential
component, which dominates the light at $r \la 2.7$ arcsec, is in fact
highly elliptical -- almost as elliptical as the outer disk. This is
very similar to the disky pseudobulges identified in our other galaxies.
On the other hand, the stellar kinematics in this region (panel~f) are
rather dispersion-dominated, with $\vdivsigma \sim 0.7$; even at the
outer edge of this region, \vdivsigma{} is never larger than $\sim 0.9$.
So this structure is both flattened like a disk and at the same time
kinematically \textit{hot}, making it unlike any of the disky
pseudobulges in the other galaxies of our sample.

The innermost (S\'ersic) component is associated with rounder isophotes
and a lower \vdivsigma{} value, but it is so small\footnote{The S\'ersic
$R_{e}$ is $\approx 9$~pc, but this does not account for PSF
convolution.} that it could more plausibly be classified as a nuclear
star cluster than as a genuine classical bulge.

We are thus left with a curiosity: our revised analysis identifies a
structure in NGC~3489 which morphologically resembles a disky pseudobulge,
but is kinematically hot, along with a distinct nuclear component that might
best be classified as a nuclear star cluster. NGC~3489 may thus be a
kind of transition object, and an indication that the central regions of
early-type disk galaxies can be even more diverse and complicated than our
main ``composite-bulge'' argument suggests.

\clearpage

\section{WHT-ISIS Spectroscopy}\label{sec:data-spec}

\begin{table}
\caption{WHT-ISIS Instrumental Setup}
\label{tab:isis-setup}
\begin{tabular}{@{}lll}
\hline
Parameter & NGC~2859 & NGC~4371 \\
\hline
Date                                           & 2000 Dec 13    &  2001 Jun 4 \\ 
Grating ($\rm lines\;mm^{-1}$)                 & R600B          & R600B \\ 
Central wavelength (\AA)                       & 5498           & 5498  \\
Detector                                       & EEV12          & EEV12 \\
Pixel binning                                  & $1\times1$     & $1\times1$ \\  
Scale (arcsec pix$^{-1}$)                   & 0.36           & 0.36 \\ 	   
Reciprocal dispersion (\AA{} pix$^{-1}$)       & 0.45           & 0.45 \\ 
Slit width (arcsec)                           & 0.93           & 1.2 \\
Slit length (\arcmin)                          & 4.0            & 4.0 \\
Slit position angle (\degr)                    & 85             & 92 \\
Instrumental $\sigma$ (\kms)                   & 60             & 60  \\ 
Seeing FWHM (arcsec)                          & 1.0            & 1.0 \\ 
Galaxy exposure                                & 3$\times$1200s      & 3000s + \\
                                               &                & 2$\times$2400s \\
Sky exposure                                   & 900s           & 1200s\\
\hline 
\end{tabular}

\medskip
Instrumental setup for the blue arm of the WHT-ISIS spectrograph, as used for
observations of NGC~2859 and NGC~4371.
\end{table}

We obtained long-slit spectroscopy of NGC~2959 with the ISIS
spectrograph of the William Herschel Telescope on 2000 December 13,
using the blue arm with the R600B grating; further details of the
observing setup are listed in Table~\ref{tab:isis-setup}. Because the
galaxy was large enough to fill most of the slit, we obtained a separate
900s exposure of the nearby blank sky, offset 6\arcmin{} to the east of
the galaxy centre. We also observed two kinematic standard stars (HR941
and HR2660) with the same setup.

NGC~4371 was observed with an almost identical instrumental setup as
part of ING service-time observations on 2001 June 4 (see
Table~\ref{tab:isis-setup}). A separate 1200s sky exposure was taken
with a 3\arcmin{} offset to the north of the galaxy centre, and observations
of four kinematic standard stars (HR5200, HR5966, HD5340, HR5340) were
also obtained.

Following standard \textsc{midas}\footnote{\textsc{midas} is developed and maintained by
the European Southern Observatory.} reduction of the ISIS observations,
including bias-subtraction, flat-fielding and wavelength calibration
using CuAr and CuNe+CuAr lamp exposures, the extracted galaxy spectra
were analysed using the Fourier Correlation Quotient method
\citep{bender90,bender94} in order to determine the stellar kinematics. The
resulting kinematic values -- radial velocity, velocity dispersion and
the Gauss-Hermite coefficients $h_{3}$ and $h_{4}$ -- are presented in
Table~\ref{tab:isis-kinematics}.

\begin{table*}
\begin{minipage}{126mm}
\caption{WHT-ISIS Major-Axis Stellar Kinematics for NGC~2859 and NGC~4371}
\label{tab:isis-kinematics}
\begin{tabular}{@{}lrrcccc}
\hline
Galaxy   & PA      & $R$      &  $V$            & $\sigma$       & $h_{3}$          & $h_{4}$ \\
         & \degr{} & arcsec   & \kms{}          & \kms{}         &                  &  \\
\hline
NGC~2859 & 85      & $-$14.70 & $1554.6 \pm 39.5$ & $203.1 \pm 35.2$ & $-0.124 \pm 0.081$ & $-0.072 \pm 0.149$ \\
NGC~2859 & 85      & $-$9.03  & $1586.1 \pm 32.6$ & $194.3 \pm 28.1$ & $-0.125 \pm 0.067$ & $-0.026 \pm 0.120$ \\
NGC~2859 & 85      & $-$6.03  & $1552.8 \pm 13.4$ & $160.4 \pm 10.3$ & $-0.083 \pm 0.050$ & $-0.019 \pm 0.065$ \\
NGC~2859 & 85      & $-$4.50  & $1575.0 \pm 11.2$ & $151.2 \pm  8.2$ & $-0.076 \pm 0.044$ &  $0.085 \pm 0.057$ \\
\hline 
\end{tabular}

\medskip
Stellar kinematics along the major axes of NGC~2859 and NGC~4371, as determined from spectra
obtained with WHT-ISIS; see Table~\ref{tab:isis-setup} for the instrumental setup.
This table is published in its entirety in the online edition of this paper; a sample is
provided here for guidance in terms of form and content.

\end{minipage}
\end{table*}

\end{document}